\documentclass[11pt]{LMA}

\usepackage{physics}      
\usepackage{slashed}
\usepackage{booktabs}
\usepackage{multirow}
\usepackage{xcolor}

\allowdisplaybreaks

\newcommand{\sw}{s_W}
\newcommand{\cw}{c_W} 
\newcommand{\MZ}{m_Z}
\newcommand{\MW}{m_W}
\newcommand{\MH}{m_H}
\newcommand{\gL}{g_L}
\newcommand{\gR}{g_R} 
\newcommand{\klam}{\kappa_\lambda}
\newcommand{\Ma}{\emm_{a}}
\newcommand{\Mb}{\emm_{b}}
\newcommand{\Mbt}{\emm_{\tilde b}}
\newcommand{\Mhhh}{\emm_{hhh}}
\newcommand{\Msq}{\overline{\abs{\emm}^2}}  
\newcommand{\Msqa}{\overline{\abs{\Ma}^2}}
\newcommand{\lam}{\lambda} 
\newcommand{\eps}{\varepsilon} 
\newcommand{\GeV}{\,\text{GeV}}
\newcommand{\fb}{\,\text{fb}} 
\newcommand{\PL}{P_L} 
\newcommand{\PR}{P_R}

\numberwithin{equation}{section}
\numberwithin{table}{section} 
\numberwithin{figure}{section}

\begin{document}

\title{Quantum tomography of Higgsstrahlung: Higgs boson self-coupling and $Z$ boson anomalous couplings}

\author[a]{Clelia Altomonte
\thanks{clelia.altomonte@kbfi.ee}
}

\author[b]{Marco Fabbrichesi
\thanks{marco.fabbrichesi@ts.infn.it}
} 

\author[a]{Niko Koivunen
\thanks{niko.koivunen@kbfi.ee}
}

\author[a c]{Luca Marzola
\thanks{luca.marzola@cern.ch}
}

\author[a]{Daniele Rizzo
\thanks{daniele.rizzo@kbfi.ee}
}

\affil[a]{\KBFI}

\affil[b]{INFN, Sezione di Trieste, Via Valerio 2, I-34127 Trieste, Italy}

\affil[c]{Institute of Computer Science, University of Tartu, Narva mnt 18, 51009 Tartu, Estonia}

\date{\today}
\maketitle


\begin{abstract}
The Higgs boson self-coupling enters the Higgsstrahlung process $e^+e^-\to ZH$ at one loop. The three diagrams that carry it project onto two Lorentz structures: the tree-level one, which rescales the rate, and that of the CP-even anomalous coupling $b_Z$, which changes the polarization of the final-state $Z$ boson.

To assess how well quantum state tomography constrains the $hZZ$ anomalous couplings and the trilinear coupling at FCC-ee, we compute the helicity density matrix of the $Z$ boson, for an effective vertex that also includes a CP-odd anomalous coupling $\tilde b_Z$, together with its complete one-loop correction in the Standard Model. With these corrections and no other new physics in the effective vertex, quantum state tomography determines $\klam$ to $\pm 3.9$, or $\pm 2.9$ with the full angular information, independently of the rate and of the normalization of the vertex, and not far from the limit $\delta\klam \simeq 1$ set by the quantum Fisher information. Without assumptions on the new physics, quantum tomography directly bounds the anomalous couplings at the $10^{-3}$ level, improving on the existing projections. 

The density matrix also characterizes the $Z$ boson as a qutrit. At $365\GeV$ its state in the central region is pure enough to violate the Klyachko--Can--Binicio\u{g}lu--Shumovsky contextuality inequality, $\mathbb{CNTXT}_5 = 2.06$ at $\theta = \pi/2$ against the non-contextual bound $2$. We also quantify the purity and the amount of magic of the $Z$ boson state, using the mana, the stabilizer R\'enyi entropy and the robustness of magic for the latter. At fixed angle, purity, contextuality and magic are robust against anomalous couplings of the size probed by the tomography. The inclusive state, instead, is neither contextual in this test nor magic in the helicity basis. 
\end{abstract} 

\newpage
\tableofcontents
\newpage

\section{Introduction} 
\label{sec:intro}

The trilinear coupling of the Higgs boson is the last parameter of the Standard Model (SM) Lagrangian that still lacks a sharp experimental determination. Its deviation from the SM value is typically parametrized by
\begin{equation}
	\klam \deq \frac{\lambda_{hhh}}{\lambda_{hhh}^{\rm SM}},
\end{equation}
where $\lambda_{hhh}$ is the Higgs boson self-coupling and $\lambda_{hhh}^{\rm SM}$ its SM value. A direct measurement requires the production of Higgs boson pairs, which is difficult at the LHC because of the small cross section, about 30 fb, and of the QCD multijet background to the dominant $4b$ final state. An indirect route, proposed in Ref.~\cite{McCullough:2013rea} and developed in Refs.~\cite{Degrassi:2016wml,DiVita:2017vrr}, exploits the fact that $\klam$ enters every single-Higgs process at one loop. At an $e^-e^+$ collider such as the FCC-ee, the cross section of Higgsstrahlung, $e^-e^+\to ZH$, grows by about $1.7\%$ per unit of $\klam$ at $\sqrt s = 240\GeV$, and the same rate is sensitive to the one-loop effects of new particles, such as vector-like fermions or scotogenic Yukawa interactions~\cite{Marzo:2023uxv,Marzo:2026ykm}. Since this cross section will be measured at the per-mille level, the rate alone determines $\klam$ to a precision of $28\%$, or $18\%$ in combination with the High-Luminosity LHC (HL-LHC), and to about $15\%$ if the luminosity at $240$ and $365\GeV$ is doubled~\cite{FCC:2025lpp,terHoeve:2025gey}. These figures come from global fits, with a sensitivity that depends on how the other operators are treated, and that can be up to three times weaker when they are marginalized over~\cite{terHoeve:2025gey} or, with flavor and electroweak inputs included, better than that of the HL-LHC~\cite{Maura:2025rcv}. In the longer term, Higgs boson pair production at the FCC-hh would determine $\klam$ to a few percent~\cite{FCC:2025lpp,Mastrapasqua:2026hh}.

The rate, however, is not the only observable of the process. The $Z$ boson is produced polarized, and its polarization can be reconstructed from the angular distributions of its leptonic and hadronic decays through a procedure known as \textit{quantum state tomography}~\cite{Fabbrichesi:2024wcd}. The result is the $3\times 3$ helicity density matrix of the $Z$ boson, which can be computed from the Feynman diagrams of the process and compared directly with the data. Being normalized, the density matrix does not depend on the luminosity, on the efficiencies, or on any correction that merely rescales the tree-level vertex. It depends instead on the \emph{Lorentz structure} of the $hZZ$ vertex, and therefore carries information complementary to that of the rate.

In this paper we use the tomography of the $Z$ boson helicity state to test the Lorentz structure of the $hZZ$ vertex. We describe the vertex with three form factors: $a_Z$, which multiplies the SM structure $g^{\mu\nu}$, and the CP-even and CP-odd anomalous couplings $b_Z$ and $\tilde b_Z$ of Refs.~\cite{Hagiwara:1993sw,Hagiwara:1993ck}. We then compute the one-loop SM corrections in this language. The three one-loop diagrams that contain the Higgs boson self-coupling add up to a correction of the vertex with only two structures: the tree-level one, with coefficient $c(s)$, which changes the rate, and the structure $q\vdot k\, g^{\mu\nu} - q^\nu k^\mu$ of $b_Z$, with coefficient $\beta(s)$, which changes the polarization. In other words, \emph{the Higgs boson self-coupling induces at one loop an effective anomalous coupling $b_Z = \klam\, \beta(s)$}, and the tomographic process can probe it.

To assess the reach of the method at the FCC-ee, we then set up a $\chi^2$ test that targets deviations in the Fano coefficients of the density matrix with respect to the SM predictions. If $\klam$ is the only new physics source in the vertex, the tomographic process determines it to $\pm 3.9$ at $68\%$ confidence level (CL), or $\pm 2.9$ with the full angular information. This result is independent of the normalization of the rate and not far from the limit set by the quantum Fisher information, but it is still much less precise than the determination from the rate itself. Without assumptions on the new physics, instead, the quantum state tomography yields bounds more competitive than the current projections obtained with other methods. At $68\%$ CL we have $\delta\Re b_Z=1.1\times 10^{-3}$, $\delta\Im b_Z=5.2\times 10^{-3}$,  $\delta\Re\tilde b_Z=1.6\times 10^{-3}$, and $\delta\Im\tilde b_Z=3.2\times 10^{-3}$, which further improve once full angular information is retained.   

The reconstruction of the $Z$ boson helicity density matrix also gives access to the quantum properties of this system, described as a qutrit. We investigate its purity and coherence, its contextuality, and its magic. The purity measures how close the state is to a pure one and, in the SM, it is largest at $365\GeV$ in the central region, where the longitudinal polarization dominates. We test contextuality with the five-projector inequality of Klyachko, Can, Binicio\u{g}lu and Shumovsky (KCBS)~\cite{Klyachko:2008} and with a nine-projector inequality, following Refs.~\cite{Fabbrichesi:2025ywl,Fabbrichesi:2025aqp}. Once the orientation of the projectors is optimized, contextuality manifests in the KCBS inequality violation observed at $365\GeV$ for $\abs{\cos\theta} < 0.69$, but neither at $240\GeV$ nor by the inclusive state. To quantify magic, the resource that allows quantum computers to outperform their classical counterparts, we compare three measures: the mana~\cite{Veitch:2014}, the stabilizer R\'enyi entropy~\cite{Leone:2022} and the robustness of magic~\cite{Howard:2016lyb}. The inclusive state carries no magic. At fixed angle the mana is non-zero only in part of the angular range, while the robustness, which also detects the bound magic that the mana misses, finds magic at almost every angle. These values depend on the basis in which the $Z$ boson helicity is described. Among the anomalous couplings, $\Re b_Z$ and, partly, $\Im\tilde b_Z$ change these properties at linear order. Couplings of the size probed by the tomographic process, however, induce deviations from the SM values of at most a few units in $10^{-3}$. Hence, these properties are robust features of the $Z$ boson produced in Higgsstrahlung, but not sensitive probes of the anomalous structures in the vertex. 

The polarization of the $Z$ boson produced in Higgsstrahlung has already been proposed as a probe of the $hZZ$ vertex: Ref.~\cite{Rao:2019hsp} computed its spin density matrix at tree level in the presence of the anomalous couplings, whereas Refs.~\cite{Nakamura:2018bli,Rao:2021eer} showed that the Higgs boson self-coupling induces, through the absorptive part of the one-loop amplitude, $\hat T$-odd asymmetries in the decays of the $Z$ boson. With respect to these works, we include the complete one-loop electroweak correction and reconstruct the full density matrix, from the leptonic as well as the hadronic decays of the $Z$ boson.

The paper is organized as follows. Section~\ref{sec:setup} sets up the conventions and the parametrization of the $hZZ$ vertex, and Section~\ref{sec:bounds} reviews the present bounds and the projections. Section~\ref{sec:loop} computes the SM amplitude at tree level and at one loop, with emphasis on the three diagrams that contain the Higgs boson self-coupling and on the form factors $c(s)$ and $\beta(s)$ they produce. Section~\ref{sec:density} computes the helicity density matrix of the $Z$ boson, and Section~\ref{sec:QIT} studies its purity, contextuality and magic. Section~\ref{sec:stat} contains the statistical analysis that yields the bounds on $\klam$ and on the form factors, and compares them with the quantum Fisher information. Section~\ref{sec:outlook} summarizes the results. The appendices collect the dictionary between the normalizations used in the literature, the conventions for the density matrix, the contextuality tests and the definitions of the magic of a qutrit, the loop functions, the form factors in terms of scalar integrals and the validation of the loop computation.


\section{The process and the effective vertex} 
\label{sec:setup}
We start by characterizing the process we consider and by introducing the effective vertex containing the form factors that our analysis will target.

\subsection{The amplitude}  
\label{sub:amplitude}

\begin{figure}[h]
\centering
\includegraphics[width=0.48\textwidth]{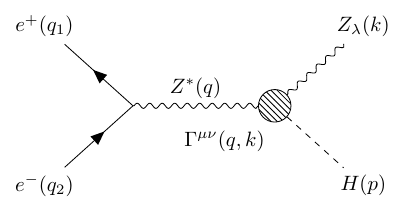}
\caption{The process $e^-e^+\to ZH$ with the effective $hZZ$ vertex $\Gamma^{\mu\nu}(q, k)$ of Eq.~\eqref{eq:vertex}. The labels indicate the four-momentum of each particle and the helicity $\lambda$ of the $Z$ boson. We indicate with $q = q_1 + q_2$ the momentum of the virtual $Z$ boson, whose index $\mu$ is contracted with the electron current, while $\nu$ is that of the outgoing $Z$.}
\label{fig:process}
\end{figure}

The process we consider is 
\begin{equation}
	e^-(q_2)\, e^+(q_1) \ra Z(k)\, H(p), \qquad q \deq q_1 + q_2 = k + p,
\end{equation}
with massless electrons,  depicted in Figure~\ref{fig:process}. In the center-of-mass (CM) frame the positron momentum $q_1$ defines the $z$ axis and the $Z$ boson is emitted at the polar angle $\theta$ with a three-momentum of magnitude $\vb*k = \sqrt{\lam}/(2\sqrt s)$, given by the K\"all\'en function
\begin{equation}
	\lam \deq \lam(s, \MZ^2, \MH^2) = s^2 + \MZ^4 + \MH^4 - 2 s \MZ^2 - 2 s \MH^2 - 2 \MZ^2 \MH^2 .
	\label{eq:lambda}
\end{equation}
The Mandelstam variables are $s = q^2$, $t = (q_2 - k)^2$, $u = (q_2 - p)^2$, with $s + t + u = \MZ^2 + \MH^2$.

The electron current couples to the $Z$ through the vertex 
\begin{equation}
	V^\mu  = i \frac{g}{\cw}\, \gamma^\mu (\gL \PL + \gR \PR)
\end{equation} 
with $\cw = \MW/\MZ$ being the cosine of the Weinberg angle, $g = e/\sw$, and
\begin{equation}
	\gL = \sw^2 - \frac{1}{2}, \qquad \gR = \sw^2, \qquad \PL = \frac{1 - \gamma^5}{2}, \qquad \PR = \frac{1 + \gamma^5}{2}.
\end{equation}
The on-shell relation $\cw = \MW/\MZ$ is not a choice of scheme only: at one loop it is required by the gauge cancellations between the $Z$ and the neutral Goldstone boson. We use it throughout the paper, together with $\alpha(0)$ for the couplings, and we comment on the $G_\mu$ scheme when comparing with the literature.

The $hZZ$ interaction is described by the effective Lagrangian
\begin{equation}
	\begin{gathered}
		\mathcal L_{hZZ} = \frac{h}{v}\qty[ a_Z\, \MZ^2\, Z_\mu Z^\mu + \frac{b_Z}{2}\, Z_{\mu\nu} Z^{\mu\nu} - \frac{\tilde b_Z}{2}\, Z_{\mu\nu}\tilde Z^{\mu\nu} ], \\
		Z_{\mu\nu} = \partial_\mu Z_\nu - \partial_\nu Z_\mu, \qquad \tilde Z^{\mu\nu} = \frac12\, \epsilon^{\mu\nu\rho\sigma} Z_{\rho\sigma},
	\end{gathered}
	\label{eq:Leff}
\end{equation}
with $h$ the Higgs boson, $v$ its vacuum expectation value (vev), $Z_{\mu\nu}$ the field strength of the $Z$ boson and $\tilde Z^{\mu\nu}$ its dual\footnote{In our conventions $\epsilon^{0123} = +1$.}. The three terms are, respectively, the SM interaction rescaled by $a_Z$ and the two anomalous structures, CP-even and CP-odd, which arise from operators of dimension six in the SM effective field theory (SMEFT). The corresponding effective $hZZ$ vertex is~\cite{Hagiwara:1993sw,Hagiwara:1993ck}
\begin{equation}
	\Gamma^{\mu\nu}(q, k) = \frac{g \MZ}{\cw}\qty[ a_Z\, g^{\mu\nu} + \frac{b_Z}{\MZ^2}\qty(q\vdot k\, g^{\mu\nu} - q^\nu k^\mu) + \frac{\tilde b_Z}{\MZ^2}\, \epsilon^{\mu\nu\rho\sigma} q_\rho k_\sigma ],
	\label{eq:vertex}
\end{equation}
where momenta and Lorentz indices follow the conventions shown in Figure~\ref{fig:process}. We report in Table~\ref{tab:parities} the parities of the three terms under the usual discrete symmetries. 

A hermitian Lagrangian forces the three couplings $a_Z$, $b_Z$ and $\tilde b_Z$ to be real at tree level. Loop contributions turn these numbers into complex functions of $s$, with imaginary parts brought by on-shell intermediate states (in our conventions, the phases of $b_Z$ and $\tilde b_Z$ are defined relative to that of $a_Z$, which is kept real). These functions are the form factors that we study in the following. 

As with the vertex~\eqref{eq:vertex}, the amplitude of the process splits into three terms, each with a characteristic Lorentz structure multiplied by the corresponding form factor. Explicitly we have
\begin{align}
	\emm = a_Z\, \Ma + b_Z\, \Mb + \tilde b_Z\, \Mbt ,
	\label{eq:ampl_vertex}
\end{align}
with
\begin{align}
	\Ma &= \frac{g^2 \MZ}{\cw^2\, (s - \MZ^2)}\; \bar v(q_1)\, \slashed{\eps}^{*}\, \qty(\gL \PL + \gR \PR)\, u(q_2) , 
	\label{eq:Mtree}
	\\
	\Mb &= 
	\frac{g^2}{\cw^2\, \MZ\, (s - \MZ^2)} 
	\bar v(q_1) 
	\qty[ \qty(q\vdot k) \slashed{\eps}^{*}  \qty(\gL \PL + \gR \PR)  
	- \qty(q\vdot\eps^*) \slashed{k} \qty(\gL \PL + \gR \PR) ]u(q_2), 
	\label{eq:Mb}\\
	\Mbt &= 
	-\frac{i\, g^2}{\cw^2\, \MZ\, (s - \MZ^2)}
	\bar v(q_1)
	\qty[ \qty(\Delta q\vdot k) \slashed{\eps}^{*} (\gR\PR - \gL\PL) 
	- \qty(\Delta q\vdot\eps^*) \slashed{k} (\gR\PR - \gL\PL)]
	 u(q_2),
	\label{eq:ampl_bt}
\end{align}
and $\Delta q\deq q_1-q_2$. The factor $i$, with opposite signs for the two chiralities, makes the last term CP-odd. 

\begin{table}[htbp]
\centering
\renewcommand{\arraystretch}{1.35}
\begin{tabular}{@{}lllccc@{}}
\toprule
 form factor & term of $\mathcal L_{hZZ}$ & structure in $\Gamma^{\mu\nu}$ & C & P & T \\
\midrule
$a_Z$ & $\MZ^2\, Z_\mu Z^\mu $ & $g^{\mu\nu}$ & $+$ & $+$ & $+$  \\
$b_Z$ & $\frac12\, Z_{\mu\nu} Z^{\mu\nu} $ & $q\vdot k\, g^{\mu\nu} - q^\nu k^\mu$ & $+$ & $+$ & $+$  \\
$\tilde b_Z$ & $-\frac12\, Z_{\mu\nu} \tilde Z^{\mu\nu}$ & $\epsilon^{\mu\nu\rho\sigma} q_\rho k_\sigma$ & $+$ & $-$ & $-$  \\
\bottomrule
\end{tabular}
\caption{Discrete symmetries of the three terms of the effective interaction~\eqref{eq:Leff}, with the structures they give in the vertex~\eqref{eq:vertex}.}
\label{tab:parities}
\end{table}

\subsection{The squared amplitude}  
\label{sub:treeparam}

Summing over the polarizations of the $Z$ boson and averaging over the spins of the initial-state particles, the squared amplitude obtained from Eq.~\eqref{eq:ampl_vertex} becomes a quadratic form in the three couplings. To first order in the anomalous ones,
\begin{equation}
	\Msq = a_Z^2\, \Msqa + \Re\qty(a_Z b_Z^*)\, X_b - \Im\qty(a_Z \tilde b_Z^*)\, \tilde X_{\tilde b} + \order{b_Z^2,\, \tilde b_Z^2,\, b_Z\tilde b_Z},
	\label{eq:M2param}
\end{equation}
where the interferences of $\Ma$ with the two anomalous structures are
\begin{equation}
	X_b \deq \frac14\sum_{\rm spins} 2 \Re\qty(\Ma \Mb^*), \qquad \tilde X_{\tilde b} \deq \frac14\sum_{\rm spins} 2 \Im\qty(\Ma \Mbt^*).
	\label{eq:Xdef}
\end{equation}
Equation~\eqref{eq:M2param} contains two of the four real combinations of the couplings that can appear at linear order. Written out, the full cross terms are
\begin{align}
	\frac14\sum_{\rm spins} 2\Re\qty[a_Z b_Z^*\, \Ma\Mb^*] &= \Re\qty(a_Z b_Z^*)\, X_b - \Im\qty(a_Z b_Z^*)\, \tilde X_b , \nn\\
	\frac14\sum_{\rm spins} 2\Re\qty[a_Z \tilde b_Z^*\, \Ma\Mbt^*] &= \Re\qty(a_Z \tilde b_Z^*)\, X_{\tilde b} - \Im\qty(a_Z \tilde b_Z^*)\, \tilde X_{\tilde b} ,
	\label{eq:crossterms}
\end{align}
with $\tilde X_b \deq \frac14\sum 2\Im(\Ma\Mb^*)$ and $X_{\tilde b} \deq \frac14\sum 2\Re(\Ma\Mbt^*)$ completing the definitions~\eqref{eq:Xdef}. These two functions, however, vanish identically as a consequence of \textit{naive time reversal} $\hat T$, which reverses the momenta and the spins without exchanging the initial and the final state. To see this explicitly, consider the tree-level amplitude $\emm_i$, the matrix element of a hermitian operator $\mathcal O_i$, built from the electron vertex of the SM, the propagator of the $Z$ boson and the $i$-th term of Eq.~\eqref{eq:Leff}, $i\in\{a, b, \tilde b\}$. As reported in Table~\ref{tab:parities}, the first two terms are $T$-even, whereas the one involving $\tilde b_Z$ is $T$-odd. With $T$ being the antiunitary operator of time reversal, $T\mathcal O_i T^{-1} = \eta_i\,\mathcal O_i$ with $\eta_a = \eta_b = +1$ and $\eta_{\tilde b} = -1$. Antiunitarity, $\braket{T x}{T y} = \braket{y}{x}$, and then hermiticity give
\begin{equation}
	\mel{f}{\mathcal O_i}{i} = \eta_i\, \mel{T i}{\mathcal O_i}{T f} = \eta_i\, \mel{T f}{\mathcal O_i}{T i}^* ,
\end{equation}
implying
\begin{equation}
	\emm_i(\hat T c) = \eta_i\, \emm_i(c)^* ,
	\label{eq:Tparity}
\end{equation}
where $c$ denotes the momenta and the spins of a kinematic configuration and $\hat T c$ the same configuration with momenta and spins reversed. This defines the action of the naive time reversal operator $\hat T$. By Eq.~\eqref{eq:Tparity}, $(\Ma\emm_j^*)(\hat Tc) = \eta_j\,[(\Ma\emm_j^*)(c)]^*$, hence the real and imaginary parts of the interference of $\Ma$ with $\emm_j$ have opposite $\hat T$-parities. Thus, $X_b$ and $\tilde X_{\tilde b}$ are spin sums of $\hat T$-even functions, $\tilde X_b$ and $X_{\tilde b}$ of $\hat T$-odd ones.

As the beams are unpolarized, the sum over all the spins is unchanged when they are reversed and what remains is a Lorentz-invariant function of the momenta $q_1$, $q_2$ and $k$ alone. Scalar products do not change when the three-momenta are reversed, and the only $\hat T$-odd invariant (a Levi-Civita tensor contracted with four independent vectors) cannot be formed. In other words, all the momenta of a $2\to 2$ process with unpolarized initial state lie in the production plane, and no triple product exists (a transverse polarization of the beams is required to supply the missing vector). As a consequence, $\hat T$-odd spin sums necessarily vanish, so that $\tilde X_b = X_{\tilde b} = 0$.

We anticipate that these two terms still affect the density matrix of the $Z$ boson because the spin of this particle provides the required $\hat T$-odd correlation. In fact, on the triad $\{\hat r, \hat n, \hat k\}$ of Eq.~\eqref{eq:polvectors} which we will introduce to describe the $Z$ spin vector, $\hat T$ reverses $\hat r$, $\hat k$ and the spin, but not the normal to the production plane: $\hat n \propto \hat z \times \hat k$. The components of the density matrix with one index along $\hat n$ are therefore $\hat T$-odd, allowing for the measurement of $\Im b_Z$ and $\Re\tilde b_Z$ as we show in Section~\ref{sub:rhoshift}.

Coming back to the squared amplitude in Eq.~\eqref{eq:M2param}, the first term is 
\begin{equation}
	\Msqa = \frac{g^4 \qty(\gL^2 + \gR^2)}{4 \cw^4\, (s - \MZ^2)^2}\; \qty[\MZ^4 + \MZ^2 \qty(s - t - u) + t u]
	= \frac{g^4 \qty(\gL^2 + \gR^2)}{16 \cw^4\, (s - \MZ^2)^2}\; \qty(\lam \sin^2\theta + 8 \MZ^2 s),
	\label{eq:M2tree}
\end{equation}
where the second form holds in the CM frame. The angular dependence is the familiar one: the $\sin^2\theta$ term, proportional to $\vb*k^2$, is brought by the longitudinal $Z$ boson mode and dominates at high energy. Notice that the squared amplitude is even under $\theta\to\pi - \theta$ although the couplings violate parity, because the parity-odd part, proportional to $\gL^2 - \gR^2$, cancels in the sum over the polarizations of the $Z$. 

The interference of $\Ma$ with the $b_Z$ structure is remarkably simple,
\begin{equation}
	X_b = \frac{g^4 \qty(\gL^2 + \gR^2)}{\cw^4\, (s - \MZ^2)^2}\; s\; \qty(q\vdot k) = \frac{g^4 \qty(\gL^2 + \gR^2)}{2\cw^4\, (s - \MZ^2)^2}\; s\, \qty(s + \MZ^2 - \MH^2) ,
	\label{eq:Xb}
\end{equation}
and, importantly, in the CM frame does not depend on $\theta$. Comparing with Eq.~\eqref{eq:M2tree},
\begin{equation}
	\frac{X_b}{\Msqa} = \frac{16\, s\, \qty(q\vdot k)}{\lam \sin^2\theta + 8 \MZ^2 s} ,
\end{equation}
which is largest where $\Msqa$ is smallest: along the beam. The $b_Z$ term then does not simply rescale the angular distribution, but it changes the balance between its isotropic and $\sin^2\theta$ contributions, due respectively to the transverse and the longitudinal $Z$ boson polarizations. 

The CP-odd coupling does not interfere with $\Ma$ in the spin-summed squared amplitude, $X_{\tilde b} = 0$, so that a real $\tilde b_Z$ changes neither the rate nor the angular distribution at linear order. An imaginary one, CP-odd and $\hat T$-even, gives
\begin{equation}
	\frac{\tilde X_{\tilde b}}{\Msqa} = \frac{8\, s\sqrt\lam\, A_e\cos\theta}{\lam \sin^2\theta + 8 \MZ^2 s}, \qquad A_e \deq \frac{\gL^2 - \gR^2}{\gL^2 + \gR^2} = \frac{1 - 4\sw^2}{1 - 4\sw^2 + 8\sw^4} = 0.213 ,
	\label{eq:Xtilde}
\end{equation}
inducing the forward--backward asymmetry of the $Z$. For a CP-symmetric initial state this integrates to zero, but it leaves an imprint in the angular distribution. 

For the sake of completeness, we also give the squares of the two anomalous amplitudes, needed at second order
\begin{align}
	\overline{\abs{\Mb}^2} &= \frac{g^4 \qty(\gL^2 + \gR^2)}{16 \cw^4\, \MZ^2\, (s - \MZ^2)^2}\; s\, \qty[\lam \qty(1 + \cos^2\theta) + 8 \MZ^2 s] , \nn\\
	\overline{\abs{\Mbt}^2} &= \frac{g^4 \qty(\gL^2 + \gR^2)}{16 \cw^4\, \MZ^2\, (s - \MZ^2)^2}\; s\, \lam \qty(1 + \cos^2\theta),
	\label{eq:Mb2}
\end{align}
with the $(1 + \cos^2\theta)$ dependence of transversely polarized bosons in place of the $\sin^2\theta$ of Eq.~\eqref{eq:M2tree}. The CP-odd structure then produces transverse $Z$ bosons only. The rate-based bounds on $\tilde b_Z$, presented below in Table~\ref{tab:bounds}, constrain $\abs{\tilde b_Z}^2$ through Eq.~\eqref{eq:Mb2}.

\section{Present bounds and projections}  
\label{sec:bounds}

Before proceeding with the analysis, we briefly review the present bounds and the projections concerning the four parameters of this study: $\klam$, $a_Z$, $b_Z$ and $\tilde b_Z$. The literature uses half a dozen normalizations for the same vertex and Tables~\ref{tab:bounds} and~\ref{tab:projections} convert everything to Eq.~\eqref{eq:vertex} through the dictionary of Appendix~\ref{app:dictionary}.

\emph{Present bounds.} The $hZZ$ coupling itself is known to $6$--$7\%$ from the Run-2 combinations~\cite{ATLAS:2022vkf,CMS:2022dwd}. In our parametrization, this bounds $a_Z$. The CP-even structure $b_Z$ is constrained at the $10^{-1}$ level by the CMS analyses of $H\to 4\ell$ with vector boson fusion (VBF) and $VH$ production~\cite{CMS:2021nnc,CMS:2022uox}, and more recently by the production-side analysis in the $\gamma\gamma$ channel~\cite{CMS:2026gg}. The Higgs-basis fit of Ref.~\cite{CMS:2021nnc} with the four $HVV$ couplings varied gives $c_{zz} = 0.01^{+0.11}_{-0.10}$, i.e.\ $\delta b_Z \simeq 0.03$ at $68\%$ CL. All the analyses assume $a_i^{WW} = a_i^{ZZ}$  and no anomalous $H\gamma\gamma$, $HZ\gamma$ couplings. The CP-odd $\tilde b_Z$ is known to a few $10^{-2}$: the cleanest $hZZ$-only number is the ATLAS $\tilde d$ from the $H\to 4\ell$ decay angles~\cite{ATLAS:2023xmo}, $\abs{\tilde b_Z} < 0.05$, with the CMS $H\to\gamma\gamma$ and $4\ell+\tau\tau$ results close by~\cite{CMS:2026gg,CMS:2022uox}. The ATLAS combination of five channels~\cite{ATLAS:2026cp} is dominated by VBF and constrains $c_{H\widetilde W}$, i.e.\ a combination of $hWW$ and $hZZ$. The self-coupling, finally, is bounded by the $HH$ searches to $-0.7 < \klam < 6.1$ in the ATLAS$+$CMS combination~\cite{ATLASCMS:2026hh}. The ATLAS $H+HH$ combination gives $-0.4 < \klam < 6.3$~\cite{ATLAS:2022jtk}, $[-1.5, 6.5]$ in its 2026 update~\cite{ATLAS:2026comb}.

\begin{table}[!htb]
\centering
\footnotesize
\setlength{\tabcolsep}{4pt}
\renewcommand{\arraystretch}{1.15}
\begin{tabular}{@{}l >{\raggedright\arraybackslash}p{3.3cm} >{\raggedright\arraybackslash}p{5.3cm} >{\raggedright\arraybackslash}p{4.6cm} c@{}}
\toprule
 & Bound (this normalization) & Measured quantity & Data, assumptions & Ref. \\
\midrule
$a_Z$ & $0.99 \pm 0.06$ ($68\%$) & $\kappa_Z$, effective loops, no BSM decays & ATLAS, $139\fb^{-1}$ & \cite{ATLAS:2022vkf}\\
$a_Z$ & $1.04 \pm 0.07$ ($68\%$) & $\kappa_Z$, effective loops, $B_{\rm BSM} = 0$ & CMS, $138\fb^{-1}$ & \cite{CMS:2022dwd}\\
$b_Z$ & $[-0.17, +0.11]$ & $f_{a2} \in [-1.0, 2.5]\times 10^{-3}$, one coupling at a time & CMS $4\ell + \tau\tau$, $a_i^{WW} = a_i^{ZZ}$ & \cite{CMS:2022uox}\\
$b_Z$ & $[-0.04, +0.08]$ & $f_{a2} \in [-5.5, 1.2]\times 10^{-4}$, one at a time & CMS $\gamma\gamma$ (VBF, $VH$), $a_i^{WW} = a_i^{ZZ}$ & \cite{CMS:2026gg}\\
$b_Z$ & $+0.003^{+0.030}_{-0.027}$ ($68\%$) & $c_{zz} = 0.01^{+0.11}_{-0.10}$, four $HVV$ couplings floated & CMS $4\ell$, SU(2)$\times$U(1) & \cite{CMS:2021nnc}\\
$\tilde b_Z$ & $\abs{\tilde b_Z} < 0.05$ & $\tilde d \in [-0.026, 0.025]$, decay angles only & ATLAS $H\to 4\ell$, $\tilde d = \tilde d_B$ & \cite{ATLAS:2023xmo}\\
$\tilde b_Z$ & $\abs{\tilde b_Z} < 0.06$ & $f_{a3} \in [-1.5, 1.5]\times 10^{-4}$, one at a time & CMS $\gamma\gamma$ (VBF, $VH$), $a_i^{WW} = a_i^{ZZ}$ & \cite{CMS:2026gg}\\
$\tilde b_Z$ & $[-0.02, +0.15]$ & $f_{a3} \in [-0.01, 0.88]\times 10^{-3}$, one at a time & CMS $4\ell + \tau\tau$, $a_i^{WW} = a_i^{ZZ}$ & \cite{CMS:2022uox}\\
$\tilde b_Z$ & $\abs{\tilde b_Z} \lesssim 0.06$ & $c_{H\widetilde W} \in [-0.13, 0.60]$, $c_{H\widetilde B} = c_{H\widetilde W B} = 0$ & ATLAS, five channels (mostly VBF) & \cite{ATLAS:2026cp}\\
$\klam$ & $[-0.71, 6.1]$ & $HH$ searches & ATLAS $+$ CMS, $126$--$140\fb^{-1}$ & \cite{ATLASCMS:2026hh}\\
$\klam$ & $[-1.2, 7.2]$; $[-1.35, 6.37]$ & $HH$ searches & ATLAS; CMS & \cite{ATLAS:2024ish,CMS:2026hh}\\
$\klam$ & $[-0.4, 6.3]$; $[-1.2, 7.5]$ & $HH$ $+$ single-$H$ (loop effects) & ATLAS; CMS & \cite{ATLAS:2022jtk,CMS:2024hhc}\\
$\klam$ & $[-1.5, 6.5]$ & $HH$ $+$ single-$H$, 2026 update & ATLAS & \cite{ATLAS:2026comb}\\
\bottomrule
\end{tabular}
\caption{Present bounds on the parameters of the vertex~\eqref{eq:vertex} and on $\klam$ from the LHC Run 2, converted to our conventions with the dictionary of Appendix~\ref{app:dictionary}. The intervals are observed at $95\%$ CL unless stated. The $f_{ai}$ conversions use one anomalous coupling at a time, $a_1 = 2$, $b_Z = -a_2$ and $\tilde b_Z = a_3$. The Warsaw-basis row converts $c_{H\widetilde W}$ alone and is therefore a combination of $hWW$ and $hZZ$. The CMS $\gamma\gamma$ result is from a 2026 preprint.}
\label{tab:bounds}
\end{table}

\emph{Projections.} At the FCC-ee the $hZZ$ coupling is a per-mille measurement, $0.10\%$ with four interaction points (IP)~\cite{FCC:2025lpp} ($0.17$--$0.21\%$ in the two-IP global fits~\cite{deBlas:2019rxi}), to be compared with $0.25\%$ at CEPC~\cite{CEPCStudyGroup:2018ghi} and $0.4$--$0.5\%$ at the ILC at $250\,$GeV~\cite{ILCInternationalDevelopmentTeam:2022izu}. The same cross-section measurements give $\klam$ to $28\%$ stand-alone and $18\%$ with HL-LHC~\cite{FCC:2025lpp}, through the rate parameter $C_1$ discussed in Section~\ref{sub:rate}. For $b_Z$ the projections split into two scenarios: the rate measures $b_Z$ to $\simeq 10^{-3}$ (Ref.~\cite{Craig:2015wwr}, CEPC $5\,{\rm ab}^{-1}$: $\delta\hat\alpha_{ZZ} = 6.4\times 10^{-4}$), but only in combination with $a_Z$ and with potential further factors that rescale the cross section. The angular distributions alone give $\delta\hat\alpha_{ZZ} = 0.016$ at $5\,{\rm ab}^{-1}$ in the same analysis, $\delta b_Z = 0.03$, free of the rate normalization. The ILC three-parameter fits of Ref.~\cite{Ogawa:2017bsx}, angles plus rate, with polarized beams and the $500\,$GeV run, reach $\delta b_Z = 1.7\times 10^{-3}$ and $\delta\tilde b_Z = 1.9\times 10^{-3}$. For the CP-odd coupling, which does not affect the rate, the CEPC~\cite{Sha:2022bkt} and FCC-ee~\cite{Atwal:2025cpv} angular analyses reach $\abs{\tilde b_Z} \simeq 0.01$--$0.02$. The tomographic approach presented in the present paper belongs to the second category, with the full density matrix in place of a few angular asymmetries. Its results are collected in the same normalization in Table~\ref{tab:results} of Section~\ref{sub:formfactors} and improve on the present projections.

\begin{table}[!htb]
\centering
\footnotesize
\setlength{\tabcolsep}{4pt}
\renewcommand{\arraystretch}{1.15}
\begin{tabular}{@{}l >{\raggedright\arraybackslash}p{3.3cm} >{\raggedright\arraybackslash}p{5.3cm} >{\raggedright\arraybackslash}p{4.6cm} c@{}}
\toprule
 & Bound (this normalization) & Measured quantity & Data, assumptions & Ref. \\
\midrule
$a_Z$ & $\delta a_Z = 1.0\times 10^{-3}$ & $\kappa_Z$, global $\kappa$ fit with HL-LHC & FCC-ee $240+365\,$GeV, 4 IPs & \cite{FCC:2025lpp}\\
$a_Z$ & $\delta a_Z = 1.7$--$2.1\times 10^{-3}$ & $\kappa_Z$ & FCC-ee, 2 IPs, $5 + 1.5\,{\rm ab}^{-1}$ & \cite{deBlas:2019rxi}\\
$a_Z$ & $\delta a_Z = 2.5\times 10^{-3}$; $3.8$--$4.9\times 10^{-3}$ & $\kappa_Z$; $g_{hZZ}$ & CEPC $5.6\,{\rm ab}^{-1}$; ILC $250\,$GeV & \cite{CEPCStudyGroup:2018ghi,ILCInternationalDevelopmentTeam:2022izu}\\
$b_Z$ & $\delta b_Z = 1.3\times 10^{-3}$ & $\delta\hat\alpha_{ZZ} = 6.4\times 10^{-4}$, rate, one parameter & CEPC $240\,$GeV, $5\,{\rm ab}^{-1}$, $Z\to\ell\ell$, $H\to b\bar b$ & \cite{Craig:2015wwr}\\
$b_Z$ & $\delta b_Z = 0.032$ & $\delta\hat\alpha_{ZZ} = 0.016$, angles only, one parameter & same & \cite{Craig:2015wwr}\\
$b_Z$ & $\delta b_Z = 1.4\times 10^{-3}$ & $\delta c_{ZZ} = 0.0051$, 12-parameter EFT fit, rates $+$ angles & CEPC $240\,$GeV, $5.6\,{\rm ab}^{-1}$ & \cite{CEPCStudyGroup:2018ghi}\\
$b_Z$ & $\delta b_Z = 2.1\times 10^{-3}$ ($1.7\times 10^{-3}$) & $\delta b_Z^{\rm ILC} = 0.0087$ ($0.0070$ with $ZZ$ fusion), 3-parameter fit, angles $+$ rate & ILC H20, polarized, $250 + 500\,$GeV & \cite{Ogawa:2017bsx}\\
$\tilde b_Z$ & $\delta\tilde b_Z = 2.0\times 10^{-3}$ ($1.9\times 10^{-3}$) & $\delta\tilde b_Z^{\rm ILC} = 0.0083$ ($0.0078$), same fit & same & \cite{Ogawa:2017bsx}\\
$\tilde b_Z$ & $\abs{\tilde b_Z} < 0.016$ ($68\%$), $0.033$ ($95\%$) & $\tilde c_{ZZ} \in [-0.06, 0.06]$, angles ($\phi$, optimal observables) & CEPC $240\,$GeV, $5.6\,{\rm ab}^{-1}$, $Z\to\ell\ell$ & \cite{Sha:2022bkt}\\
$\tilde b_Z$ & $\abs{\tilde b_Z} < 0.012$ ($95\%$) & $c_{H\widetilde W} \in [-0.13, 0.13]\,{\rm TeV}^{-2}$, NN optimal observable & FCC-ee $240\,$GeV, $10.8\,{\rm ab}^{-1}$, $Z\to\ell\ell$ & \cite{Atwal:2025cpv}\\
$\klam$ & $\delta\klam = 0.28$ ($0.18$ with HL-LHC) & $\sigma(ZH)$, $\sigma(\nu\bar\nu H)$ at $240$ and $365\,$GeV, SMEFT global fit & FCC-ee, 4 IPs & \cite{FCC:2025lpp}\\
$\klam$ & $\delta\klam = 0.27$; $0.35$ & $ZHH$ at $500\,$GeV; $\sigma(ZH)$ at $240\,$GeV & ILC H20; CEPC & \cite{ILCInternationalDevelopmentTeam:2022izu,CEPCStudyGroup:2018ghi}\\
$\klam$ & $\delta\klam \simeq 0.15$ & same, luminosity at $240$ and $365\,$GeV doubled (HL-4IP), with HL-LHC & FCC-ee, 4 IPs & \cite{FCC:2025lpp}\\
$\klam$ & $\delta\klam \simeq 0.15$ & SMEFT global fit, other operators marginalized & FCC-ee $+$ HL-LHC & \cite{terHoeve:2025gey}\\
$\klam$ & $\delta\klam \simeq 0.03$--$0.04$ & $HH\to b\bar b\gamma\gamma$, $\klam$ only & FCC-hh $84\,$TeV, $30\,{\rm ab}^{-1}$ & \cite{FCC:2025lpp,Mastrapasqua:2026hh}\\
\bottomrule
\end{tabular}
\caption{Projected sensitivities ($1\sigma$, statistical unless stated) on the parameters of the vertex~\eqref{eq:vertex} and on $\klam$ at future colliders, in the conventions of Table~\ref{tab:bounds}.}
\label{tab:projections}
\end{table}

\section{The SM tree-level and one-loop contributions}  
\label{sec:loop}

We analyze here how the SM contributions project on the effective vertex structure in Eq.~\eqref{eq:vertex}. After showing the tree-level results, we focus on the three one-loop diagrams that contain the Higgs boson self-coupling and that, by themselves, yield a finite correction to the $hZZ$ vertex. The rest of the electroweak corrections, which need renormalization and depend on the chirality of the electron and on the scattering angle, are treated in a dedicated subsection. At the level of the amplitude they project on the two CP-even structures of the effective vertex, with chirality- and angle-dependent coefficients, and on one more structure, produced by the box diagrams, which no $hZZ$ vertex generates. 

\subsection{Tree level}  
\label{sub:treeSM}

At tree level, the SM gives:
\begin{equation}
	\emm_{ 0} \equiv \Ma = \frac{g^2 \MZ}{\cw^2\, (s - \MZ^2)}\; \bar v(q_1)\, \slashed{\eps}^{*}(k)\, \qty(\gL \PL + \gR \PR)\, u(q_2), 
	\label{eq:MtreeSM}
\end{equation}
corresponding to
\begin{equation}
	a_Z = 1, \qquad b_Z = \tilde b_Z = 0.
\end{equation}
The amplitudes $\Mb$ and $\Mbt$ of the two anomalous structures, Eqs.~\eqref{eq:Mb} and~\eqref{eq:ampl_bt}, do not vanish, but their coefficients do: the $hZZ$ vertex of the SM, $i (g\MZ/\cw)\, g^{\mu\nu}$, is the first structure of Eq.~\eqref{eq:vertex} with unit coefficient, and no tree-level diagram generates the other two: in the SM, the CP-even $b_Z$ appears first at one loop, whereas the CP-odd $\tilde b_Z$ is induced by the Cabibbo--Kobayashi--Maskawa (CKM) phase at three loops. Notice that we omitted the diagram proceeding via the  exchange of a Higgs boson in the $s$ channel, as it is proportional to the electron mass that we neglect.

The squared amplitude then matches that of Eq.~\eqref{eq:M2tree}, and the differential and total cross sections follow as:
\begin{align}
	\dv{\sigma_{ 0}}{\Omega} &=  \frac{\vb*k \Msqa}{32\pi^2 s^{3/2}} =  \frac{g^4 \qty(\gL^2 + \gR^2)\, \sqrt{\lam}}{1024\, \pi^2\, \cw^4\, s^2\, (s - \MZ^2)^2}\qty[\lam \sin^2\theta + 8 \MZ^2 s], \label{eq:dsigma}\\
	\sigma_{ 0} &= \frac{g^4 \qty(\gL^2 + \gR^2)\, \sqrt{\lam}\, \qty(\lam + 12 \MZ^2 s)}{384\, \pi\, \cw^4\, s^2\, (s - \MZ^2)^2} . \label{eq:sigma}
\end{align}
Numerically\footnote{With $\alpha = 1/137.036$, $\MZ = 91.1876\GeV$, $\MW = 80.377\GeV$, $\MH = 125.25\GeV$ and $\sw^2 = 1 - \MW^2/\MZ^2 = 0.2231$. These are the $\alpha(0)$-scheme values without initial-state radiation. In the $G_\mu$ scheme, $\alpha \to \alpha_{G_\mu} = \sqrt 2 G_\mu \MW^2 \sw^2/\pi$, larger by $3.6\%$, and the cross section is larger by $7.5\%$.}, the tree-level cross section is $222.0\fb$ at $\sqrt s = 240\GeV$ and $108.9\fb$ at $365\GeV$. For later use we also record the fraction of longitudinally polarized $Z$ bosons, integrated over the angles,
\begin{equation}
	f_L = \frac{\lam + 4 \MZ^2 s}{\lam + 12 \MZ^2 s},
	\label{eq:fL}
\end{equation}
which is the $(0,0)$ element of the angle-averaged density matrix of the $Z$ boson.

\subsection{Corrections depending on the Higgs boson self-coupling}  
\label{sub:kdiagrams}
\subsubsection{Diagrams and computation}  
\label{sub:diagrams}

\begin{figure}[h]
\centering
\includegraphics[width=\textwidth]{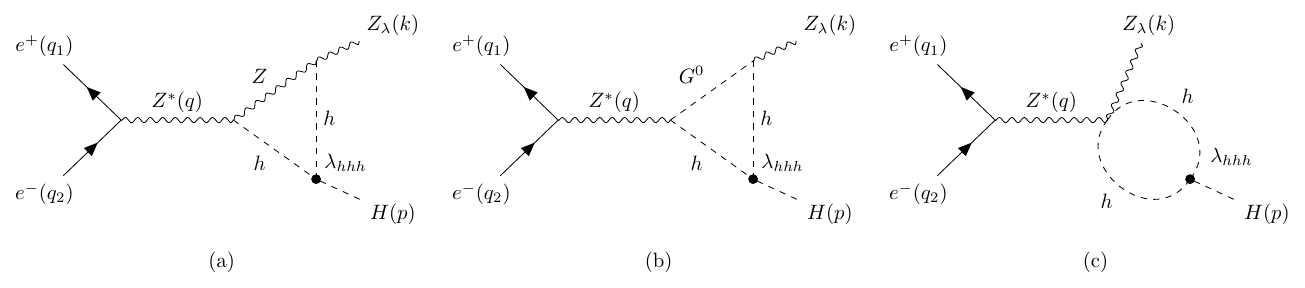}
\caption{The three one-loop corrections to $e^-e^+\to ZH$ that contain the Higgs boson self-coupling $\lambda_{hhh}$ (dot): (a) the triangle with a $Z$ and two Higgs bosons in the loop, (b) the same with the neutral Goldstone boson $G^0$ in place of the internal $Z$, (c) the bubble with the $ZZhh$ seagull. The external momenta are as in Figure~\ref{fig:process}.}
\label{fig:hhhdiagrams}
\end{figure}

We scale the SM Higgs self-coupling by $\klam$, $\lambda_{hhh} = \klam\, \lambda_{hhh}^{\rm SM}$ with $\lambda^{\rm SM}_{hhh} = 3 \MH^2 / v$, and retain here only the one-loop corrections to $e^-e^+\to ZH$ that contain the $hhh$ vertex. These three diagrams are shown in Figure~\ref{fig:hhhdiagrams} and, given that the vertex is contained in each of them exactly once, the resulting amplitude splits as
\begin{equation}
	\emm = \Ma + \klam\, \Mhhh ,
	\label{eq:ampl_kappa}
\end{equation}
where $\Mhhh$ is the sum of the three diagrams computed with the SM coupling, which we study below.

The computation is done with {\tt FeynArts}~\cite{Hahn:2000kx} and {\tt FeynCalc}~\cite{Shtabovenko:2016sxi,Shtabovenko:2023idz}, reducing the loop integrals to the Passarino--Veltman basis~\cite{Passarino:1978jh}. The Dirac algebra is addressed in $D = 4 - 2\epsilon$ dimensions and the ultraviolet pole of each loop function is separated from its finite part. Importantly, the sum of the three diagrams is ultraviolet finite: the $1/\epsilon$ poles cancel between the $B_0$ of the bubble and the $C_{00}$ of the triangles, and with them the dependence on the renormalization scale. The three diagrams are also the complete set at this order as far as the $\klam$-linear dependence is concerned: the remaining $\klam$ dependence sits in the Higgs wave-function renormalization, which is quadratic in $\klam$ and universal. We consider this contribution when computing the rate in Section~\ref{sub:rate} and conclude that, at this level, no renormalization of the $hZZ$ vertex is needed.

\subsubsection{Projection on the effective vertex structures}  
\label{sub:projection}

The finite part of $\Mhhh$ contains only four spinor chains,
\begin{equation}
	\qty[\bar v(q_1)\, \slashed{\eps}^{*}(k)\, P_{L,R}\, u(q_2)] \qquad\text{and}\qquad \qty[\bar v(q_1)\, \slashed{k}\, P_{L,R}\, u(q_2)]\; \qty(q\vdot\eps^*) ,
\end{equation}
the latter arising with equal coefficients for $q_1\vdot\eps^*$ and $q_2\vdot\eps^*$. These are precisely the structures in Eq.~\eqref{eq:ampl_vertex}, and the coefficients of the left- and right-handed chains are in the same ratio $\gL : \gR$ as at tree level (as they must, for the correction to the $hZZ$ vertex cannot affect the electron chirality). We can then write 
\begin{equation}
	\Mhhh = c(s)\, \Ma + \beta(s)\, \Mb ,
	\label{eq:decomposition}
\end{equation}
so that the one-loop amplitude has the form of the tree-level one, with corrected form factors:
\begin{equation}
	a_Z = 1 + \klam\, c(s), \qquad b_Z = \klam\, \beta(s) .
	\label{eq:effective}
\end{equation}
Explicitly, we have
\begin{align}
	\beta(s) &= \frac{3 g^2 \MH^2}{32 \pi^2 \cw^2}\; \qty(C_1 + C_{11} + C_{12}), \label{eq:beta}\\
	c(s) &= -\frac{3 g^2 \MH^2}{128 \pi^2 \cw^2 \MZ^2}\qty(B_0 - 4 C_{00} + 4 \MZ^2 C_0) + \frac{3 g^2 \MH^2\, \qty(\MH^2 - \MZ^2 - s)}{64 \pi^2 \cw^2 \MZ^2}\; \qty(C_1 + C_{11} + C_{12}), \label{eq:c}
\end{align}
where $B_0 = B_0(\MH^2; \MH^2, \MH^2)$ is the two-point function of the bubble, while the $C$ functions are those of the triangle with two Higgs bosons and a $Z$ boson in the loop,
\begin{equation}
	C_{i} = C_{i}(\MH^2, s, \MZ^2; \MH^2, \MH^2, \MZ^2), \qquad C_{0,00} = C_{0,00}(\MH^2, \MZ^2, s; \MH^2, \MH^2, \MZ^2),
\end{equation}
in the conventions of Appendix~\ref{app:loopfunctions}. Three features of Eqs.~\eqref{eq:beta} and~\eqref{eq:c} stand out. Firstly, the tensor coefficients enter only through the combination $C_1 + C_{11} + C_{12}$: the pieces of the $Z$ and $G^0$ triangles that do not have this form cancel against each other, provided $\cw = \MW/\MZ$. Secondly, the combination $B_0 - 4 C_{00}$ in $c(s)$ is finite: $B_0$ and $C_{00}$ carry the ultraviolet poles $\Delta$ and $\Delta/4$, respectively. Thirdly, $\beta(s)$ is finite, as it must be because no counterterm has the $b_Z$ structure at this order ($b_Z$ corresponds to a dimension-six operator).

\subsubsection{Numerical values}  
\label{sub:numbers}

The reduction of the tensor coefficients to scalar integrals, which gives $\beta(s)$ and $c(s)$ in terms of $A_0$, $B_0$ and $C_0$ only, is reported in Appendix~\ref{app:reduced} and it was used for the numerical evaluation reported in Table~\ref{tab:numbers}.

\begin{table}[h]
\centering
\begin{tabular}{lcccc}
\toprule
$\sqrt s$ & $\Re c$ & $\Im c$ & $\Re\beta$ & $\Im\beta$ \\
\midrule
$240\GeV$ & $7.17\times 10^{-3}$ & $5.19\times 10^{-3}$ & $5.61\times 10^{-4}$ & $1.72\times 10^{-4}$ \\
$250\GeV$ & $6.22\times 10^{-3}$ & $5.51\times 10^{-3}$ & $5.41\times 10^{-4}$ & $2.35\times 10^{-4}$ \\
$365\GeV$ & $1.70\times 10^{-3}$ & $4.70\times 10^{-3}$ & $1.94\times 10^{-4}$ & $3.91\times 10^{-4}$ \\
$500\GeV$ & $3.71\times 10^{-4}$ & $3.37\times 10^{-3}$ & $2.91\times 10^{-5}$ & $2.96\times 10^{-4}$ \\
\bottomrule
\end{tabular}
\caption{The form factor corrections $c(s)$ and $\beta(s)$ of Eqs.~\eqref{eq:beta} and~\eqref{eq:c} at the FCC-ee energies, at $250\GeV$ and at $500\GeV$. The one-loop amplitude is $\emm = \qty[1 + \klam c(s)]\Ma + \klam \beta(s) \Mb$.}
\label{tab:numbers}
\end{table}

Above the $Zh$ threshold both form factors are complex, the imaginary parts coming from the $Zh$ cut of the triangles. Table~\ref{tab:numbers} gives their values at the energies of interest, evaluated with {\tt LoopTools}~\cite{Hahn:1998yk}. The dependence of the form factor corrections on $\sqrt s$ is shown in Figure~\ref{fig:formfactors} of Appendix~\ref{app:validation}, together with the checks of the numerical evaluation. These two complex functions of the CM energy encapsulate the Higgs boson self-coupling contribution to the process under analysis: $\beta(s)$ enters the $Z$ boson helicity density matrix, in Section~\ref{sub:rhoSM}, and $c(s)$, together with $\beta(s)$, the rate shown in Section~\ref{sub:rate}.

\subsection{The remaining one-loop corrections}  
\label{sub:fullloop}

At particle level there are $175$ one-loop diagrams for $e^-e^+\to ZH$ without tadpoles and external self-energies, bosonic and fermionic loops included, and $7$ counterterm diagrams, with the counterterm Lagrangian of {\tt FeynArts}~\cite{Hahn:2000kx} in the on-shell scheme of Ref.~\cite{Denner:1991kt}. The renormalization constants, $\delta\MZ^2$, $\delta\MW^2$, $\delta Z_{ZZ}$, $\delta Z_{AA}$, $\delta Z_{AZ}$, $\delta Z_{ZA}$, $\delta Z_H$, $\delta Z_e^{L,R}$, and from them $\delta Z_e$ and $\delta\sw$, are computed from the self-energies generated in the same run, with massless electrons, as well as massless light quarks in the loops of the $2\to 2$ amplitude but carrying, in the self-energies only, the effective masses that reproduce $\Delta\alpha_{\rm had}(\MZ^2) = 0.0277$ in the charge renormalization. The couplings are $\alpha(0)$ and the on-shell value of $\sw$. Two checks fix the conventions. First, the $1/\epsilon$ poles of the $175$ diagrams cancel identically against those of the counterterms, structure by structure, once the on-shell relation $\MW = \cw\MZ$ is imposed, and for no other assignment of the relative signs between the vector, scalar and fermion self-energies. Second, the value of the $\Delta r$ parameter affecting the muon decay assembled from the same constants is $0.0303$, and the complete weak correction to the rate reproduces the result of Refs.~\cite{Bondarenko:2018sgg,Arbuzov:2021fmd} in both the $\alpha(0)$ and the $G_\mu$ scheme (Appendix~\ref{app:validation}).

One class of diagrams is left out of the computation on purpose: the ones with a photon exchanged between the incoming leptons and the photonic part of $\delta Z_e^{L,R}$, which together are UV finite, and the real emission that cancels their infrared divergence. For massless electrons the photonic vertex correction is independent of chirality and proportional to the tree-level vertex, so it multiplies $\Ma$ by a function of $s$. Hence, the contribution will cancel once the $Z$ boson density matrix is normalized. The soft and collinear real emission does the same. Hard initial-state radiation, instead, lowers the effective $\sqrt s$ of the process and boosts the $ZH$ system event by event and, as such, can be kept under control at the level of reconstruction and event selection.

After considering momentum conservation, the Dirac algebra and the Chisholm identity for the Levi-Civita terms, the whole renormalized one-loop amplitude collapses onto four spinor chains, 
\begin{equation}
\bar v(q_1)\slashed\eps^*P_\chi u(q_2),\qquad \text{and }\qquad	\bar v(q_1)\slashed k P_\chi u(q_2),
\end{equation}
with $\chi = L, R$, the latter multiplied by $\eps^*\vdot q_1$ or $\eps^*\vdot q_2$, with no three-gamma chain and no explicit $\epsilon$ tensor surviving. These are the chains of $\Ma$ and $\Mb$, Eqs.~\eqref{eq:Mtree} and~\eqref{eq:Mb}, with one more combination, and the one-loop amplitude can be written as
\begin{equation}
	\emm_1 = \sum_{\chi = L,R}\qty[ c_\chi(s,t)\, \Ma^\chi + b_\chi(s,t)\, \Mb^\chi + d_\chi(s,t)\, \emm_d^\chi ],
	\label{eq:M1structures}
\end{equation}
where $\Ma^\chi$ and $\Mb^\chi$ are the parts of $\Ma$ and $\Mb$ with the electron chirality $\chi$ (i.e.\ with $\gL\PL + \gR\PR$ replaced by $g_\chi P_\chi$), and
\begin{equation}
	\emm_d^\chi = -\frac{g^2\, g_\chi}{\cw^2\, \MZ\, (s - \MZ^2)}\; (q_1 - q_2)\vdot\eps^*\; \bar v(q_1)\, \slashed{k}\, P_\chi\, u(q_2)
	\label{eq:Md}
\end{equation}
is a structure that the $hZZ$ vertex does not produce. Eq.~\eqref{eq:M1structures} then \emph{defines} the six form factors of the complete correction, which are dimensionless and normalized to the tree level so that a pure vertex correction (such as the one induced by the three diagrams analyzed in isolation before), $c\,\Ma + b\,\Mb$, has $c_L = c_R = c$, $b_L = b_R = b$ and $d_\chi = 0$. The form factors depend on the scattering angle through the box diagrams and the $t$-channel propagators: $c_\chi$ and $b_\chi$ are even under $\cos\theta \to -\cos\theta$ and $d_\chi$ is odd.

\begin{table}[h]
\centering
\small
\setlength{\tabcolsep}{5pt}
\begin{tabular}{@{}llccc@{}}
\toprule
$\sqrt s$ & & $c_\chi$ [\%] & $b_\chi$ [$10^{-3}$] & $d_\chi(\cos\theta = 0.8)$ [$10^{-3}$] \\
\midrule
$240\GeV$ & $\klam$ diagrams, $\chi = L, R$ & $0.72 + 0.52\, i$ & $0.56 + 0.17\, i$ & $0$ \\
 & rest, $\chi = L$ & $-1.28 - 5.58\, i$ & $-9.52 - 1.99\, i$ & $1.44 + 1.19\, i$ \\
 & rest, $\chi = R$ & $8.58 - 0.47\, i$ & $-2.60 + 0.70\, i$ & $-0.01 - 0.01\, i$ \\
$365\GeV$ & $\klam$ diagrams, $\chi = L, R$ & $0.17 + 0.47\, i$ & $0.19 + 0.39\, i$ & $0$ \\
 & rest, $\chi = L$ & $-1.91 - 6.89\, i$ & $-3.31 - 2.29\, i$ & $1.41 + 3.46\, i$ \\
 & rest, $\chi = R$ & $9.33 + 0.67\, i$ & $-2.46 - 0.13\, i$ & $-0.05 - 0.02\, i$ \\
\bottomrule
\end{tabular}
\caption{The form factors of the one-loop amplitude, Eq.~\eqref{eq:M1structures}, at $\theta = \pi/2$ (for $d_\chi$, which vanishes there, at $\cos\theta = 0.8$): the contribution of the three diagrams with the Higgs boson self-coupling, $c(s)$ and $\beta(s)$ of Table~\ref{tab:numbers} per unit of $\klam$, and that of the rest of the weak correction, $\klam$ independent ($\alpha(0)$ scheme, photonic corrections to the initial state excluded). The complete SM value at $\klam = 1$ is the sum of the two lines of each chirality. The angular dependence is in Figure~\ref{fig:ffloop}.}
\label{tab:ffloop}
\end{table}

\begin{figure}[!htp]
\centering
\includegraphics[width=\textwidth]{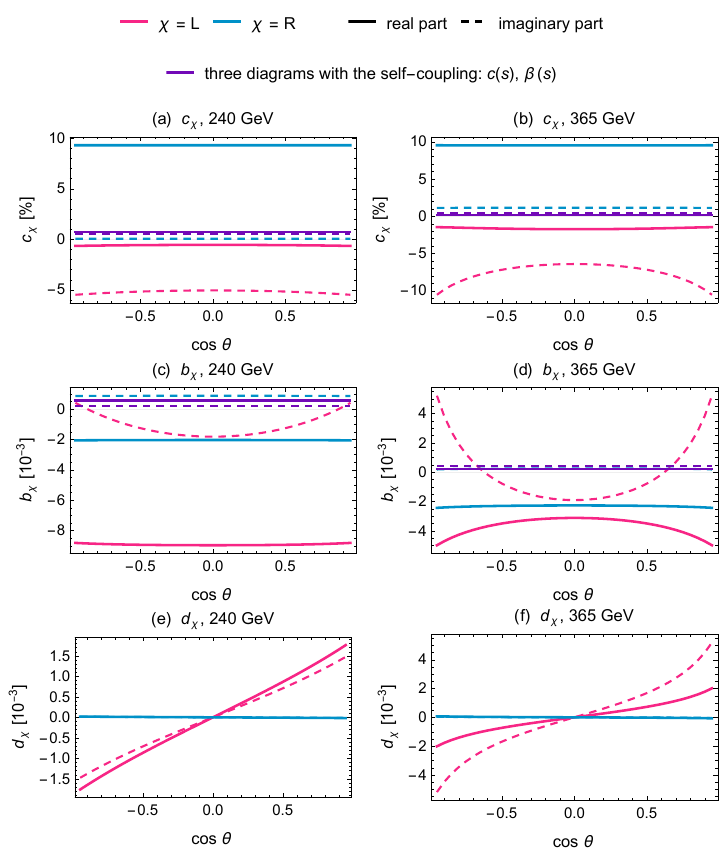}
\caption{The form factors of the complete one-loop amplitude at $\klam = 1$, Eq.~\eqref{eq:M1structures}, as functions of the scattering angle at $\sqrt s = 240\GeV$ (left column) and $365\GeV$ (right column), for left-handed (pink) and right-handed (cyan) electrons, real parts solid and imaginary parts dashed. The violet lines in panels (a)--(d) are the contributions of the three diagrams containing the Higgs boson self-coupling, $c(s)$ and $\beta(s)$, which do not depend on the chirality and on the angle.}
\label{fig:ffloop}
\end{figure}

Table~\ref{tab:ffloop} compares the form factors of the complete correction with those of the three diagrams with the Higgs boson self-coupling, and Figure~\ref{fig:ffloop} shows their dependence on the scattering angle. These loop corrections differ from the ones containing the self-coupling in three ways. First, they distinguish between the two chiralities. The chirality-independent part of $c_\chi$, a correction to $a_Z$ of a few percent, comes with a difference $\Re(c_R - c_L) \simeq 0.10$ at both energies, a correction of ten percent to the ratio $\gR/\gL$ of the electron couplings, which is the running of the effective weak mixing angle above the $Z$ pole and the $WW$ and $ZZ$ box diagrams that act on the left-handed electrons only. The parameter $c_L$ acquires an imaginary part of $-5\%$ at $240\GeV$ and $-6$ to $-9\%$ at $365\GeV$, from the $WW$ cut, which only the left-handed electrons source. Second, they produce the $b_Z$ structure with both chiralities, with the sign opposite to that of $\beta(s)$ and an order of magnitude larger: $b_L = -9.0\times 10^{-3}$ and $b_R = -2.0\times 10^{-3}$ at $240\GeV$, $-3.1\times 10^{-3}$ and $-2.3\times 10^{-3}$ at $365\GeV$ at a scattering angle of $\pi/2$. Their mean, $\bar b = (b_L + b_R)/2 = -5.5\times 10^{-3}$ at $240\GeV$, is the $b_Z$ of the SM at one loop, ten times the $\beta(s)$ induced by the Higgs self-coupling, and will affect the polarization of the $Z$ boson. Third, the box diagrams add the structure $\emm_d^\chi$, at the level of $10^{-3}$ for the left-handed electrons and two orders of magnitude below for the right-handed ones. The real parts of $c_\chi$ and $b_\chi$ are almost flat in $\cos\theta$ at $240\GeV$, while the imaginary part of $b_L$, and at $365\GeV$ also its real part, vary substantially over the angular range. These six functions are the whole content of the one-loop correction as far as the $Z$ boson is concerned and fully determine the density matrix of the $Z$ boson within the SM at one loop.

\subsection{The rate}  
\label{sub:rate}

To first order in $\klam \Mhhh$, which is the accuracy of a one-loop computation, Eq.~\eqref{eq:M2param} with the form factors~\eqref{eq:effective} gives the spin-summed squared amplitude
\begin{equation}
	\Msq = \Msqa\, \qty[1 + 2 \klam \Re c(s)] + \klam \Re \beta(s)\; X_b ,
	\label{eq:M2loop}
\end{equation}
in which the imaginary parts of the form factors do not appear. We checked Eq.~\eqref{eq:M2loop} against the full $\abs{\Ma + \klam \Mhhh}^2$ evaluated numerically with the complex form factors: the difference is the $\klam^2$ term, which we also computed and which matches its expected value. Integrating over the angles, the cross section becomes $\sigma = \sigma_{ 0}\qty(1 + \klam C_1)$, with
\begin{equation}
	C_1 = 2 \Re c(s) + \Re\beta(s)\; \frac{\sigma_b}{\sigma_{ 0}} ,
	\label{eq:C1}
\end{equation}
where $\sigma_b$ is the cross section contribution sourced by $X_b$. Numerically, the parameter amounts to $1.74\%$ at $240\GeV$ and $0.50\%$ at $365\GeV$. This is the coefficient introduced in Ref.~\cite{Degrassi:2016wml} and tabulated for Higgsstrahlung in Ref.~\cite{DiVita:2017vrr}, which we reproduce at the level of the quoted digits over the whole energy range once the Higgs wave-function renormalization is included (see Appendix~\ref{app:validation}). We take this as a further validation of our loop computation. Furthermore, the parameter fixes the normalization of $\beta(s)$ independently of any tomographic consideration: the same three diagrams that produce the well-known $1.7\%$ shift of the rate produce the $b_Z$ of Eq.~\eqref{eq:beta}. The rate and the polarization are sensitive to different combinations of the two form factors: the rate $C_1\klam$ is dominated by $c(s)$ and degenerate with everything that rescales the vertex. The polarization depends on $\beta(s)\klam$ alone and will be central to the analysis in Section~\ref{sub:rhoSM}.

The rest of the one-loop correction changes the rate as well, and to a larger extent. Integrated over the angles, the complete weak correction at $\klam = 1$ is $\delta_{\rm weak} = +3.2\%$ at $240\GeV$ and $+3.0\%$ at $365\GeV$ in the $\alpha(0)$ scheme. Of this, $C_1 + \delta Z_h = 1.6\%$ and $0.35\%$ is the Higgs boson self-coupling, with the wave-function renormalization of Eq.~\eqref{eq:dZh}, and the rest, amounting to $1.6\%$ and $2.7\%$, does not depend on $\klam$:
\begin{equation}
	\sigma = \sigma_{ 0}\qty[1 + \delta_{\rm rest} + \klam\, C_1 + \klam^2\, \delta Z_h], \qquad \delta_{\rm rest} = 1.6\%\ (240\GeV),\ 2.7\%\ (365\GeV) .
	\label{eq:sigmaloop}
\end{equation}
The correction can also be read form factor by form factor. At $240\GeV$, the chirality-blind part of $c_\chi$ gives $+8.7\%$ and its chirality-odd part $-2.1\%$. The mean $\bar b$ of the two $b_\chi$, the $b_Z$ of the SM at one loop, gives $-3.0\%$, and the remaining terms (the difference of the two $b_\chi$ and the boxes) $-0.4\%$. The three diagrams with the Higgs boson self-coupling, with their $1.6\%$, are thus not a minor part of the rate correction: at $240\GeV$ they account for half of it. The opposite holds for the polarization, studied in Section~\ref{sub:rhoSM}, where their contribution is an order of magnitude smaller than that of the remaining loops.
\section{The helicity density matrix of the $Z$ boson}  
\label{sec:density}

The polarization of the $Z$ boson is described by its helicity density matrix, which follows from the amplitude and, experimentally, can be reconstructed from the $Z$ boson decays. Section~\ref{sub:rhogeneral} sets up the computation for the general vertex of Section~\ref{sec:setup}, Section~\ref{sub:rhotree} gives analytical expressions for the SM result at tree level and Section~\ref{sub:rhoSM} details the SM one-loop corrections which the amplitude of Section~\ref{sec:loop} supports.

\subsection{From the vertex to the density matrix}  
\label{sub:rhogeneral}

\subsubsection{Projector and spin basis}  
\label{sub:projector}

The polarization of the $Z$ boson is described by the $3\times 3$ density matrix $\rho = R/\Tr R$ in the basis of its helicity states $\lambda = +1, 0, -1$. Writing the amplitude as $\emm = \eps^*_\mu(\lambda)\, \emm^\mu$, the unnormalized matrix $R$ is obtained as
\begin{equation}
	R_{\lambda\lambda'} = \sum_{\rm spins} \emm^\mu\, \emm^{\nu *}\; \mathcal P_{\mu\nu}(\lambda, \lambda'), \qquad \mathcal P_{\mu\nu}(\lambda, \lambda') \deq \eps^*_\mu(\lambda)\, \eps_\nu(\lambda'), 
	\label{eq:projector}
\end{equation}
where the sum runs over the spins of the initial-state particles. The projector $\mathcal P_{\mu\nu}(\lambda, \lambda')$ is given by tensor products of polarization vectors, which are the helicity vectors  describing the spin vector of the $Z$ boson. To give their explicit expressions, we consider the $\{\hat r, \hat n, \hat k\}$ triad commonly utilized in these studies. In the CM frame, $\hat k$ is given by the direction of the $Z$ boson and the remaining versors can be built with the help of an auxiliary direction, $\hat z$, which we take to be that of the positron, yielding
\begin{equation}
	\hat r = \frac{\cos\theta\, \hat k - \hat z}{\sin\theta}, \qquad \hat n = \frac{\hat z \times \hat k}{\sin\theta} = \hat k \times \hat r.
\end{equation}
As we can see, $\hat r$ lies in the production plane, $\hat n$ is normal to it, and $\{\hat r, \hat n, \hat k\}$ is right-handed. The corresponding polarization vectors are then
\begin{equation}
	\eps^\mu(\pm 1) = \mp\frac{1}{\sqrt 2}\qty(0,\, \hat r \pm i\, \hat n), \qquad
	\eps^\mu(0) = \frac{1}{\MZ}\qty(\vb*k,\, E_Z\, \hat k),
	\label{eq:polvectors}
\end{equation}
with $E_Z = (s + \MZ^2 - \MH^2)/(2\sqrt s)$ and $\vb*k = \sqrt\lam/(2\sqrt s)$. The overall phases of $\eps(\pm1)$ are subject to conventions and fix the signs of the coherences in the density matrix. Opposite signs for $\eps(\pm1)$ are  also commonly considered in the literature, and yield $\rho_{+0}$ and $\rho_{+-}$ with opposite sign with respect to our convention,  while $\rho_{0-}$ does not change.

The projector can be given in closed form in terms of the rotation generators. Let $n_1^\mu = (0, \hat r)$, $n_2^\mu = (0, \hat n)$ and $n_3^\mu = \eps^\mu(0)$ be the triad recast as four-vectors, orthogonal to $k$ and to each other, and $S^i$ the spin-one matrices in the basis $\ket{+1}, \ket{0}, \ket{-1}$ quantized along $n_3$, with $S^1$ and $S^2$ along $n_1$ and $n_2$:
\begin{equation}
	S^1 = \frac{1}{\sqrt2}\mqty(0 & 1 & 0 \\ 1 & 0 & 1 \\ 0 & 1 & 0), \qquad S^2 = \frac{1}{\sqrt2}\mqty(0 & -i & 0 \\ i & 0 & -i \\ 0 & i & 0), \qquad S^3 = \mqty(1 & 0 & 0 \\ 0 & 0 & 0 \\ 0 & 0 & -1).
\end{equation}
We also define $S^{ij} \deq S^i S^j + S^j S^i - \frac43\, \delta^{ij}\, \ii$, the traceless symmetric tensors. Then~\cite{Choi:1989yf}
\begin{equation}
	\mathcal P^{\mu\nu}(\lambda, \lambda') = \eps^{\mu *}(\lambda)\, \eps^{\nu}(\lambda') = \frac13\qty[ \qty(-g^{\mu\nu} + \frac{k^\mu k^\nu}{\MZ^2})\, \delta_{\lambda\lambda'} - \frac{3 i}{2\MZ}\, \epsilon^{\mu\nu\sigma\tau} k_\sigma\, n_{i\tau}\, \qty(S^i)_{\lambda\lambda'} - \frac32\, n_i^\mu n_j^\nu\, \qty(S^{ij})_{\lambda\lambda'} ] ,
	\label{eq:projectorS}
\end{equation}
which reduces to the usual polarization sum 
\begin{equation}
	\sum_{\lambda} \mathcal P_{\mu\nu}(\lambda, \lambda) = -g_{\mu\nu} + \frac{k_\mu k_\nu}{\MZ^2},
\end{equation}
when summed over the repeated polarizations. More details about the underlying conventions are discussed in Appendix~\ref{app:conventions}. 

Contracting the projector with $\emm^\mu \emm^{\nu*}$ and normalizing gives the density matrix in an explicitly covariant form
\begin{align}
	\rho &= \frac13\qty[\ii + \frac32\, P_i\, S^i + \frac34\, T_{ij}\, S^{ij}], \nn\\
	P_i &= \ev{S^i} = -\frac{i}{\MZ}\, \frac{\sum \epsilon^{\mu\nu\sigma\tau} \emm_\mu \emm^*_\nu\, k_\sigma\, n_{i\tau}}{\sum \abs{\emm}^2}, \nn\\
	T_{ij} &= \ev{S^{ij}} = \frac23\,\delta_{ij} - \frac{\sum\qty[(n_i\vdot\emm)(n_j\vdot\emm^*) + (n_j\vdot\emm)(n_i\vdot\emm^*)]}{\sum\abs{\emm}^2},
	\label{eq:rhoPT}
\end{align}
with $P_i$ the polarization vector of the $Z$ boson and $T_{ij}$ the traceless alignment tensor along the triad. The $\delta_{ij}$ term in the definition of the latter removes the trace of the bilinear in the second term, ensuring that $T_{ij}$ is traceless as well. All the sums are over the electron and positron spins and $\sum\abs{\emm}^2 = \sum_\lambda\abs{\eps^*(\lambda)\vdot\emm}^2$. 

Equation~\eqref{eq:rhoPT} makes it evident that $P_i$ is built from the antisymmetric part of $\emm_\mu \emm^*_\nu$, and its component along $\hat n$, as well as the $\hat n\hat k$ and $\hat n\hat r$ components of $T_{ij}$, need a non-vanishing relative phase between the components of the amplitude. The real couplings of the two $T$-even structures cannot provide it, but an absorptive part can, and so can the $T$-odd structure. 

For massless electrons the two chiralities do not interfere, and the sums over the electron spins in Eq.~\eqref{eq:rhoPT} reduce to sums over $\chi = L, R$, involving a single amplitude $\emm_\chi$ each. Since the vectors $n_i$ are orthonormal and orthogonal to $k$, we can make use of the projections $m_i^\chi \deq n_i\vdot\emm_\chi$ to write:
\begin{equation}
	P_i = -\frac{\epsilon_{ijl}\sum_\chi \Im\qty(m_j^\chi m_l^{\chi *})}{\sum_\chi \abs{\vb*m^\chi}^2}, \qquad
	T_{ij} = \frac23\,\delta_{ij} - \frac{2\sum_\chi \Re\qty(m_i^\chi m_j^{\chi *})}{\sum_\chi \abs{\vb*m^\chi}^2}, \qquad
	\abs{\vb*m^\chi}^2 = \sum_i \abs{m_i^\chi}^2 .
	\label{eq:rhom}
\end{equation}
The projections are scalar products and, as such, can be evaluated in any frame. In particular, in the CM frame the chiral current is 
\begin{equation}
	J^\mu_\chi=\bar v(q_1)\gamma^\mu P_\chi u(q_2) = \sqrt s\,(0, 1, i\sigma, 0) ,
\end{equation}
with $\sigma = +1$ for $\chi = L$ and $\sigma = -1$ for $\chi = R$, up to a phase for each chirality which cancels in Eq.~\eqref{eq:rhom}. The three structures of the vertex, restricted to the chirality $\chi$ as in Eq.~\eqref{eq:M1structures}, have then the following projections
\begin{equation}
	\begin{aligned}
		n_i\vdot\Ma^\chi &= -N_\chi\sqrt s\,\qty(\cos\theta,\; i\sigma,\; \frac{E_Z}{\MZ}\sin\theta), \\
		n_i\vdot\Mb^\chi &= -N_\chi\sqrt s\,\qty(\frac{q\vdot k}{\MZ^2}\cos\theta,\; i\sigma\,\frac{q\vdot k}{\MZ^2},\; \frac{\sqrt s}{\MZ}\sin\theta), \\
		n_i\vdot\Mbt^\chi &= N_\chi\sqrt s\,\frac{\sqrt\lam}{2\MZ^2}\qty(i\sigma,\; -\cos\theta,\; 0),
	\end{aligned}
	\qquad N_\chi = \frac{g^2\, g_\chi\, \MZ}{\cw^2\,(s - \MZ^2)} ,
	\label{eq:mproj}
\end{equation}
where the components on the right-hand sides are given for $i = 1, 2, 3$. 

In the literature, the density matrix describing the $Z$ boson helicity is often given in terms of its Fano decomposition 
\begin{equation}
	\rho = \frac13[\ii + \sum_a f_a \Lambda^a],
\end{equation}
where the parameters $P_i$ and $T_{ij}$ are collected in the Fano coefficients $f_a$ of the expansion in the orthogonal basis $\Lambda = \{S^1, S^2, S^3, S^{12}, S^{13}, S^{23},$ $\frac12(S^{11} - S^{22}), \frac{\sqrt3}{2} S^{33}\}$, with $\Tr\Lambda^a\Lambda^b = 2\delta^{ab}$. The coefficients are extracted through the projections
\begin{equation}
	f_a = \frac32\Tr[\rho\Lambda^a],
\end{equation}
and map onto the polarization and alignment tensor according to:
\begin{equation}
	\begin{aligned}
		f_1 &= \tfrac32\, P_1, & f_2 &= \tfrac32\, P_2, & f_3 &= \tfrac32\, P_3, \\
		f_4 &= \tfrac32\, T_{12}, & f_5 &= \tfrac32\, T_{13}, & f_6 &= \tfrac32\, T_{23}, \\
		f_7 &= \tfrac34\, \qty(T_{11} - T_{22}), & f_8 &= \tfrac{3\sqrt3}{4}\, T_{33} .
	\end{aligned}
	\label{eq:fanoPT}
\end{equation}

\subsubsection{The impact of anomalous couplings}  
\label{sub:rhoshift}

With the amplitude $a_Z \Ma + b_Z \Mb + \tilde b_Z \Mbt$ of Eq.~\eqref{eq:ampl_vertex}, the unnormalized density matrix can be written as
\begin{equation}
	R = \abs{a_Z}^2 R_{tt} + \qty( a_Z^* b_Z\, R_{bt} + a_Z^* \tilde b_Z\, R_{\tilde b t} + {\rm h.c.} ) + \order{b_Z^2,\, \tilde b_Z^2,\, b_Z\tilde b_Z} ,
\end{equation}
where $R_{tt}$ is built from $\Ma=\emm_{ 0}$ alone through Eq.~\eqref{eq:projector}, and determines the tree-level expression of the density matrix
\begin{equation}
	\rho_{ 0} = \frac{R_{tt}}{\Tr R_{tt}}. 
\end{equation}

The matrices $R_{bt}$ and $R_{\tilde b t}$ are built in similar fashion, from the products of $\Mb$ and $\Mbt$ with $\Ma^*$. To first order in the ratios $b_Z/a_Z$ and $\tilde b_Z/a_Z$ we then have
\begin{equation}
		\rho = \rho_{ 0} + \Re\frac{b_Z}{a_Z}\; \partial_R \rho + \Im\frac{b_Z}{a_Z}\; \partial_I \rho + \Re\frac{\tilde b_Z}{a_Z}\; \tilde\partial_R \rho + \Im\frac{\tilde b_Z}{a_Z}\; \tilde\partial_I \rho ,
	\label{eq:rhoexp}
\end{equation}
in which we introduce the notation $\partial_{R(I)}$ and $\tilde\partial_{R(I)}$ to indicate differentiation with respect to the real (imaginary) part of $b_Z/a_Z$ and $\tilde b_Z/a_Z$, respectively, both evaluated at zero\footnote{Two derivatives per coupling are necessary because $\rho$ is not an analytic function of the couplings: with $a_Z = 1$, $R$ depends on $b_Z$ and $b_Z^*$ through $b_Z R_{bt} + b_Z^* R_{bt}^\dagger = \Re b_Z\, H + \Im b_Z\, A$. }. More explicitly, we have  
\begin{equation}
	\partial_R\rho = \frac{H - \rho_{ 0} \Tr H}{\Tr R_{tt}}, \qquad \partial_I\rho = \frac{A - \rho_{ 0} \Tr A}{\Tr R_{tt}} ,
\end{equation}
where we defined the two hermitian matrices $H = R_{bt} + R_{bt}^\dagger$ and $A = i\qty(R_{bt} - R_{bt}^\dagger)$. The matrices $\tilde H$ and $\tilde A$, entering  $\tilde\partial_R$ and $\tilde\partial_I$, are built from $R_{\tilde b t}$ in a similar way. Notice that $a_Z$ enters the above expressions only through ratios, so the normalization of the vertex drops out of the density matrix. Notice also that $\rho$ is sensitive to $\Im b_Z$ and $\Re\tilde b_Z$, through $A$ and $\tilde H$. Importantly, these parameters do not affect the production rate and, consequently, cannot be constrained by measuring the total cross section of the process.  

The four derivatives have a closed form: in components, $\lambda, \lambda' = +1, 0, -1$, we have
\begin{align}
		\qty(\partial_R\rho)_{\lambda\lambda'} &= \kappa\, \qty(2\qty(\rho_0)_{00} - \delta_{\lambda 0} - \delta_{\lambda' 0})\, \qty(\rho_0)_{\lambda\lambda'} , 
		\\
		\qty(\partial_I\rho)_{\lambda\lambda'} &= -i\kappa\, \qty(\delta_{\lambda 0} - \delta_{\lambda' 0})\, \qty(\rho_0)_{\lambda\lambda'} , 
		\\
		\qty(\tilde\partial_R\rho)_{\lambda\lambda'} &= -i r\, \qty(\lambda - \lambda')\, \qty(\rho_0)_{\lambda\lambda'} , 
		\\
		\qty(\tilde\partial_I\rho)_{\lambda\lambda'} &= r\, \qty[\lambda + \lambda' - 2\qty(\qty(\rho_0)_{++}-\qty(\rho_0)_{--})]\, \qty(\rho_0)_{\lambda\lambda'} ,	\label{eq:dops}
\end{align}
where we defined 
\begin{equation}
	\kappa = \frac{\lam}{4\MZ^2\; q\vdot k} , \qquad r = \frac{\sqrt\lam}{2\MZ^2},
\end{equation}
with $\lambda$ here given as in Eq.~\eqref{eq:lambda}: $\lam  = s^2 + \MZ^4 + \MH^4 - 2 s \MZ^2 - 2 s \MH^2 - 2 \MZ^2 \MH^2$.

Focusing for the moment on the CP-even coupling, we see that the transverse block of $\partial_R\rho$ is given by a rescaling of the tree-level one:
\begin{equation}
	\qty(\partial_R \rho)_{\lambda\lambda'} = 2\kappa\,(\rho_0)_{00}\,(\rho_0)_{\lambda\lambda'}, \quad \lambda,\lambda' = \pm1, \qquad 2\kappa\,(\rho_0)_{00} = \frac{2(q\vdot k)\,\lam\sin^2\theta}{\MZ^2 \qty(\lam \sin^2\theta + 8 \MZ^2 s)}.
	\label{eq:dRtrans}
\end{equation} 
The longitudinal entry compensates
\begin{equation}
	(\partial_R\rho)_{00} = -2\kappa\,(\rho_0)_{00}\qty[1 - (\rho_0)_{00}] = -\frac{8 s\,(q\vdot k)\,\lam\sin^2\theta\,(1 + \cos^2\theta)}{\qty(\lam \sin^2\theta + 8 \MZ^2 s)^2},
\end{equation}
whereas the $\pm 0$ coherences receive a real shift from $\Re b_Z$ and an imaginary one from $\Im b_Z$:
\begin{align}
	\qty(\partial_R\rho)_{+0} &= \frac{\sqrt s\, \lam\, \qty(\cos\theta + A_e)\sin\theta\, \qty(8\MZ^2 s \cos^2\theta - \lam\sin^2\theta)}{\sqrt2\, \MZ\, \qty(\lam \sin^2\theta + 8 \MZ^2 s)^2}, 
	\\
	\qty(\partial_I\rho)_{+0} &= -i\, \frac{\sqrt s\, \lam\, \qty(\cos\theta + A_e)\sin\theta}{\sqrt2\, \MZ\, \qty(\lam \sin^2\theta + 8 \MZ^2 s)}, \\
	\qty(\partial_R\rho)_{0-} &= -\frac{\sqrt s\, \lam\, \qty(\cos\theta - A_e)\sin\theta\, \qty(8\MZ^2 s \cos^2\theta - \lam\sin^2\theta)}{\sqrt2\, \MZ\, \qty(\lam \sin^2\theta + 8 \MZ^2 s)^2}, 
	\\
	\qty(\partial_I\rho)_{0-} &= -i\, \frac{\sqrt s\, \lam\, \qty(\cos\theta - A_e)\sin\theta}{\sqrt2\, \MZ\, \qty(\lam \sin^2\theta + 8 \MZ^2 s)},
	\label{eq:dIrho}
\end{align}
with $A_e$ given in Eq.~\eqref{eq:Xtilde}. The remaining entries of $\partial_I\rho$ all vanish. 

Coming to the CP-odd coupling, a real $\tilde b_Z$ shifts the imaginary parts of the coherences as follows
\begin{align}
	\qty(\tilde\partial_R\rho)_{+0} &= i\,\frac{\sqrt2\,(q\vdot k)\sqrt{s\lam}\,(\cos\theta + A_e)\sin\theta}{\MZ \qty(\lam \sin^2\theta + 8 \MZ^2 s)}, 
	\\
	\qty(\tilde\partial_R\rho)_{0-} &= -i\,\frac{\sqrt2\,(q\vdot k)\sqrt{s\lam}\,(\cos\theta - A_e)\sin\theta}{\MZ \qty(\lam \sin^2\theta + 8 \MZ^2 s)}, \\
	\qty(\tilde\partial_R\rho)_{+-} &= -2i\,\frac{s\sqrt\lam\,\sin^2\theta}{\lam \sin^2\theta + 8 \MZ^2 s},
	\label{eq:dRhoBT}
\end{align}
while an imaginary component produces real shifts:
\begin{align}
	\qty(\tilde\partial_I\rho)_{++} &= \frac{2\, s\sqrt\lambda\, \qty(1 + \cos^2\theta + 2 A_e\cos\theta)\qty[\lambda\sin^2\theta + 8 m_Z^2 s \qty(1 - A_e\cos\theta)]}{\qty(\lambda\sin^2\theta + 8 m_Z^2 s)^2}, 
	\\
	\qty(\tilde\partial_I\rho)_{--} &= -\frac{2\, s\sqrt\lambda\, \qty(1 + \cos^2\theta - 2 A_e\cos\theta)\qty[\lambda\sin^2\theta + 8 m_Z^2 s \qty(1 + A_e\cos\theta)]}{\qty(\lambda\sin^2\theta + 8 m_Z^2 s)^2}, 
	\\
	\qty(\tilde\partial_I\rho)_{00} &= -\frac{32\, A_e\, s\sqrt\lambda\, (q\vdot k)^2 \sin^2\theta\cos\theta}{\qty(\lambda\sin^2\theta + 8 m_Z^2 s)^2}, 
	\\
	\qty(\tilde\partial_I\rho)_{+0} &= -\frac{\sqrt{2 s\lambda}\, (q\vdot k)\qty(\cos\theta + A_e)\sin\theta\, \qty[\lambda\sin^2\theta + 8 m_Z^2 s \qty(1 - 2 A_e\cos\theta)]}{m_Z \qty(\lambda\sin^2\theta + 8 m_Z^2 s)^2}, 
	\\
	\qty(\tilde\partial_I\rho)_{0-} &= -\frac{\sqrt{2 s\lambda}\, (q\vdot k)\qty(\cos\theta - A_e)\sin\theta\, \qty[\lambda\sin^2\theta + 8 m_Z^2 s \qty(1 + 2 A_e\cos\theta)]}{m_Z \qty(\lambda\sin^2\theta + 8 m_Z^2 s)^2}, 
	\\
	\qty(\tilde\partial_I\rho)_{+-} &= -\frac{16\, A_e\, m_Z^2\, s^2\sqrt\lambda\, \sin^2\theta\cos\theta}{\qty(\lambda\sin^2\theta + 8 m_Z^2 s)^2}.
	\label{eq:dImBt}
\end{align}

The shifts in the remaining entries of the density matrix can be obtained from the ones presented above through the hermiticity condition $\rho=\rho^\dagger$.

\subsection{The SM at tree level}  
\label{sub:rhotree}

At tree level the SM gives $a_Z = 1$, $b_Z = \tilde b_Z = 0$, and the amplitude is $\emm_{ 0} = \Ma$, Eq.~\eqref{eq:MtreeSM}. Eq.~\eqref{eq:projector} then gives 
\begin{equation}
	\rho_{ 0} = \frac{1}{\lam \sin^2\theta + 8 \MZ^2 s}\mqty( T_+ & -C_+ & P \\ -C_+ & L & C_- \\ P & C_- & T_- ),
	\label{eq:rhotree}
\end{equation}
with 
\begin{align}
		T_\pm &= 2\MZ^2 s\, \qty(1 + \cos^2\theta \pm 2 A_e \cos\theta), 
		\\ 
		L &= 4 (q\vdot k)^2 \sin^2\theta, 
		\\
		C_\pm &= 2\sqrt2\, \MZ\sqrt s\, (q\vdot k)\qty(\cos\theta \pm A_e)\sin\theta, 
		\\ 
		P &= 2 \MZ^2 s \sin^2\theta,
\end{align}
and the electron asymmetry $A_e$ of Eq.~\eqref{eq:Xtilde}. All the entries are real. The longitudinal entry carries the $\sin^2\theta$ of the squared amplitude, the two transverse ones the $1 + \cos^2\theta$ of a spin-one particle produced by a vector current, split by the parity-violating $A_e \cos\theta$ into a net helicity along $\hat k$. The $+-$ coherence is the alignment in the production plane, the $\pm 0$ coherences the transverse polarization in the plane, which vanishes along the beam. The density matrix measured when all the events are used, $\bar\rho = \int R\, \dd\!\cos\theta / \int \Tr R\, \dd\!\cos\theta$, is
\begin{equation}
	\bar\rho_{ 0} = \frac{1}{\lam + 12 \MZ^2 s}\mqty( 4\MZ^2 s & X & 2\MZ^2 s \\ X & 4(q\vdot k)^2 & X \\ 2\MZ^2 s & X & 4\MZ^2 s ), \qquad X = -\frac{3\pi}{2\sqrt2}\, A_e\, \MZ \sqrt s\, (q\vdot k), 
	\label{eq:rhotreeav}
\end{equation}
whose $00$ entry is the longitudinal fraction of Eq.~\eqref{eq:fL}. Notice that the $\pm0$ coherences survive the average only through $A_e$: the term odd in $\cos\theta$ integrates to zero, and what is left is proportional to the parity-violating coupling. Figure~\ref{fig:density}(a) shows the entries of $\bar\rho_{ 0}$ as functions of the energy.

In terms of the Fano coefficients of Section~\ref{sub:projector}, the state at fixed scattering angle is
\begin{equation}
	\begin{gathered}
		f_1 = -\frac{12\, A_e\, \MZ\sqrt s\, (q\vdot k)\sin\theta}{\lam \sin^2\theta + 8 \MZ^2 s}, \qquad f_3 = \frac{12\, A_e\, \MZ^2\, s\cos\theta}{\lam \sin^2\theta + 8 \MZ^2 s}, \qquad f_5 = -\frac{12\, \MZ\sqrt s\, (q\vdot k)\sin\theta\cos\theta}{\lam \sin^2\theta + 8 \MZ^2 s}, \\
		f_7 = \frac{6\, \MZ^2\, s\sin^2\theta}{\lam \sin^2\theta + 8 \MZ^2 s}, \qquad f_8 = \frac{\sqrt 3}{2}\qty[1 - \frac{12\, (q\vdot k)^2\sin^2\theta}{\lam \sin^2\theta + 8 \MZ^2 s}], \qquad f_2 = f_4 = f_6 = 0 ,
	\end{gathered}
	\label{eq:fanotree}
\end{equation}
and the inclusive state
\begin{equation}
	\begin{gathered}
	\bar f_1 = -\frac{9\pi}{2}\, \frac{A_e\, \MZ\sqrt s\, (q\vdot k)}{\lam + 12 \MZ^2 s}, \qquad \bar f_7 = \frac{6\, \MZ^2\, s}{\lam + 12 \MZ^2 s}, \qquad \bar f_8 = -\frac{\sqrt 3\, \lam}{\lam + 12 \MZ^2 s}, 
	\\
	\bar f_2 = \bar f_3 = \bar f_4 = \bar f_5 = \bar f_6 = 0 .
	\end{gathered}
	\label{eq:fanotreeav}
\end{equation}

\subsection{The SM at one loop}  
\label{sub:rhoSM}

At one loop, and strictly to first order in the anomalous coupling expansion, the density matrix of the SM is
\begin{equation}
	\rho_{\rm SM}(\klam) = \rho_{ 0} + \klam\, \delta\rho_{hhh} + \delta\rho_{\rm rest},
	\label{eq:rhoSM}
\end{equation}
with $\delta\rho_{hhh}$ the shift produced by the three diagrams of Section~\ref{sub:kdiagrams} per unit of $\klam$ and $\delta\rho_{\rm rest}$ the rest of the one-loop correction in Section~\ref{sub:fullloop}.

For each chirality, the one-loop amplitude of Eq.~\eqref{eq:M1structures} is $\emm_1^\chi = c_\chi\Ma^\chi + b_\chi\Mb^\chi + d_\chi\emm_d^\chi$, and the shift of the unnormalized matrix is
\begin{equation}
	\delta R_{\lambda\lambda'} = \sum_{\chi = L,R} \eps^*_\mu(\lambda)\, \eps_\nu(\lambda')\qty( \Ma^{\chi\mu}\, \emm_1^{\chi\nu *} + \emm_1^{\chi\mu}\, \Ma^{\chi\nu *} ),
	\label{eq:dRloop}
\end{equation}
with $\delta\rho = (\delta R - \rho_{ 0}\Tr\delta R)/\Tr R_{tt}$, as in Eq.~\eqref{eq:rhoexp}. Through the projector, $\delta R$ is the bilinear of Eq.~\eqref{eq:rhom} with $m_j m_l^*$ replaced by $m_j\,\delta m_l^* + \delta m_j\, m_l^*$, where $m_i^\chi = n_i\vdot\Ma^\chi$ and $\delta m_i^\chi = n_i\vdot\emm_1^\chi$. The projections of $\Ma^\chi$ and $\Mb^\chi$ are in Eq.~\eqref{eq:mproj}, and those of the box structure of Eq.~\eqref{eq:Md} are
\begin{equation}
	n_i\vdot\emm_d^\chi = N_\chi\, \frac{\sqrt{s\lam}}{2\MZ^2}\qty(\sin^2\theta,\; 0,\; -\frac{E_Z}{\MZ}\sin\theta\cos\theta),
	\label{eq:mbox}
\end{equation}
where the right-hand side is again explicitly given for $i=1,2,3$. It gives no component along $\hat n$ and, apart from the coupling $g_\chi$ in $N_\chi$, the same shift for the two chiralities. 

\subsubsection{The Higgs self-coupling}  
\label{sub:size}
\begin{figure}[h]
\centering
\includegraphics[width=\textwidth]{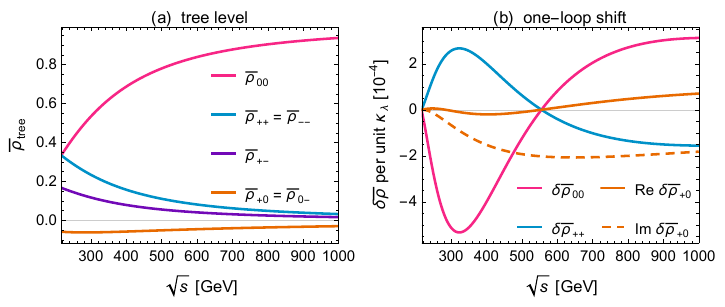}
\caption{(a) The entries of the angle-averaged density matrix of the $Z$ at tree level, Eq.~\eqref{eq:rhotreeav}. (b) Their shift per unit of $\klam$, due to the correction $\delta\bar\rho_{hhh}$, with $\Re\beta(s)$ and $\Im\beta(s)$ of Figure~\ref{fig:formfactors}. The shifts $\delta\bar\rho_{--} = \delta\bar\rho_{++}$, $\delta\bar\rho_{+-} = \delta\bar\rho_{++}/2$ and $\delta\bar\rho_{0-} = \delta\bar\rho_{+0}^{\,*}$ are not shown.}
\label{fig:density}
\end{figure}

For the three diagrams with the Higgs self-coupling, $c_\chi = c(s)$, $b_\chi = \beta(s)$ and $d_\chi = 0$, so $b_Z/a_Z = \klam\beta(s) + \order{\klam^2}$, and Eq.~\eqref{eq:rhoexp} gives $\delta\rho_{hhh} = \Re\beta\, \partial_R\rho + \Im\beta\, \partial_I\rho$. The form factor $c(s)$ drops out in the expansion, and with it the whole dependence of the rate on $\klam$. Figure~\ref{fig:density} and Table~\ref{tab:rho} give the numbers for the inclusive matrix.

The shifts are of order $10^{-4}$: $\delta\bar\rho_{00} = -2\times 10^{-4}$ per unit of $\klam$ at $240\GeV$ and $-5\times 10^{-4}$ at $365\GeV$. The real shift of the coherences is suppressed by $A_e$ and by the zero of $\lam - 4\MZ^2 s$ at $\sqrt s \simeq 284\GeV$, while the imaginary one, due to $\Im\beta$, reaches $1.3\times 10^{-4}$ at $365\GeV$ and is the largest off-diagonal effect. The same quantities for a genuine anomalous coupling are obtained by replacing $\klam \Re\beta$ with $b_Z$. 

\begin{table}[h]
\centering
\begin{tabular}{lcccc}
\toprule
 & $\bar\rho_{++} = \bar\rho_{--}$ & $\bar\rho_{00}$ & $\bar\rho_{+-}$ & $\bar\rho_{+0} = \bar\rho_{0-}^{\,*}$ \\
\midrule
$240\GeV$, tree & $0.301$ & $0.397$ & $0.151$ & $-0.0614$ \\
\quad shift per unit $\klam$ & $+0.98\times 10^{-4}$ & $-1.95\times 10^{-4}$ & $+0.49\times 10^{-4}$ & $(+0.05 - 0.08\, i)\times 10^{-4}$ \\
$365\GeV$, tree & $0.179$ & $0.641$ & $0.090$ & $-0.0602$ \\
\quad shift per unit $\klam$ & $+2.43\times 10^{-4}$ & $-4.86\times 10^{-4}$ & $+1.21\times 10^{-4}$ & $(-0.18 - 1.28\, i)\times 10^{-4}$ \\
\bottomrule
\end{tabular}
\caption{The angle-averaged density matrix of the $Z$ at tree level and its shift per unit of $\klam$ at the FCC-ee energies, using the form factors of Table~\ref{tab:numbers}. The shift of $\bar\rho_{0-}$ is the complex conjugate of that of $\bar\rho_{+0}$.}
\label{tab:rho}
\end{table}

\subsubsection{The complete correction}  
\label{sub:baseline_shift}

The complete inclusive one-loop density matrix follows from the average over the scattering angle of the unnormalized matrices, $\bar R_{tt} = \int R_{tt}\,\dd\!\cos\theta$ and $\delta\bar R = \int\delta R\,\dd\!\cos\theta$ with $\delta R$ of Eq.~\eqref{eq:dRloop}, summed over the initial spins:
\begin{equation}
\delta\bar\rho_{\rm SM} = \frac{\delta\bar R - \bar\rho_{ 0}\Tr\delta\bar R}{\Tr\bar R_{tt}} .	
\end{equation}
Table~\ref{tab:baseline} gives the shift of the modified Fano coefficients, next to the shift per unit of $\klam$. Four coefficients, omitted in the table, do not move after the average as their shifts are odd in $\cos\theta$: they are the CP-odd coefficients sourced by $\tilde b_Z$, which the SM, CP even, does not populate.

\begin{table}[h]
\centering
\small
\begin{tabular}{@{}llcccc@{}}
\toprule
$\sqrt s$ & $f_a$ & tree & SM one-loop shift & per unit $\klam$ & $10\%$ of $\delta f_{\rm rest}$ \\
\midrule
$240\GeV$ & $\frac32 P_1$ & $-0.261$ & $+0.136$ & $0.2\times 10^{-4}$ & $1.4\times 10^{-2}$ \\
 & $\frac32 T_{23}$ & $0$ & $-1.4\times 10^{-3}$ & $0.3\times 10^{-4}$ & $1.4\times 10^{-4}$ \\
 & $\frac34(T_{11} - T_{22})$ & $0.452$ & $-1.3\times 10^{-3}$ & $1.5\times 10^{-4}$ & $1.5\times 10^{-4}$ \\
 & $\frac{3\sqrt3}{4}T_{33}$ & $-0.165$ & $-6.3\times 10^{-3}$ & $5.1\times 10^{-4}$ & $6.8\times 10^{-4}$ \\
$365\GeV$ & $\frac32 P_1$ & $-0.255$ & $+0.134$ & $-0.8\times 10^{-4}$ & $1.3\times 10^{-2}$ \\
 & $\frac32 T_{23}$ & $0$ & $-7.4\times 10^{-3}$ & $5.4\times 10^{-4}$ & $7.9\times 10^{-4}$ \\
 & $\frac34(T_{11} - T_{22})$ & $0.269$ & $-5.0\times 10^{-3}$ & $3.6\times 10^{-4}$ & $5.3\times 10^{-4}$ \\
 & $\frac{3\sqrt3}{4}T_{33}$ & $-0.800$ & $-2.35\times 10^{-2}$ & $1.26\times 10^{-3}$ & $2.5\times 10^{-3}$ \\
\bottomrule
\end{tabular}
\caption{SM one-loop shift of the Fano coefficients of the inclusive density matrix (weak corrections, $\alpha(0)$ scheme), next to the tree-level value, the shift per unit of $\klam$ and the theory uncertainty assumed in Section~\ref{sec:stat}, $10\%$ of the $\klam$-independent shift. The coefficients not listed vanish at tree level and at one loop. The shifts are compared with the statistical errors of the tomography in Table~\ref{tab:baselineErr}.}
\label{tab:baseline}
\end{table}


\section{The $Z$ boson as a qutrit} 
\label{sec:QIT}

The density matrix of the $Z$ boson describes a qutrit, whose properties can be characterized with the tools of quantum information theory. In this section we consider the purity and the coherence of the state, in Section~\ref{sub:purity}, its quantum contextuality, in Section~\ref{sub:contextuality}, and its magic, in Section~\ref{sub:magic}. We then study how they are modified by the anomalous couplings of the vertex~\eqref{eq:vertex}. All results presented here are at tree level and exact in the couplings.

The purity and the contextuality witnesses depend only on the spectrum of the density matrix, and their dependence on the couplings is made evident in the helicity basis. On every helicity amplitude, for both chiralities of the electron, the vertex~\eqref{eq:vertex} acts multiplicatively, $\emm(\lambda) = w_\lambda\, \Ma(\lambda)$, with
\begin{equation}
	w_{\pm1} = a_Z + \frac{q\vdot k}{\MZ^2}\, b_Z \mp i\, r\, \tilde b_Z , \qquad w_0 = a_Z + \frac{s}{q\vdot k}\, b_Z ,
\end{equation}
so that the density matrix is 
\begin{equation}
	\rho = \frac{W \rho_0 W^\dagger}{\Tr\qty[W \rho_0 W^\dagger]}, \qquad W = {\rm diag}\qty(w_{+1}, w_0, w_{-1}),
\end{equation}
both at fixed angle and for the inclusive matrix. A diagonal $W$ whose entries have equal moduli is unitary, hence leaves the spectrum unchanged. The couplings $\Im b_Z$ and $\Re\tilde b_Z$ only change the phases of the $w_\lambda$ at first order and, therefore, enter the purity and the contextuality witnesses at second order. The coupling $\Re b_Z$, which changes the relative weight of the longitudinal and transverse amplitudes, and $\Im\tilde b_Z$, which changes that of the two transverse helicities, instead enter linearly.

\subsection{Purity and coherence}
\label{sub:purity}

\begin{figure}[!htb]
\begin{center}
\includegraphics[width=0.5\textwidth]{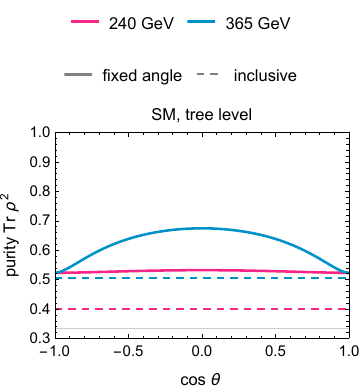}\\[2mm]
\includegraphics[width=\textwidth]{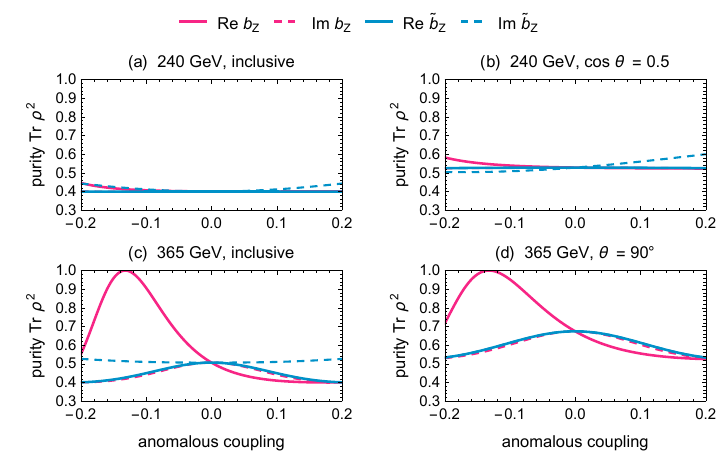}
\caption{The purity $\Tr\rho^2$ of the $Z$ boson state. Upper panel: in the SM at tree level, as a function of $\cos\theta$ for $\sqrt s=240$ and $365\GeV$, with the inclusive values as dashed lines. Lower panels: as a function of the real and imaginary parts of $b_Z$ and $\tilde b_Z$, varied one at a time with $a_Z = 1$, for the inclusive state and at fixed angle.}
\label{fig:purity}
\end{center}
\end{figure}

The purity of the state,
\begin{equation}
	P = \Tr\rho^2 = \sum_{k=1}^3 \rho_k^2 ,
\end{equation}
equals $1$ for a pure state and $1/3$ for the maximally mixed state of a qutrit. At fixed angle the state is of rank two, since each chirality of the electron prepares a $Z$ boson in a pure state and the two add incoherently. As a consequence the purity is larger, as shown in the upper panel of Figure~\ref{fig:purity}. At $240\GeV$ it depends little on the angle, $P \simeq 0.52$--$0.53$, whereas at $365\GeV$ it grows from $0.52$ along the beam to $0.674$ at $\theta = \pi/2$, where the longitudinal polarization dominates. The average over the scattering angle combines states with different polarizations and therefore lowers the purity. In the SM, the inclusive density matrix has $P = 0.400$ at $240\GeV$ and $0.506$ at $365\GeV$.  

The lower panels of Figure~\ref{fig:purity} show the purity as a function of the anomalous couplings, varied one at a time with $a_Z = 1$. The largest effect is again due to $\Re b_Z$. For $\Re b_Z = -\MZ^2/(q\vdot k)$, that is $-0.13$ at $365\GeV$ and $-0.33$ at $240\GeV$, the transverse amplitudes vanish, the $Z$ boson is produced in the pure longitudinal state and $P = 1$. Around the SM point the slope is $\partial P/\partial\Re b_Z = -2.0$ at $365\GeV$, for the inclusive state as well as at $\theta=\pi/2$, and only $-0.02$ for the inclusive state at $240\GeV$. The CP-odd $\Im\tilde b_Z$ changes the balance between the two transverse helicities and enters linearly only at fixed scattering angle away from $\pi/2$, $\partial P/\partial\Im\tilde b_Z = 0.27$ at $240\GeV$ and $\cos\theta = 0.5$. Its effect on the inclusive state and at $\theta=\pi/2$ is of second order, as is that of $\Im b_Z$ and $\Re\tilde b_Z$ everywhere. For couplings of order $10^{-3}$ the purity changes by at most $2\times 10^{-3}$.

Besides the purity, the mixedness of the state is measured by the linear entropy, the von Neumann entropy and the R\'enyi-2 entropy,
\begin{equation}
S_L=\tfrac32\,(1-P)\in[0,1],\qquad S_{\rm vN}=-\Tr\rho\ln\rho\in[0,\ln3],\qquad S_2=-\ln P.
\end{equation}
All three depend only on the spectrum. $S_L$ and $S_2$ are functions of $P$, while $S_{\rm vN}$ depends on the full spectrum. $S_{\rm vN}$ vanishes for a pure state and equals $\ln 3\simeq1.10$ for the maximally mixed qutrit. 

\begin{figure}[!htb]
\begin{center}
\includegraphics[width=\textwidth]{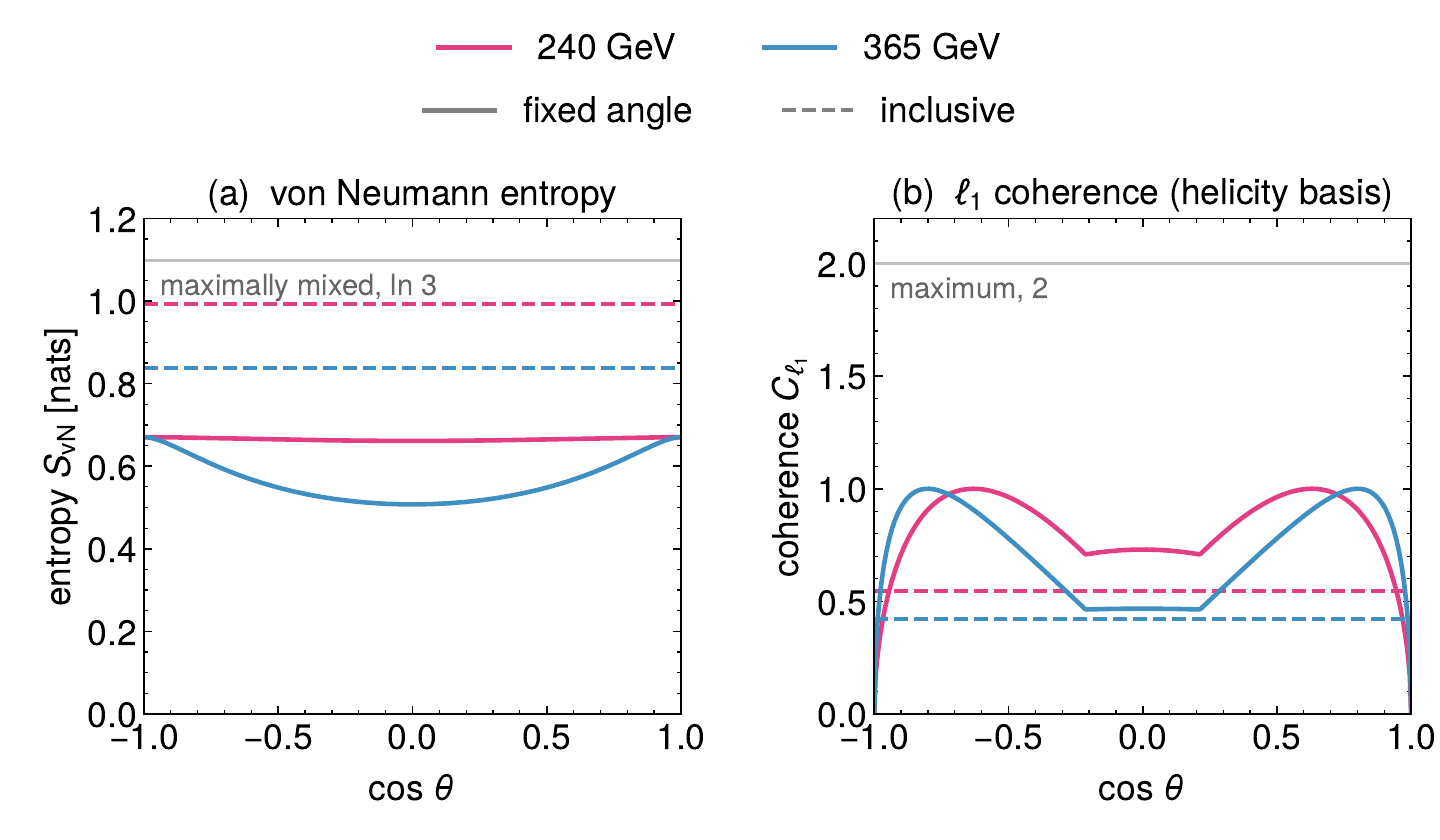}
\caption{(a) The von Neumann entropy $S_{\rm vN}=-\Tr\rho\ln\rho$ and (b) the $\ell_1$-norm coherence $C_{\ell_1}=\sum_{i\neq j}|\rho_{ij}|$ in the helicity basis of the $Z$ boson, as computed at tree level in the SM, as functions of $\cos\theta$ for $\sqrt s=240$\,GeV (pink) and 365\,GeV (blue), at fixed angle (solid) and for the inclusive state (dashed). The gray lines mark the maxima: $\ln 3$ for the maximally mixed qutrit and $2$ for $C_{\ell_1}$.}
\label{fig:entropy_coherence_angle}
\end{center}
\end{figure}

\begin{figure}[!htbp]
\begin{center}
\includegraphics[width=1\textwidth]{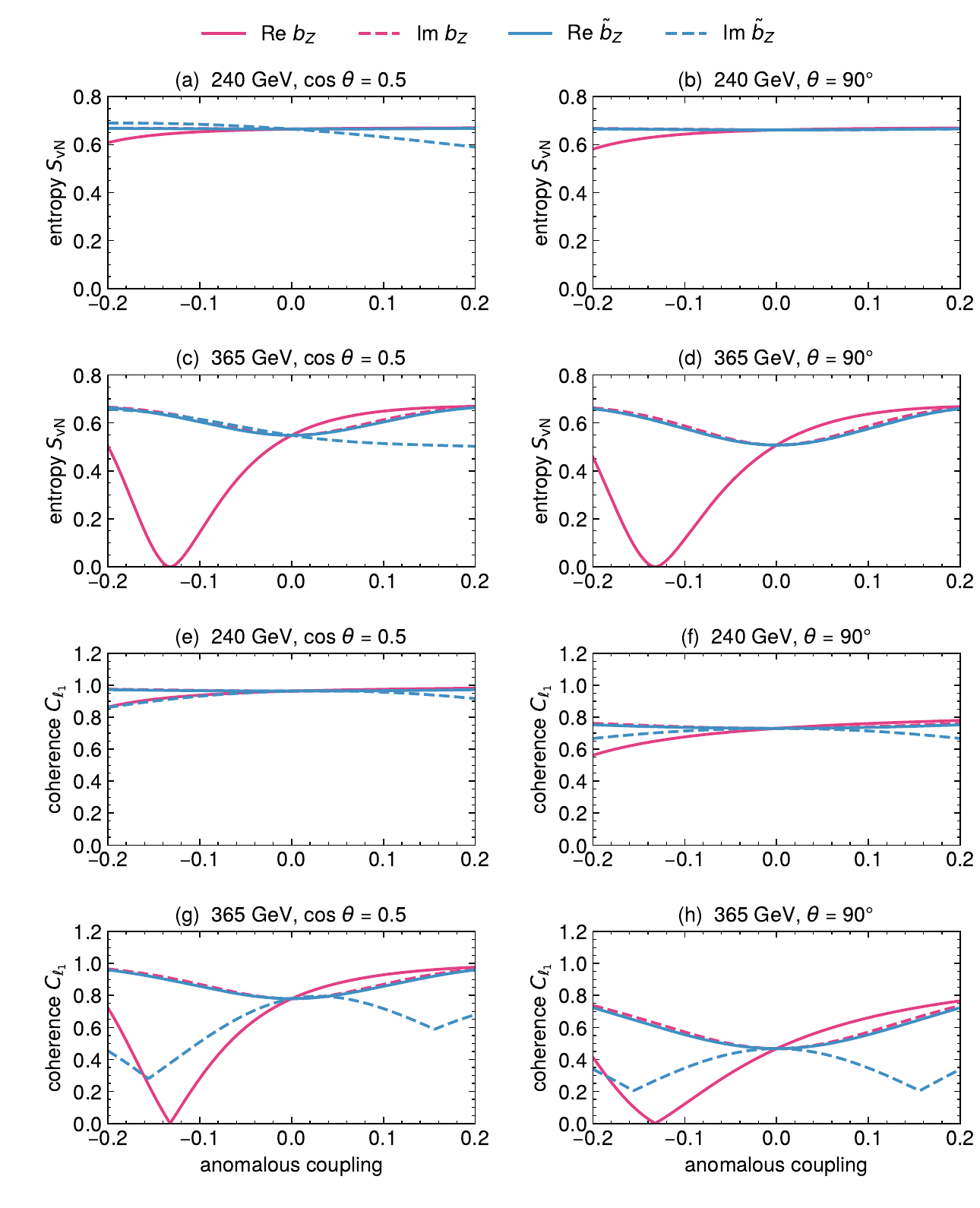}
\caption{The von Neumann entropy $S_{\rm vN}$, panels (a)--(d), and the $\ell_1$-norm coherence $C_{\ell_1}$, panels (e)--(h), of the $Z$ boson at fixed angle, as functions of the real and imaginary parts of $b_Z$ and $\tilde b_Z$ varied one at a time with $a_Z=1$, at $\sqrt s =240$ and $365$\,GeV, for $\cos\theta=0.5$ (left column) and $\theta=\pi/2$ (right column).}   
\label{fig:entropy} 
\end{center}
\end{figure}

The coherence of the state in the helicity basis is measured by the $\ell_1$-norm of Baumgratz, Cramer and Plenio~\cite{Baumgratz:2014coh},
\begin{equation}
C_{\ell_1}(\rho)=\sum_{i\neq j}|\rho_{ij}|\in[0,2].
\end{equation}
It quantifies the interference between helicity amplitudes, transverse--longitudinal ($\rho_{\pm0}$) and transverse--transverse ($\rho_{+-}$), and is invariant under rotations about the $Z$ momentum that only change the phases of the $\rho_{ij}$. Figures~\ref{fig:entropy_coherence_angle} and~\ref{fig:entropy} show $S_{\rm vN}$ and $C_{\ell_1}$ as functions of the angle and of the couplings.

The behavior of the entropy, shown in Figure~\ref{fig:entropy_coherence_angle}, follows from that of the purity. The inclusive states are strongly mixed, with $S_{\rm vN}=0.993$ at $240\GeV$ and $0.837$ at $365\GeV$, close to the maximum $\ln3=1.099$. At fixed scattering angle the entropy is lower, $0.66$--$0.67$ at $240\GeV$ and $0.51$--$0.67$ at $365\GeV$. It is lowest at $\theta=\pi/2$, where the longitudinal polarization dominates. 

Considering the coherence, the inclusive values are $0.547$ and $0.420$, at the two CM energies considered. At fixed angle the coherence is small along the beam ($0.07$ at $240\GeV$, $0.12$ at $365\GeV$), reaches its maximum, $1.00$, at $|\cos\theta|\simeq0.63$ at $240\GeV$ and at $|\cos\theta|\simeq0.80$ at $365\GeV$, and is $0.730$ and $0.468$ at $\theta=\pi/2$.

Figure~\ref{fig:entropy} shows that the coherence is even in $\Im b_Z$ and $\Re\tilde b_Z$, because it depends only on the moduli $|\rho_{ij}|$. Its linear response to $\Re b_Z$ is larger than that of the spectral quantities: $\partial C_{\ell_1}/\partial\Re b_Z=1.62$ (inclusive) and $2.63$ ($\pi/2$) at $365\GeV$, $0.14$ and $0.16$ (inclusive and $\cos\theta=0.5$) at $240\GeV$. $\Im\tilde b_Z$ enters linearly away from $\pi/2$, $\partial C_{\ell_1}/\partial\Im\tilde b_Z=0.12$ at $240\GeV$ and $\cos\theta=0.5$.

\subsection{Contextuality}
\label{sub:contextuality}

A qutrit is the smallest system that exhibits quantum contextuality: the outcomes of projective measurements cannot be assigned in advance independently of the set of compatible measurements they are performed with~\cite{Kochen:1967,Klyachko:2008}. 

The two tests we employ to study the contextuality of the $Z$ boson produced in Higgsstrahlung are described in Appendix~\ref{app:contextuality}. Their non-contextual bounds are $\mathbb{CNTXT}_5 \leq 2$ and $\mathbb{CNTXT}_9 \leq 3$, and the maximal values achieved by quantum mechanics are $\sqrt5$ and $10/3$, respectively. Once the orientation of the projectors is optimized, both the contextuality witnesses depend only on the spectrum $\rho_1\leq\rho_2\leq\rho_3$ of the density matrix, Eq.~\eqref{eq:ctxclosed}, and the results of the tests can then be predicted by tracking the purity. For a given largest eigenvalue $\rho_3$, the purity is smallest when the other two eigenvalues are equal, so that $P \geq \rho_3^2 + (1-\rho_3)^2/2$. The violation of the KCBS inequality requires $\rho_3 > (5+\sqrt5)/10$ and, in turn, $P > 0.562$. We conclude that the inclusive states, with $P \leq 0.51$, cannot exhibit contextuality, while the state at $365\GeV$ in the central region can. The nine-projector test is instead violated by every state but the maximally mixed one.

An explicit calculation shows that in the SM, at tree level, the largest eigenvalue of the inclusive matrix is $0.52$ at $240\GeV$ and $0.66$ at $365\GeV$, and the KCBS inequality is not violated: $\mathbb{CNTXT}_5 = 1.82$ and $1.95$, respectively. For fixed scattering angle, at $365\GeV$ the largest eigenvalue reaches $0.795$ at $\theta = \pi/2$, where $\mathbb{CNTXT}_5 = 2.06$, and the inequality is violated for $\abs{\cos\theta} < 0.69$. At $240\GeV$ the state is never pure enough and $\mathbb{CNTXT}_5 \leq 1.92$ at all angles. The nine-projector witness takes the values $\mathbb{CNTXT}_9 = 3.12$ and $3.19$ for the inclusive matrix and $3.21$ and $3.27$ at $\pi/2$, above the bound as it must be.

\begin{figure}[!htb]
\begin{center}
\includegraphics[width=\textwidth]{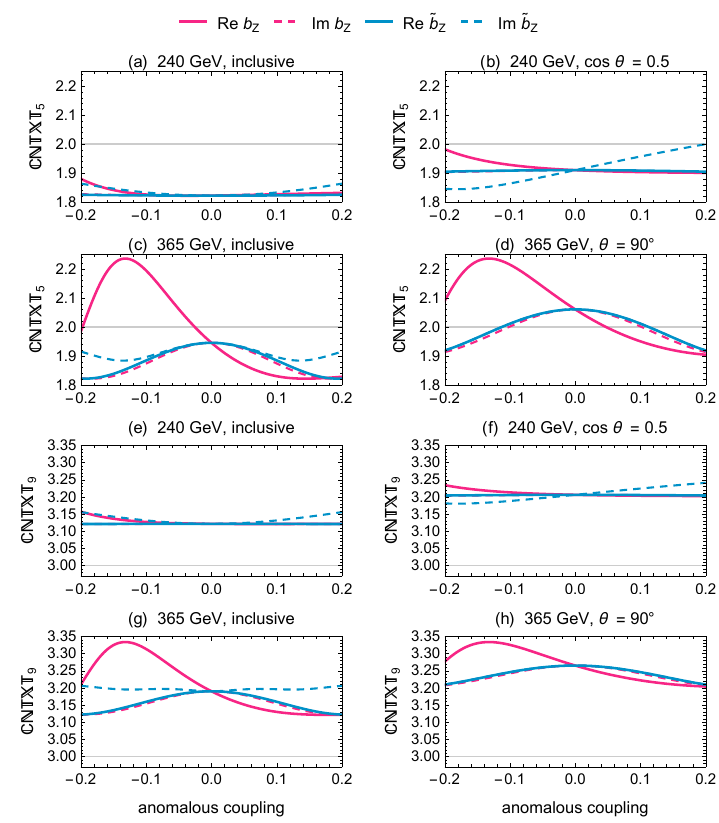}
\caption{The optimized contextuality witnesses $\mathbb{CNTXT}_5$ (KCBS, upper rows) and $\mathbb{CNTXT}_9$ (lower rows) of the $Z$ boson as functions of the real and imaginary parts of $b_Z$ and $\tilde b_Z$, varied one at a time with $a_Z = 1$, for the inclusive density matrix and at fixed angle at $\sqrt s = 240$ and $365\GeV$. Tree level, exact in the couplings. The horizontal lines are the non-contextual bounds.}
\label{fig:ctx}
\end{center}
\end{figure}

\begin{figure}[!htb]
\begin{center}
\includegraphics[width=\textwidth]{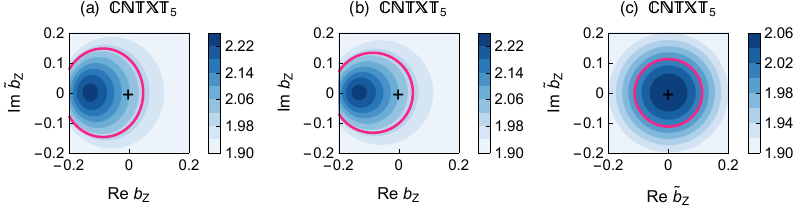}
\caption{The KCBS witness $\mathbb{CNTXT}_5$ at $\sqrt s = 365\GeV$ and $\theta = \pi/2$ in the planes $(\Re b_Z, \Im\tilde b_Z)$, $(\Re b_Z, \Im b_Z)$ and $(\Re\tilde b_Z, \Im\tilde b_Z)$, with $a_Z = 1$. The magenta line is the non-contextual bound $\mathbb{CNTXT}_5 = 2$, the cross the SM point.}
\label{fig:ctxmaps}
\end{center}
\end{figure}

Figure~\ref{fig:ctx} shows the two witnesses as functions of the four real parameters of the anomalous couplings, varied one at a time with $a_Z = 1$. The largest effect is that of $\Re b_Z$. A negative value reduces the transverse amplitudes and, for $\Re b_Z = -\MZ^2/(q\vdot k)$, that is $-0.13$ at $365\GeV$ and $-0.33$ at $240\GeV$, cancels them: the $Z$ boson is then produced in the pure longitudinal state and both witnesses reach their quantum maximum, $\sqrt5$ and $10/3$. At $365\GeV$ the inclusive state violates the KCBS inequality for $-0.20 < \Re b_Z < -0.026$. At $\theta = \pi/2$ contextuality is lost for $\Re b_Z > 0.047$, for $\abs{\Im b_Z} > 0.10$ and for $\abs{\tilde b_Z} > 0.11$, regardless of the value of the phase. Away from $\theta=\pi/2$, $\Im\tilde b_Z$ enters linearly through the helicity asymmetry it induces, with $\partial\mathbb{CNTXT}_5/\partial\Im\tilde b_Z \simeq 0.4$--$0.5$ at $\cos\theta = 0.5$. The two-dimensional dependence of $\mathbb{CNTXT}_5$ at $\sqrt s = 365\GeV$ and $\theta = \pi/2$ is shown in Figure~\ref{fig:ctxmaps}.

For couplings of the size probed by the tomographic process presented in Section~\ref{sec:stat}, $\order{10^{-3}}$, the witnesses change by at most a few units in $10^{-3}$, $\partial\mathbb{CNTXT}_5/\partial\Re b_Z = -1.9$ for the inclusive state and $-1.4$ at $\pi/2$ at $365\GeV$, $0.03$ and $-0.16$ at $240\GeV$. Contextuality is therefore a robust property of the $Z$ boson produced at $365\GeV$ in the central region, but not a sensitive probe of the anomalous couplings: being a function of the spectrum alone, it is not sensitive at linear order to $\Im b_Z$ and $\Re\tilde b_Z$, which are measured instead through the coherences of the density matrix. The results above are at tree level, but the one-loop corrections of Section~\ref{sub:rhoSM}, which shift the Fano coefficients by a few units in $10^{-3}$, are not expected to change these conclusions.

\subsection{Magic}
\label{sub:magic}

Magic, or non-sta\-bi\-li\-zer\-ness, measures how far a state is from the stabilizer states, which can be prepared and transformed by Clifford operations and are efficiently simulated on a classical computer. For a qutrit we quantify it with the mana~\cite{Veitch:2014}, $\mathbb{M}$, built on the discrete Wigner function of Gross~\cite{Gross:2006}, as defined in Appendix~\ref{app:magic}. The mana vanishes for every mixture of stabilizer states and can be computed exactly for mixed states. Unlike the purity and the optimized contextuality witnesses, it depends on the basis that defines the stabilizer states. Minimizing the mana over the rotations of the spin of the $Z$ boson gives vanishing values for all the states we consider. The results below then hold only for the helicity basis employed in the tomographic reconstruction, Eq.~\eqref{eq:polvectors}.  Moreover, even within the helicity basis, the mana depends on the phases of the helicity states: a change of phases is not a Clifford operation, hence it changes the Wigner function.  

\begin{figure}[!htb]
\begin{center}
\includegraphics[width=0.85\textwidth]{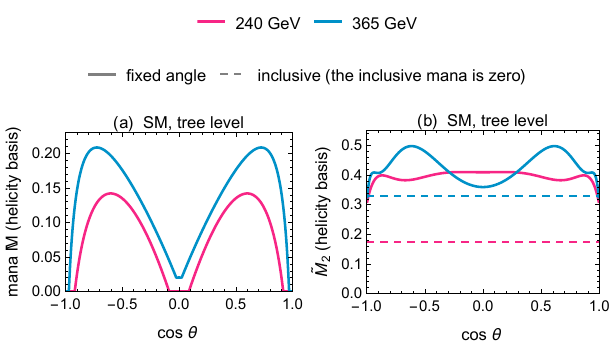}
\caption{Magic of the $Z$ boson state in the helicity basis, in the SM at tree level: (a) the mana $\mathbb{M}$ and (b) the stabilizer R\'enyi entropy $\tilde M_2$ as functions of $\cos\theta$ at $240$ and $365\GeV$. The inclusive mana vanishes, while the inclusive $\tilde M_2$ is shown by the dashed lines.}
\label{fig:magic}
\end{center}
\end{figure}

\begin{figure}[!htp]
\begin{center}
\includegraphics[width=\textwidth]{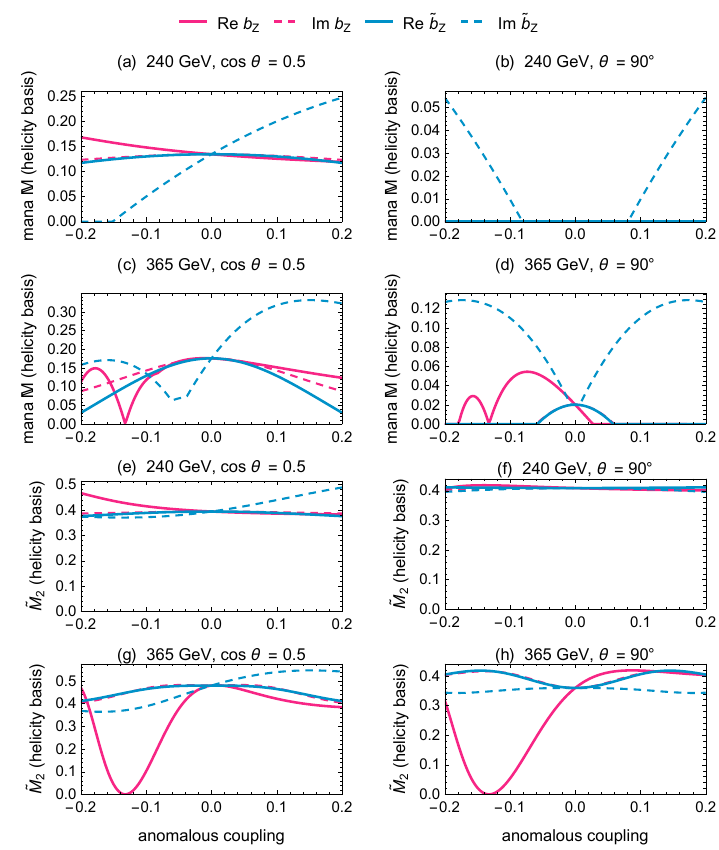}
\caption{The mana $\mathbb{M}$, panels (a)--(d), and the stabilizer R\'enyi entropy $\tilde M_2$, panels (e)--(h), of the $Z$ boson at fixed angle, as functions of the real and imaginary parts of $b_Z$ and $\tilde b_Z$ varied one at a time with $a_Z = 1$, at $240$ and $365\GeV$, for $\cos\theta = 0.5$ (left column) and $\theta = \pi/2$ (right column). }
\label{fig:magiccpl}
\end{center}
\end{figure}

The inclusive density matrix has a non-negative Wigner function at both energies, and therefore $\mathbb{M} = 0$. Its robustness, defined below, is also equal to one, so that the inclusive state is a mixture of stabilizer states: the magic of the $Z$ boson only appears at fixed angle. As shown in Figure~\ref{fig:magic}(a), in the SM the mana is non-zero in two bands symmetric in $\cos\theta$. It reaches $0.142$ at $\abs{\cos\theta} \simeq 0.60$ at $\sqrt s =240\GeV$ and $0.208$ at $\abs{\cos\theta} \simeq 0.72$ at $365\GeV$. It vanishes near the beam axis and, at $240\GeV$, also for $\abs{\cos\theta} < 0.087$. At $\sqrt s = 365\GeV$ and $\theta = \pi/2$ it is small, $\mathbb{M} = 0.020$. The dependence of these values on the phase convention of the helicity states is discussed in Appendix~\ref{app:magic}. For comparative purposes, we show in Figure~\ref{fig:magic}(b) magic as evaluated with the stabilizer R\'enyi entropy $\tilde M_2$~\cite{Leone:2022}, often used in the literature as a measure of this quantity. More details can be found in Appendix~\ref{app:magic}. 

The dependence on the anomalous couplings is shown in Figure~\ref{fig:magiccpl}. As for the other analyzed quantities, $\Im b_Z$ and $\Re\tilde b_Z$ enter only at second order, but for a different reason, since the mana does not depend on the spectrum alone. The SM density matrix is real, and reversing the sign of either coupling replaces $\rho$ with its complex conjugate, which only permutes the points of the Wigner function: the mana is therefore even in these two couplings. The coupling $\Im\tilde b_Z$, instead, enters linearly at fixed angle away from $\pi/2$, with $\partial\mathbb{M}/\partial\Im\tilde b_Z = 0.75$ at $\sqrt s = 240\GeV$ and $2.3$ at $365\GeV$, for $\cos\theta = 0.5$. This is the largest linear response among the quantities of this section, but couplings of order $10^{-3}$ still change the mana by at most $2.3\times 10^{-3}$. The effect of $\Re b_Z$ is smaller, $\partial\mathbb{M}/\partial\Re b_Z \simeq -0.1$ at $\cos\theta = 0.5$. For $\Re b_Z = -\MZ^2/(q\vdot k)$ the $Z$ boson is in the longitudinal state $\ket{1}$, a stabilizer state, and its mana vanishes. Figure~\ref{fig:magiccpl}(e)--(h) shows the dependence of $\tilde M_2$ on the same couplings.

\begin{figure}[!htbp]
\begin{center}
\includegraphics[width=0.75\textwidth]{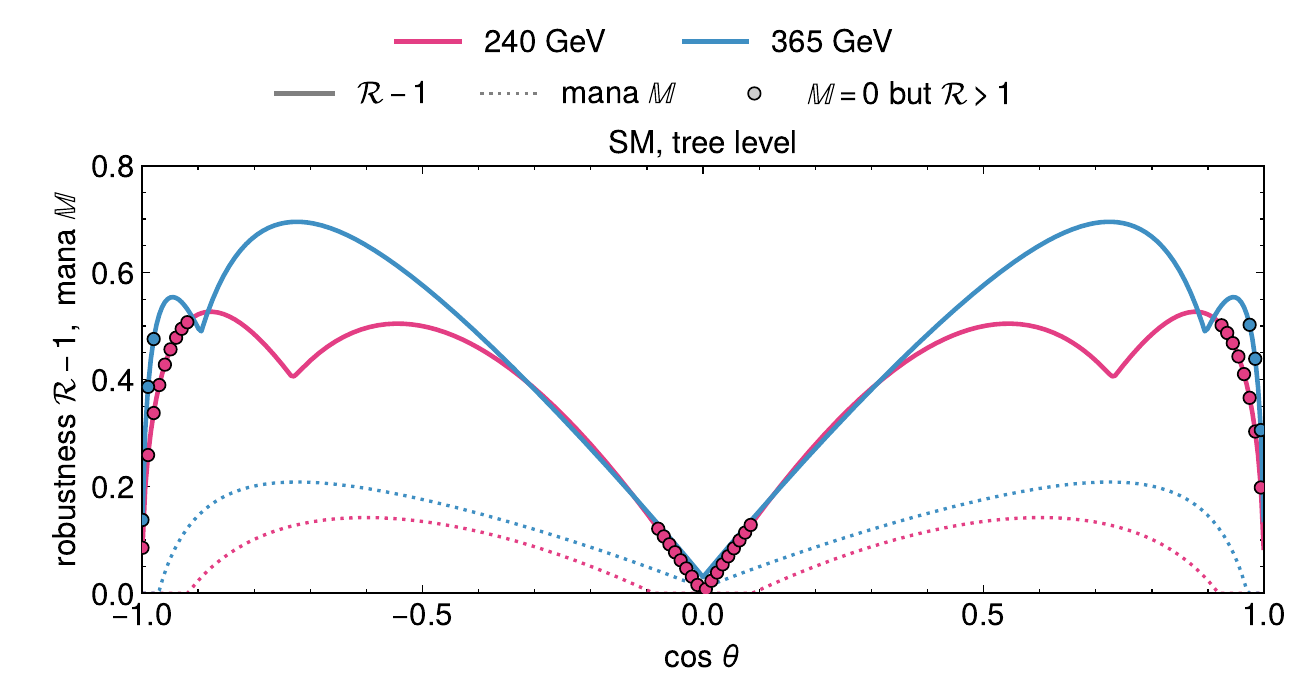}\\[2mm]
\includegraphics[width=0.85\textwidth]{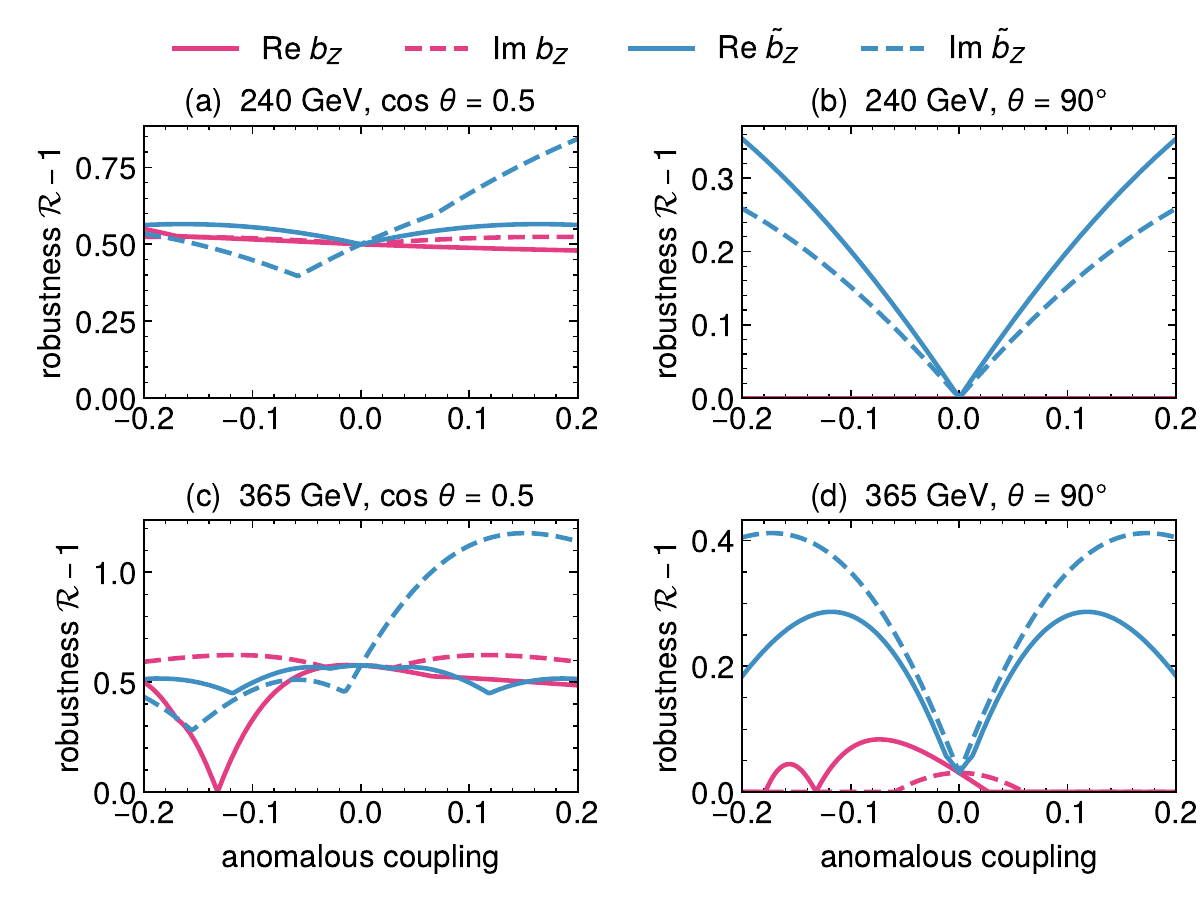}
\caption{The robustness of magic $\mathcal{R}-1$ of the $Z$ boson state. Upper panel: SM at tree level as a function of $\cos\theta$ at 240 and 365\,GeV, compared with the mana. The dots mark states with zero mana that are still magic ($\mathbb{M}=0$ but $\mathcal{R}>1$). Lower panels: dependence on the real and imaginary parts of $b_Z$ and $\tilde b_Z$, varied one at a time with $a_Z=1$, at 240 and 365\,GeV, for $\cos\theta=0.5$ (left column) and $\theta=\pi/2$ (right column).}
\label{fig:robustness}
\end{center}
\end{figure}

It is worth commenting on the cases where magic is present but mana vanishes (Figure~\ref{fig:robustness}). The mana vanishes on every mixture of stabilizer states, because their Wigner function is non-negative. A non-vanishing mana therefore always signals magic. The converse holds only for pure states, for which a vanishing mana implies a stabilizer state by the discrete Hudson theorem~\cite{Gross:2006}. For mixed states it fails: there are states with a non-negative Wigner function that are not mixtures of stabilizer states~\cite{Veitch:2012ihl}, said to carry \emph{bound magic}. The stabilizer R\'enyi entropy $\tilde M_2$ has the opposite limitation. It vanishes only on the pure stabilizer states and on the maximally mixed state, and it is positive on every other mixture of stabilizer states, where no magic is present. The states of the $Z$ boson at fixed angle are mixed, hence a vanishing mana does not exclude magic and a non-vanishing $\tilde M_2$ does not establish it. More details are given in Appendix~\ref{app:robustness}.

We therefore also look at the robustness of magic~\cite{Howard:2016lyb},
\begin{equation}
\mathcal R(\rho)=\min\Big\{\textstyle\sum_s|x_s|\ :\ \rho=\sum_sx_s\,|s\rangle\langle s|,\ x_s\in\mathbb R\Big\},\label{eq:robustness}
\end{equation}
where $|s\rangle$ runs over the twelve stabilizer states $|j\rangle$ and $(|0\rangle+\omega^a|1\rangle+\omega^b|2\rangle)/\sqrt3$. It is the minimal $\ell_1$-norm of a decomposition of $\rho$ into stabilizer states with real, possibly negative, weights. Robustness detects every magic state, including the bound magic states that the mana misses. It equals 1 if and only if $\rho$ is a mixture of stabilizer states and exceeds 1 otherwise, and $\mathbb{M} \leq \ln \mathcal{R}$. 

Figure~\ref{fig:robustness} compares the robustness with the mana as functions of the angle and anomalous couplings. At $\sqrt s = 240\GeV$, the only non-magic angle is exactly $\theta=\pi/2$. Both regions where the mana vanishes are magic. Near $\theta=\pi/2$, $\cos \theta=0.05$ gives $\mathcal{R}=1.077$. Near the beam axis, $\cos \theta=0.95$ gives $\mathcal{R}=1.454$. Over the zero-mana regions $\mathcal{R}$ reaches 1.51, and over all angles 1.53.
At $365\GeV$, no angle is free of magic. At $\theta=\pi/2$, $\mathcal{R}=1.031$. Near the beam axis, $|\cos \theta|>0.974$, the mana vanishes but $\mathcal{R}$ reaches 1.50. The maximum over the angle is $\mathcal{R}=1.69$ at $|\cos \theta| \simeq 0.72$, where the mana also peaks. The fixed-angle states are prone to bound magic because they have rank two: they lie on the boundary of the state space, where the gap between the region $W_\rho \geq 0$ and the stabilizer polytope is widest. 

\section{Tomographic bounds}  
\label{sec:stat}

The density matrix of the $Z$ boson is measured through the angular distribution of its decay products. We set up the reconstruction and employ a $\chi^2$ test centered on the SM prediction at one loop, Eq.~\eqref{eq:rhoSM},  to assess i) how well quantum state tomography determines $\klam$ when everything else in the vertex is as in the SM, Section~\ref{sub:kappa}, and ii) how well the same methodology determines the form factors of the effective vertex when nothing is assumed about the new physics, Section~\ref{sub:formfactors}. Section~\ref{sub:quantum} compares the results with the reach of polarization measurements allowed by quantum mechanics.

\subsection{Decay distribution, reconstruction and $\chi^2$}  
\label{sub:decay}
\label{sub:estimator}

The density matrix is read off the direction $\hat d = (\sin\theta^*\cos\varphi^*, \sin\theta^*\sin\varphi^*, \cos\theta^*)$ of the fermion produced by the decay $Z\to f\bar f$, in the rest frame of the $Z$ boson, with components $d_i$ along the triad $\{\hat r, \hat n, \hat k\}$ of Eq.~\eqref{eq:polvectors}. With the decay amplitude 
\begin{equation}
	\mathcal A(\lambda) = \eps_\mu(\lambda)\, \bar u(p_f)\gamma^\mu(\gL \PL + \gR \PR) v(p_{\bar f}),
\end{equation}
and utilizing the same polarization vectors of Eq.~\eqref{eq:polvectors}, the normalized angular distribution is 
\begin{equation}
W(\Omega^*) = \Tr[\rho\, \mathcal D(\Omega^*)] ,
\end{equation}
with $\mathcal D_{\lambda\lambda'} \propto \sum_{\rm spins}\mathcal A(\lambda')\mathcal A^*(\lambda)$	and $\int\mathcal D\, \dd\Omega^* = \ii$. In terms of the polarization vector and alignment tensor of Eq.~\eqref{eq:rhoPT} it is
\begin{equation}
		W(\hat d) = \frac{1}{4\pi}\qty[1 - \frac32\, A_f\, P_i\, d_i + \frac34\, T_{ij}\, d_i d_j], \qquad A_f = \frac{\gL^2 - \gR^2}{\gL^2 + \gR^2}.
	\label{eq:W}
\end{equation}
The sign of the vector term states that a left-handed fermion is emitted backward with respect to the spin of the $Z$ boson. Equation~\eqref{eq:W} makes the reach of each decay channel clear. The vector polarization comes with the analyzing power $A_f$ and is odd under $\hat d \to -\hat d$, the exchange of the fermion with the antifermion. The alignment is instead even under the sign exchange and independent of $A_f$. A decay in which the fermion charge is not measured, for instance $Z\to q\bar q$ without jet-charge tagging, gives the symmetrized distribution, $(1 + \frac34 T_{ij} d_i d_j)/(4\pi)$, and measures only the five components of $T_{ij}$. By contrast, the process $Z\to\ell^+\ell^-$ measures all the eight parameters, in particular the polarization components $P_i$ albeit with a small $A_\ell$.

In terms of the Fano coefficients, the distribution is written as 
\begin{equation}
	W = 1/(4\pi) + \sum_a f_a g_a(\Omega^*),
\end{equation}
with $g_a = \Tr[\Lambda^a\mathcal D]/3$ being eight angular functions that are mutually orthogonal: $\int g_a g_b\, \dd\Omega^* = G_a\delta_{ab}$. The coefficient $G_a$ is $A_f^2/(12\pi)$ for the vector contribution and $1/(60\pi)$ for the tensor coefficients.

As for the tomographic reconstruction, since the $g_a$ are orthogonal, the estimators
\begin{equation}
	\hat f_a = \frac{1}{G_a}\, \ev{g_a(\Omega^*)}_{\rm events}, \qquad\text{or}\qquad \hat P_i = -\frac{2}{A_f}\ev{d_i}, \qquad \hat T_{ij} = 10\ev{d_i d_j - \tfrac13\delta_{ij}},
	\label{eq:estimators}
\end{equation}
are unbiased, and their covariance for $N$ events is $C_{ab}/N$ with
\begin{equation}
	C_{ab} = \frac{1}{G_a G_b}\int g_a g_b\, W\, \dd\Omega^* - f_a f_b ,
	\label{eq:cov}
\end{equation}
a linear function of the true coefficients, which we evaluate at the SM value. For the charge-blind channels, the three vector functions vanish identically and the estimator is restricted to the five tensor coefficients. As we show in Section~\ref{sub:quantum}, this linear estimator loses nothing with respect to the full likelihood of the decay angles. The SM contribution is evaluated at one loop, Eq.~\eqref{eq:rhoSM}, with the anomalous couplings of Section~\ref{sub:rhoshift} on top of it,
\begin{equation}
	f_a(\klam, b_Z, \tilde b_Z) = f_a^{ 0}(b_Z, \tilde b_Z) + \klam\, \delta f_a^{hhh} + \delta f_a^{\rm rest},
	\label{eq:model}
\end{equation}
where $f^{ 0}_a(b_Z, \tilde b_Z)$ are the coefficients of the exact density matrix of the parametrized vertex at tree level, with $a_Z = 1$ and the four real parameters $\Re b_Z$, $\Im b_Z$, $\Re\tilde b_Z$ and $\Im\tilde b_Z$. The contributions $\delta f^{hhh}_a$, $\delta f^{\rm rest}_a$ are the corrections to the Fano coefficients sourced by the two shifts of Eq.~\eqref{eq:rhoSM}. The bounds are then given through the $\chi^2$ test built as
\begin{equation}
	\chi^2 = \sum N\, \Delta f^{\sf T} C^{-1} \Delta f + \sum_k \epsilon_k^2, \qquad \Delta f = f(\klam, b_Z, \tilde b_Z) + \sigma_{\rm th}\sum_k \epsilon_k\, v_k - f^{\rm SM},
	\label{eq:chi2}
\end{equation}
with eight terms for the leptons and five for the hadrons in the first sum, which runs over decay channels and CM energies. The pseudo-data are the SM prediction at one loop itself, $f^{\rm SM}_a = f_a(\klam = 1, b_Z = 0, \tilde b_Z = 0)$, with no statistical fluctuation, and the covariance~\eqref{eq:cov} is evaluated at the same point, so that the intervals below are the expected ones. The nuisance parameters $\epsilon_k$ model the theory uncertainty. The SM prediction is a one-loop prediction, and we assign to its $\klam$-independent part an uncertainty $\sigma_{\rm th} = 10\%$ of $\delta f^{\rm rest}$: the natural size of the missing two-loop terms and three times the difference between the $\alpha(0)$ and the $G_\mu$ normalizations of the loop. The vectors $v_k$ are the shifts $\delta f^{\rm rest}$ restricted to the vector ($P_i$) and to the tensor ($T_{ij}$) coefficients, each multiplied by a nuisance parameter for each energy. When the $\cos\theta$ distribution is used, its $\klam$-independent one-loop correction also acquires a nuisance parameter. All nuisance parameters are profiled. 

A few remarks are in order. The $\chi^2$ test uses the inclusive density matrix $\bar\rho$, given by the event-weighted average of $\rho(\theta)$, so that its dependence on the parameters contains both the local shift of the polarization and the shift of the weight $\dd\sigma/\dd\cos\theta$. Since the $b_Z$ structure has a different angular shape (see Section~\ref{sub:treeparam}), the two are separated by resolving the scattering angle\footnote{To this purpose we use the 16 Gauss–Legendre nodes in $\cos\theta$ at which the one-loop amplitude is computed.}, and we quote both. The test does not use the total rate, deliberately: the rate is enhanced by $C_1$ but it is degenerate with $a_Z$, with the luminosity and with the scheme-dependent $c(s)$. The quantity entering Eq.~\eqref{eq:chi2}, instead, is a ratio not plagued by this degeneracy. 

As benchmarks, we take the FCC-ee runs detailed in the feasibility study, yielding $10.8\,{\rm ab}^{-1}$ at $240\GeV$ and $2.7\,{\rm ab}^{-1}$ at $365\GeV$ with four interaction points. To set the sample sizes we utilize the tree-level cross sections and the branching ratios (BR) ${\rm BR}(Z\to e^+e^-, \mu^+\mu^-) = 0.067$ and ${\rm BR}(Z\to{\rm hadrons}) = 0.70$, assuming full efficiency and a perfect reconstruction of the decay angles. This gives $2.4\times 10^6$ $ZH$ events at $240\GeV$, of which $1.6\times 10^5$ leptonic and $1.7\times 10^6$ hadronic, whereas we have $2.9\times 10^5$ at $365\GeV$, of which $2.0\times 10^4$ leptonic and $2.1\times 10^5$ hadronic. Table~\ref{tab:baselineErr} gives the statistical errors obtained with these samples for the Fano coefficients reconstructed in the tomography process, together with the SM shifts in units of them. In Table~\ref{tab:coefficients} we list instead the coefficients that carry the signal at the SM point, their statistical errors, and their shifts per unit of $\klam$ and of the four couplings. 

\begin{table}[h]
\centering
\small
\begin{tabular}{@{}lllcc@{}}
\toprule
$\sqrt s$ & $f_a$ & channel & stat.\ error & SM shift/error \\
\midrule
$240\GeV$ & $\frac32 P_1$ & $Z\to\ell\ell$ & $2.1\times 10^{-2}$ & $6.4$ \\
 & $\frac32 T_{23}$ & $Z\to q\bar q$ & $2.9\times 10^{-3}$ & $-0.48$ \\
 & $\frac34(T_{11} - T_{22})$ & $Z\to q\bar q$ & $3.0\times 10^{-3}$ & $-0.44$ \\
 & $\frac{3\sqrt3}{4}T_{33}$ & $Z\to q\bar q$ & $2.9\times 10^{-3}$ & $-2.1$ \\
$365\GeV$ & $\frac32 P_1$ & $Z\to\ell\ell$ & $6.2\times 10^{-2}$ & $2.2$ \\
 & $\frac32 T_{23}$ & $Z\to q\bar q$ & $8.1\times 10^{-3}$ & $-0.92$ \\
 & $\frac34(T_{11} - T_{22})$ & $Z\to q\bar q$ & $9.1\times 10^{-3}$ & $-0.55$ \\
 & $\frac{3\sqrt3}{4}T_{33}$ & $Z\to q\bar q$ & $7.7\times 10^{-3}$ & $-3.0$ \\
\bottomrule
\end{tabular}
\caption{Statistical errors of the tomographic reconstruction at the FCC-ee on the Fano coefficients of Table~\ref{tab:baseline}, each from the channel that measures it best (leptons for the vector coefficient, hadrons for the tensor ones, Table~\ref{tab:coefficients}), and the SM one-loop shift in units of them.}
\label{tab:baselineErr}
\end{table}

\begin{table}[h]
\centering
\footnotesize
\setlength{\tabcolsep}{3.2pt}
\begin{tabular}{@{}llcccccccc@{}}
\toprule
 & & & & & \multicolumn{5}{c}{shift per unit of} \\
\cmidrule(l){6-10}
$\sqrt s$ & channel & $f_a$ & SM & stat.\ error & $\klam$ & $\Re b_Z$ & $\Im b_Z$ & $\Re\tilde b_Z$ & $\Im\tilde b_Z$ \\
\midrule
$240\GeV$ & $Z\to q\bar q$ & $\frac{3\sqrt3}{4}T_{33}$ & $-0.172$ & $2.9\times 10^{-3}$ & $5.1\times 10^{-4}$ & $0.90$ & $0$ & $0$ & $0$ \\
 & $1.68\times 10^6$ & $\frac34(T_{11} - T_{22})$ & $0.451$ & $3.0\times 10^{-3}$ & $1.5\times 10^{-4}$ & $0.26$ & $0$ & $0$ & $0$ \\
 & & $\frac32 T_{23}$ & $-1.4\times 10^{-3}$ & $2.9\times 10^{-3}$ & $0.3\times 10^{-4}$ & $0$ & $0.19$ & $0$ & $0$ \\
 & & $\frac32 T_{12}$ & $0$ & $3.0\times 10^{-3}$ & $0$ & $0$ & $0$ & $1.34$ & $0$ \\
 & & $\frac32 T_{13}$ & $0$ & $3.1\times 10^{-3}$ & $0$ & $0$ & $0$ & $0$ & $-0.39$ \\
 & $Z\to\ell\ell$ & same five & & $0.9$--$1.0\times 10^{-2}$ & same & same & same & same & same \\
 & $1.61\times 10^5$ & $\frac32 P_{1}$ & $-0.125$ & $2.1\times 10^{-2}$ & $0.2\times 10^{-4}$ & $0.04$ & $0$ & $0$ & $0$ \\
 & & $\frac32 P_{2}$ & $0$ & $2.0\times 10^{-2}$ & $0$ & $0$ & $0$ & $-0.39$ & $0$ \\
 & & $\frac32 P_{3}$ & $0$ & $2.0\times 10^{-2}$ & $0$ & $0$ & $0$ & $0$ & $2.68$ \\
$365\GeV$ & $Z\to q\bar q$ & $\frac{3\sqrt3}{4}T_{33}$ & $-0.823$ & $7.7\times 10^{-3}$ & $1.26\times 10^{-3}$ & $6.5$ & $0$ & $0$ & $0$ \\
 & $2.06\times 10^5$ & $\frac34(T_{11} - T_{22})$ & $0.264$ & $9.1\times 10^{-3}$ & $3.6\times 10^{-4}$ & $1.9$ & $0$ & $0$ & $0$ \\
 & & $\frac32 T_{23}$ & $-7.4\times 10^{-3}$ & $8.1\times 10^{-3}$ & $5.4\times 10^{-4}$ & $0$ & $1.39$ & $0$ & $0$ \\
 & & $\frac32 T_{12}$ & $0$ & $9.1\times 10^{-3}$ & $0$ & $0$ & $0$ & $3.46$ & $0$ \\
 & & $\frac32 T_{13}$ & $0$ & $8.4\times 10^{-3}$ & $0$ & $0$ & $0$ & $0$ & $-1.64$ \\
 & $Z\to\ell\ell$ & same five & & $2.5$--$2.9\times 10^{-2}$ & same & same & same & same & same \\
 & $1.98\times 10^4$ & $\frac32 P_{1}$ & $-0.122$ & $6.2\times 10^{-2}$ & $-0.8\times 10^{-4}$ & $-0.39$ & $0$ & $0$ & $0$ \\
 & & $\frac32 P_{2}$ & $0$ & $5.9\times 10^{-2}$ & $0$ & $0$ & $0$ & $-1.64$ & $0$ \\
 & & $\frac32 P_{3}$ & $0$ & $5.2\times 10^{-2}$ & $0$ & $0$ & $0$ & $0$ & $6.91$ \\
\bottomrule
\end{tabular}
\caption{Fano coefficients of the inclusive density matrix of the SM at one loop, their statistical errors with the FCC-ee samples (number of events after the channel, full efficiency), and their shifts per unit of $\klam$ and of the four real parameters of the anomalous vertex. The tree-level values, the one-loop shifts and their theory uncertainty are in Table~\ref{tab:baseline}. The coefficients not listed vanish at the SM point and do not shift at linear order.}
\label{tab:coefficients}
\end{table}

\subsection{The Higgs self-coupling alone}  
\label{sub:kappa}

\begin{figure}[h]
\centering
\includegraphics[width=0.6\textwidth]{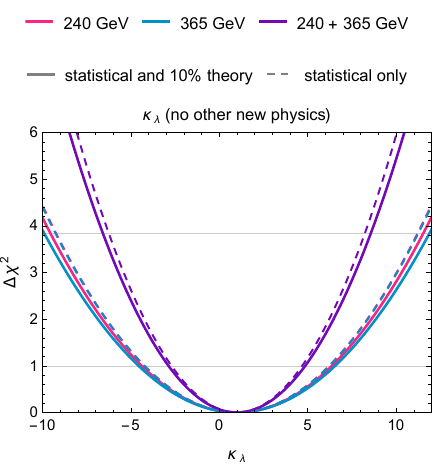}
\caption{$\Delta\chi^2$ of the inclusive density matrix against the SM at one loop, Eq.~\eqref{eq:chi2}, both channels, as a function of $\klam$ with no other new physics in the vertex, at $240\GeV$, $365\GeV$ and combined, with the $10\%$ theory uncertainty (solid) and statistical only (dashed). The horizontal lines mark the $68\%$ and $95\%$ CL.}
\label{fig:kappa}
\end{figure}

As a first exercise we perform our $\chi^2$ test supposing that the only unknown of Eq.~\eqref{eq:rhoSM} is $\klam$. The model is then $\rho_{\rm SM}(\klam) = \rho_{ 0} + \klam\,\delta\rho_{hhh} + \delta\rho_{\rm rest}$, matching Eq.~\eqref{eq:model} with $b_Z = \tilde b_Z = 0$, and the pseudo-data are $\rho_{\rm SM}(\klam=1)$. The fit compares the inclusive forms, the event average $\bar\rho_{\rm SM}(\klam)$, whose Fano coefficients and their shift per unit of $\klam$ are reported in Table~\ref{tab:coefficients}. When the scattering angle is resolved, it compares $\rho_{\rm SM}(\klam; \theta)$ and the normalized $\cos\theta$ distribution of the events. Importantly, this test is independent of the constraint derived from the rate of interaction  through $C_1$. The results are presented in Table~\ref{tab:sensitivity} and Figure~\ref{fig:kappa}. Statistically, the inclusive density matrix determines $\klam$ to $\pm 5.2$ at $240\GeV$ and at $365\GeV$, and to $\pm 3.7$ for the two runs together, basically through the hadronic channel contribution exclusively. These quotes are statistical only, and coincide with the uncertainties of a reconstruction around the tree-level point. The theory uncertainty of the one-loop prediction raises them to $5.4$, $5.6$ and $3.9$. 

As shown in the last line of the table, rate analyses achieve here a better result. With $\klam$ the only free parameter, the cross section changes by $C_1 + 2\delta Z_h = 1.44\%$ per unit of $\klam$ at $240\GeV$ and by $0.20\%$ at $365\GeV$ (Appendix~\ref{app:validation}). With the same sample of visible $Z$ decays and the same full efficiency, its statistical precision of $0.07\%$ at $240\GeV$ gives $\delta\klam = 0.05$, seventy times better than the tomographic bound. The same $10\%$ theory uncertainty on the $\klam$-independent one-loop correction to the rate, $1.6\%$ at $240\GeV$ and $2.7\%$ at $365\GeV$ (weak corrections only), degrades the bound to $\delta\klam = 0.12$, still outperforming the tomographic result. The rate is more sensitive because $C_1$ is sizeably larger than $\beta$, and quantum state tomography cannot compete with it as far as the determination of $\klam$ is concerned. What our methodology still offers is a determination that is independent of the luminosity, of the efficiencies and branching ratios, and of the normalization $a_Z$ of the vertex, with which the rate is degenerate. And, as the next section shows, it is really a measurement of $b_Z$.

\begin{table}[h]
\centering
\small
\setlength{\tabcolsep}{6pt}
\begin{tabular}{@{}lccc@{}}
\toprule
 & \multicolumn{3}{c}{$\delta\klam$} \\
\cmidrule(lr){2-4}
 & $240\GeV$ & $365\GeV$ & combination \\
\midrule
inclusive $\bar\rho$, $Z\to\ell\ell$ & $17.6$ ($17.7$) & $17.7$ ($17.8$) & \\
inclusive $\bar\rho$, $Z\to q\bar q$ & $5.5$ ($5.6$) & $5.5$ ($5.8$) & \\
inclusive $\bar\rho$, both & $5.2$ ($5.4$) & $5.2$ ($5.6$) & $3.7$ ($3.9$) \\
\quad $68\%$ CL & $[-4.4, 6.4]$ & $[-4.6, 6.6]$ & $[-2.9, 4.9]$ \\
\quad $95\%$ CL & $[-9.6, 11.5]$ & $[-9.9, 11.9]$ & $[-6.6, 8.6]$ \\
$\rho(\theta)$ resolved, polarization & $5.2$ ($5.3$) & $5.6$ ($5.9$) & $3.8$ ($3.95$) \\
$\cos\theta$ distribution only & $5.5$ ($5.7$) & $5.8$ ($6.2$) & $4.0$ ($4.2$) \\
polarization and distribution & $3.8$ ($3.9$) & $4.0$ ($4.3$) & $2.75$ ($2.9$) \\
inclusive $\bar\rho$, $\Re b_Z^{\rm NP}$ free & $84$ ($84$) & $14$ ($14$) & $7.3$ ($7.4$) \\
\quad $68\%$ and $95\%$ CL & & & $[-6.4, 8.4]$, $[-13.6, 15.6]$ \\
\midrule
quantum Cram\'er--Rao bound & $1.6$ & $1.7$ & $1.2$ \\
\quad with the scattering angle resolved & $1.6$ & $1.25$ & $0.98$ \\
\midrule
rate, same sample & $0.05$ ($0.12$) & $1.0$ ($1.7$) & $0.05$ ($0.12$) \\
\bottomrule
\end{tabular}
\caption{One-sigma sensitivity to $\klam$ from the tomography of the $Z$ at the FCC-ee, with the $\chi^2$ of Eq.~\eqref{eq:chi2} around the SM at one loop: statistical, and in parentheses with the $10\%$ theory uncertainty on the $\klam$-independent one-loop shift. The intervals include it. No other new physics in the vertex, except in the line ``$\Re b_Z^{\rm NP}$ free'', where an energy-independent anomalous coupling is fitted together with $\klam$ (intervals with $\Re b_Z^{\rm NP}$ profiled). The quantum Cram\'er--Rao bound is the one of Section~\ref{sub:quantum} for an ideal measurement of the polarization on all the $ZH$ events, of the inclusive state and with the scattering angle resolved (the information of the $\cos\theta$ distribution included). The last line is the rate alone, through $C_1 + 2\delta Z_h$, with the same visible sample and efficiency.}
\label{tab:sensitivity}
\end{table}

\subsection{The form factors}  
\label{sub:formfactors}
\label{sub:sensitivity}

\begin{figure}[!tbp]
\centering
\includegraphics[width=0.92\textwidth]{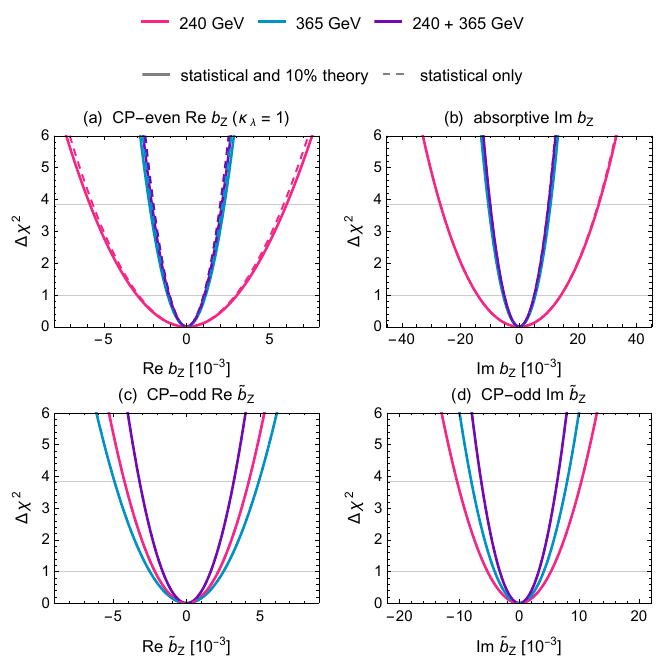}
\caption{$\Delta\chi^2$ of the inclusive density matrix against the SM at one loop, Eq.~\eqref{eq:chi2}. The result considers leptonic and hadronic channels for runs at $240\GeV$ and $365\GeV$, on top of their combination. The $\Delta\chi^2$ is plotted as a function of (a) the CP-even coupling $\Re b_Z$ at $\klam = 1$, (b) its absorptive part $\Im b_Z$, (c) the CP-odd coupling $\Re\tilde b_Z$ and (d) its imaginary part $\Im\tilde b_Z$. The baseline is provided by the SM result at one loop. The parameters are varied one at a time. The results consider an additional $10\%$ theory uncertainty (solid) over the statistical uncertainty (dashed, with no theory uncertainty for the CP-odd couplings). The horizontal lines mark the $68\%$ and $95\%$ CL.}
\label{fig:couplings}
\end{figure}

\begin{figure}[!tbp]
\centering
\includegraphics[width=0.92\textwidth]{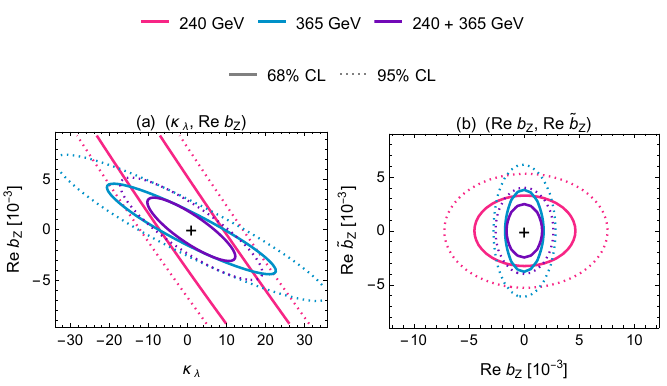}
\caption{The $68\%$ (solid) and $95\%$ CL (dotted) contours of the two-parameter fits of the inclusive density matrix, with the $10\%$ theory uncertainty, for runs at $240\GeV$ and $365\GeV$, as well as for their combination. Panel (a): the $(\klam, \Re b_Z)$ plane, with an energy-independent anomalous coupling, which shows the degeneracy between the two at a single energy. Panel (b): the $(\Re b_Z, \Re\tilde b_Z)$ plane. The couplings are deviations from the SM at one loop, and the cross marks the SM at one loop.}
\label{fig:contours}
\end{figure}

We now perform a complementary analysis supposing that the new physics imprint is solely contained in the form factors, setting the remaining parameters to the SM values. The model is then that of Eq.~\eqref{eq:model} with $\klam = 1$: the exact density matrix of the parametrized vertex at tree level (a function of the real and imaginary parts of the form factors), plus the two one-loop shifts of the SM. The test is then sensitive to the deviations that the form factors induce over the baseline provided by the SM at one loop. As in Section~\ref{sub:kappa}, the test uses the inclusive $\bar\rho$ and, where stated, $\rho(\theta)$ evaluated at the nodes and the $\cos\theta$ distribution, and is again performed for two values of the CM energy. The combination of the two runs assumes that the couplings do not depend on the energy, as for the case of local operators of dimension six generated by heavy new physics. The results obtained by varying one parameter at a time are shown in Table~\ref{tab:couplings} and Figures~\ref{fig:couplings} and~\ref{fig:contours}, and Table~\ref{tab:results} collects them in the format of Table~\ref{tab:bounds} for comparison purposes. The four parameters enter different Fano coefficients of the inclusive matrix, Table~\ref{tab:coefficients}, and correlations are numerically negligible. This justifies the choice of considering one variation at a time: the same results on the individual parameters would be obtained by profiling the other three in a joint four-parameter fit.  

For the CP-even coupling the inclusive matrix gives $\delta\Re b_Z = 2.9\times 10^{-3}$ at $240\GeV$, $1.1\times 10^{-3}$ at $365\GeV$ and $1.0\times 10^{-3}$ for the two runs together, statistical errors only. The quotes become $3.0$, $1.15$ and $1.1\times 10^{-3}$ once the theory uncertainty is included. Accounting for the full angular information and the theory uncertainty gives $2.2$, $0.87$ and $0.80\times 10^{-3}$, respectively. The intervals are symmetric to a few percent, $-2.1\times 10^{-3} < \Re b_Z - \Re b_Z^{\rm SM} < 2.1\times 10^{-3}$ at $95\%$ CL for the combination, Figure~\ref{fig:couplings}(a). The $365\GeV$ run wins despite its eight times smaller sample because the effect of $b_Z$ grows like $q\vdot k\,\lam$, Eq.~\eqref{eq:dRtrans}. Notice that the one-loop shift of the SM, projected on the direction of $\Re b_Z$ with the weights of the measurement, is $-6.7\times 10^{-3}$ at $240\GeV$ and $-3.6\times 10^{-3}$ at $365\GeV$ (see Section~\ref{sub:baseline_shift}), corresponding to $2.3$ and $3.3$ times the statistical error. The FCC-ee will then probe the one-loop $b_Z$ contribution of the SM, and the bound is on the departure from it. Placed in Table~\ref{tab:projections}, $\delta b_Z \simeq 10^{-3}$ is the level of the rate-based projections of Refs.~\cite{Craig:2015wwr,CEPCStudyGroup:2018ghi} and of the ILC angular fit~\cite{Ogawa:2017bsx}, obtained here from the polarization alone, without the rate and without its normalization.

The imaginary part of $b_Z$ is a different object. A local operator generated by heavy new physics has a real coefficient, so $\Im b_Z$ can only come from intermediate states that go on shell at the energy of the run, like the $Zh$ cut that gives $\Im\beta$, or new light particles. It is intrinsically energy dependent, and the combination of the two runs is meaningful only within a given model. The inclusive matrix gives $\delta\Im b_Z = 1.45\times 10^{-2}$ at $240\GeV$, $5.5\times 10^{-3}$ at $365\GeV$ and $5.2\times 10^{-3}$ for the combination, with the result driven by the hadronic sample. Resolving the scattering angle recovers the normal polarization in the leptonic sample and improves this only to $1.33\times 10^{-2}$, $5.2\times 10^{-3}$ and $4.9\times 10^{-3}$, Figure~\ref{fig:couplings}(b). The sensitivity is five times worse than for $\Re b_Z$ because $\Im b_Z$ affects the coherences, which carry the factor $A_e$. Unlike $\tilde b_Z$, $\Im b_Z$ also has an SM value, and not a small one: the absorptive parts of the one-loop amplitude shift $T_{23}$ by $-1.4\times 10^{-3}$ at $240\GeV$ and $-7.4\times 10^{-3}$ at $365\GeV$ (Table~\ref{tab:baseline}), which in terms of $\Im b_Z$ give $-7.3\times 10^{-3}$ and $-5.3\times 10^{-3}$, half a sigma and one sigma of the measurement. Only a few percent of this comes from the $b_Z$ structure itself: the rest is the chirality-odd part of the $\slashed k$ terms and the boxes of Eq.~\eqref{eq:M1structures}, which the inclusive $T_{23}$ cannot tell apart from a vertex correction. A bound on $\Im b_Z$ at the $5\times 10^{-3}$ level is thus a test of the absorptive part of the SM amplitude as much as of new physics. Its theory uncertainty is negligible at this precision.

The CP-odd coupling is what the method probes best, and Figure~\ref{fig:couplings}(c) and Table~\ref{tab:couplings} show the result. Its whole effect on the inclusive density matrix is the $T_{12}$ alignment and the normal polarization $P_2$, the first charge blind and measured by the hadronic sample with its $3\times 10^{-3}$ error, the second ten times less precise and visible in lepton final states only. We remark that, in this case, there is no SM value to subtract and no theory uncertainty. The averaged $P_2$ and $T_{12}$, like $P_3$ and $T_{13}$ for $\Im\tilde b_Z$, are CP-odd observables for the CP-symmetric, unpolarized initial state, and they vanish in any CP-conserving theory at any order, absorptive parts included. An absorptive phase produces the same correlations with the opposite parity in $\cos\theta$, which average to zero. The one-loop SM result of Section~\ref{sub:baseline_shift} shows this explicitly, angle by angle: it sources $P_2$ and $T_{12}$ with odd functions of $\cos\theta$ and $T_{23}$ with an even one, while the components that $\tilde b_Z$ populates, $P_2$ and $T_{12}$ even and $T_{23}$ odd, together with the even parts of $P_3$ and $T_{13}$, vanish to machine precision. The angle-resolved analysis is therefore free of the baseline as well, once the CP-odd projection $\cos\theta \to -\cos\theta$ is taken. The only source of CP violation in the SM, the CKM phase, generates a CP-odd $hZZ$ form factor first at three loops, at the $10^{-11}$ level~\cite{Hernandez-Juarez:2025hzz}. The measurement is then a null test. We also remark that the sensitivity achieved is due to the hadronic sample, which was marginal for the $\klam$ analysis. Against Table~\ref{tab:projections}, $\delta\tilde b_Z = 1.6\times 10^{-3}$ ($1.2\times 10^{-3}$) improves by a factor of three to four on the leptons-only projections for the same FCC-ee sample ($\abs{\tilde b_Z} < 0.012$ at $95\%$ CL from a neural-network optimal observable~\cite{Atwal:2025cpv}, which our leptons-only $95\%$ interval, $\pm 1.4\times 10^{-2}$ at $240\GeV$, reproduces) and matches the three-parameter ILC fit with polarized beams at $250$ and $500\GeV$~\cite{Ogawa:2017bsx}, which gives $2.0\times 10^{-3}$.

An imaginary component of $\tilde b_Z$, CP-odd and $\hat T$-even, is determined by the tomographic approach through the $P_3$ and $T_{13}$ components of the inclusive density matrix to $5.3\times 10^{-3}$ and $4.1\times 10^{-3}$, at $240\GeV$ and $365\GeV$, respectively. The combination gives $3.25\times 10^{-3}$, Figure~\ref{fig:couplings}(d). The strongest constraint, however, comes from the forward--backward asymmetry of the $Z$, Eq.~\eqref{eq:Xtilde}, which the $\cos\theta$ distribution alone measures to $\delta\Im\tilde b_Z = 2.2\times 10^{-3}$ at each energy and $1.55\times 10^{-3}$ for the two runs, and together with the polarization to $1.4\times 10^{-3}$. It is a CP-odd observable as well, and it also vanishes in the SM at any order.

The determinations of $\klam$, in Section~\ref{sub:kappa}, and that of $\Re b_Z$, presented here, are not independent. The density matrix depends on $\klam$ only through $b_Z$: at a given energy the tomographic process measures one effective coupling, $b_Z(s) = b_Z^{\rm SM}(s) + (\klam - 1)\beta(s) + b_Z^{\rm NP}$, with $b_Z^{\rm SM}$ the one-loop value (which contains $\beta(s)$ at $\klam = 1$) and $b_Z^{\rm NP}$ a genuine anomalous coupling. The $\delta\klam$ of Section~\ref{sub:kappa} assumes $b_Z^{\rm NP} = 0$, the $\delta b_Z$ above assumes $\klam = 1$. Even taking $b_Z^{\rm NP}$ energy independent, it is difficult to tell apart these two contributions through the tomographic process. One possible way is in the energy dependence of $\Re\beta(365)/\Re\beta(240) = 0.35$ against the contribution of a constant $b_Z^{\rm NP}$, another is in the absorptive part $\Im\beta$, which a real $b_Z^{\rm NP}$ does not have and which sources $T_{23}$. At $240\GeV$ alone the two parameters are basically degenerate, whereas at $365\GeV$, where $\Im\beta$ is twice $\Re\beta$, $\delta\klam = 14$. The combination of the two runs gives $\delta\klam = 7.4$ and $\delta\Re b_Z = 2.1\times 10^{-3}$ with a correlation of $-0.85$, against $3.9$ and $1.1\times 10^{-3}$ one at a time, and $\klam \in [-6.4, 8.4]$ at $68\%$ and $[-13.6, 15.6]$ at $95\%$ CL with $b_Z^{\rm NP}$ profiled. These results are shown in Figure~\ref{fig:contours}(a), where the theory uncertainty is included. What quantum state tomography measures is therefore $b_Z(240\GeV)$ and $b_Z(365\GeV)$, from which $\klam$ follows once the other sources of $b_Z$ are fixed, by the rate, by other processes in a global fit, or by assumption. The bound on $b_Z$ is robust: a $\klam$ within the HL-LHC reach, $\abs{\klam - 1} \lesssim 0.5$, shifts $b_Z$ by less than $3\times 10^{-4}$, a third of its error.

\begin{table}[h]
\centering
\small
\setlength{\tabcolsep}{6pt}
\begin{tabular}{@{}lccc@{}}
\toprule
 & $240\GeV$ & $365\GeV$ & combination \\
\midrule
\multicolumn{4}{@{}l}{$\delta\Re b_Z$ [$10^{-3}$]} \\
\quad inclusive $\bar\rho$, $Z\to\ell\ell$ / $Z\to q\bar q$ & $9.9$ / $3.1$ & $3.7$ / $1.14$ & \\
\quad inclusive $\bar\rho$, both & $2.9$ ($3.0$) & $1.09$ ($1.15$) & $1.02$ ($1.08$) \\
\quad\quad $95\%$ CL & $[-5.8, 6.0]$ & $[-2.2, 2.3]$ & $[-2.1, 2.1]$ \\
\quad $\rho(\theta)$ resolved / $\cos\theta$ distribution & $2.9$ / $3.1$ & $1.19$ / $1.12$ & $1.10$ / $1.05$ \\
\quad polarization and distribution & $2.1$ ($2.2$) & $0.82$ ($0.87$) & $0.76$ ($0.80$) \\
\quad quantum Cram\'er--Rao bound & $0.91$ & $0.35$ & $0.33$ \\
\multicolumn{4}{@{}l}{$\delta\Im b_Z$ [$10^{-3}$]} \\
\quad inclusive $\bar\rho$, $Z\to\ell\ell$ / $Z\to q\bar q$ & $49$ / $15$ & $19$ / $5.8$ & \\
\quad inclusive $\bar\rho$, both & $14.5$ ($14.5$) & $5.5$ ($5.6$) & $5.2$ ($5.2$) \\
\quad\quad $95\%$ CL & $[-27, 27]$ & $[-10.5, 10.7]$ & $[-9.9, 10.0]$ \\
\quad $\rho(\theta)$ resolved & $13.3$ ($13.3$) & $5.2$ ($5.3$) & $4.9$ ($4.9$) \\
\quad quantum Cram\'er--Rao bound, inclusive / resolved & $4.7$ / $1.6$ & $1.9$ / $0.68$ & $1.8$ / $0.63$ \\
\multicolumn{4}{@{}l}{$\delta\Re\tilde b_Z$ [$10^{-3}$]} \\
\quad inclusive $\bar\rho$, $Z\to\ell\ell$ / $Z\to q\bar q$ & $7.2$ / $2.3$ & $8.3$ / $2.6$ & \\
\quad inclusive $\bar\rho$, both & $2.2$ & $2.5$ & $1.6$ \\
\quad\quad $95\%$ CL & $\pm 4.2$ & $\pm 4.9$ & $\pm 3.2$ \\
\quad $\rho(\theta)$ resolved & $1.8$ & $1.6$ & $1.2$ \\
\quad quantum Cram\'er--Rao bound, inclusive / resolved & $0.55$ / $0.38$ & $0.46$ / $0.29$ & $0.35$ / $0.23$ \\
\multicolumn{4}{@{}l}{$\delta\Im\tilde b_Z$ [$10^{-3}$]} \\
\quad inclusive $\bar\rho$, both & $5.3$ & $4.1$ & $3.25$ \\
\quad\quad $95\%$ CL & $\pm 10.4$ & $\pm 8.0$ & $\pm 6.4$ \\
\quad $\rho(\theta)$ resolved / $\cos\theta$ distribution & $5.2$ / $2.2$ & $3.9$ / $2.2$ & $3.1$ / $1.55$ \\
\quad polarization and distribution & $2.0$ & $1.9$ & $1.4$ \\
\quad quantum Cram\'er--Rao bound & $0.28$ & $0.24$ & $0.18$ \\
\bottomrule
\end{tabular}
\caption{One-sigma sensitivity to the four real parameters of the anomalous vertex, as deviations from the SM at one loop with $\klam = 1$. The presented uncertainties are statistical, whereas the quotes in parentheses contain also the $10\%$ theory uncertainty on the one-loop shift where it matters (the CP-odd couplings have no associated theory uncertainty). The $95\%$ CL intervals include both sources and are derived from the exact density matrix by varying one parameter at a time (correlations are negligible). The rate is excluded throughout. The combination of the two runs assumes energy-independent couplings, which holds for $\Re b_Z$ and $\tilde b_Z$ at the dimension-six level but not for the absorptive $\Im b_Z$. The quantum Cram\'er--Rao bound is the one of Section~\ref{sub:quantum} for an ideal measurement of the polarization on all the $ZH$ events, of the inclusive state and, where indicated, with the scattering angle resolved.}
\label{tab:couplings}
\end{table}

\begin{table}[h]
\centering
\footnotesize
\setlength{\tabcolsep}{4pt}
\renewcommand{\arraystretch}{1.2}
\begin{tabular}{@{}l p{5.0cm} p{8.2cm}@{}}
\toprule
 & Sensitivity ($1\sigma$) & Observable, assumptions \\
\midrule
$\Re b_Z$ & \raggedright\arraybackslash $\delta\Re b_Z = 1.1\times 10^{-3}$ ($0.8\times 10^{-3}$) & deviation from the SM value at one loop, $-6.7\times 10^{-3}$ ($240\GeV$) and $-3.6\times 10^{-3}$ ($365\GeV$); $\klam = 1$ \\
$\Im b_Z$ & \raggedright\arraybackslash $\delta\Im b_Z = 5.2\times 10^{-3}$ ($4.9\times 10^{-3}$) & through $T_{23}$; SM value $-7.3\times 10^{-3}$ ($240\GeV$) and $-5.3\times 10^{-3}$ ($365\GeV$) \\
$\Re\tilde b_Z$ & \raggedright\arraybackslash $\delta\Re\tilde b_Z = 1.6\times 10^{-3}$ ($1.2\times 10^{-3}$); $\abs{\Re\tilde b_Z} < 3.2\times 10^{-3}$ at $95\%$ CL & through the $T_{12}$ alignment of the hadronic sample, Section~\ref{sub:rhoshift}; no SM value, no theory uncertainty \\
$\Im\tilde b_Z$ & \raggedright\arraybackslash $\delta\Im\tilde b_Z = 3.2\times 10^{-3}$ ($1.4\times 10^{-3}$) & through $P_3$, $T_{13}$ and the forward--backward asymmetry of the $Z$; no SM value \\
$\klam$ & \raggedright\arraybackslash $\delta\klam = 3.9$ ($2.9$); $68\%$ CL $[-2.9, 4.9]$; $\delta\klam = 7.4$ with $b_Z$ free & through $b_Z = \klam\beta(s)$, no other new physics in the vertex \\
\bottomrule
\end{tabular}
\caption{The bounds achievable with quantum state tomography of the $Z$ boson helicity at the FCC-ee, in the normalization of Table~\ref{tab:bounds}: $240 + 365\GeV$ with $10.8$ and $2.7\,{\rm ab}^{-1}$, respectively. The analysis considers the $Z\to\ell\ell$ and $q\bar q$ decay channels, full efficiency and utilizes a $\chi^2$ test targeting the Fano coefficients of the inclusive density matrix around the SM result computed at one loop. Statistical errors are supplemented with a theoretical uncertainty given by $10\%$ of the one-loop shift. The quotes in parentheses assume that the scattering angle is resolved and the $\cos\theta$ distribution is reconstructed. The details are in Tables~\ref{tab:sensitivity} and~\ref{tab:couplings}.}
\label{tab:results}
\end{table}

\subsection{Quantum Fisher information}  
\label{sub:quantum}

We conclude the statistical analysis by assessing how far the decay of the $Z$ boson is from the best measurement of its polarization that quantum mechanics allows. To this purpose, we utilize two Fisher informations~\cite{Helstrom:1976,Braunstein:1994zz}: the classical one, which pertains to the measurement performed by the decay, and the quantum one, which pertains to the state.

The decay is a fixed measurement on the spin of the $Z$ boson. Its outcome is the direction $\Omega^*$ of the fermion, distributed according to $W(\Omega^*; p) = \Tr[\rho(p)\,\mathcal D(\Omega^*)]$ of Eq.~\eqref{eq:W}, where $p$ denotes the parameter of interest: $\klam$ or one of the form factors. The information on $p$ carried by one event is the classical Fisher information of this distribution:
\begin{equation}
	I_p = \int \dd\Omega^*\, \frac{\qty(\partial_p W)^2}{W} .
	\label{eq:FI}
\end{equation}
The Cram\'er--Rao inequality then bounds the variance of any unbiased estimator of $p$ built from $N$ events, $\delta p \geq (N I_p)^{-1/2}$, and the maximum likelihood estimator saturates the bound for large $N$. The projector estimator of Section~\ref{sub:decay} carries instead the information $I_p^{\rm proj} = \partial_p f^{\sf T}\, C^{-1}\, \partial_p f$ per event, with the covariance $C$ of Eq.~\eqref{eq:cov}, which cannot exceed $I_p$. Importantly, for the inclusive density matrix at the one-loop SM point the two quantities coincide to $0.4\%$ at $240\GeV$ and $4\%$ at $365\GeV$ for $\klam$, and to $1.0\%$ and $1.6\%$ for $\tilde b_Z$. Hence, for the samples we consider, the $\chi^2$ test of Eq.~\eqref{eq:chi2} performs as well as the full likelihood of the decay angles. This holds for the ideal sample considered here: a likelihood analysis becomes the natural tool once a non-uniform acceptance, the backgrounds and a partial tagging of the quark charge, which would give access to the vector polarization in the hadronic channel, are included.

The quantum Fisher information addresses the same question for every possible measurement on the spin of the $Z$ boson, rather than for the one performed by the decay. Writing the density matrix in its eigenbasis, $\rho = \sum_i p_i \dyad{i}$, we have:
\begin{equation}
	F_Q = \sum_{i,j} \frac{2\abs{\mel{i}{\partial_p\rho}{j}}^2}{p_i + p_j} ,
	\label{eq:QFI}
\end{equation}
where the sum runs over the pairs with $p_i + p_j > 0$. Equivalently, $F_Q = \Tr[\rho L^2]$ in terms of the symmetric logarithmic derivative $L$, defined by $\partial_p\rho = (L\rho + \rho L)/2$, of which Eq.~\eqref{eq:QFI} is the solution in the eigenbasis of $\rho$. Notice that the diagonal terms, $i = j$, yield the classical information of the eigenvalues, $(\partial_p p_i)^2/p_i$, whereas the off-diagonal ones account for the change of the eigenvectors. The quantity bounds every measurement: for any POVM $\{E_\xi\}$ on the $Z$ boson, the classical Fisher information of the outcome distribution $\Tr[\rho E_\xi]$ does not exceed $F_Q$, and the bound is attained by the measurement in the eigenbasis of $L$~\cite{Braunstein:1994zz}. The decay is one such POVM, $E(\Omega^*) = \mathcal D(\Omega^*)\, \dd\Omega^*$, so that $I_p \leq F_Q$. As a result, $N$ copies of the state yield the quantum Cram\'er--Rao bound $\delta p \geq (N F_Q)^{-1/2}$, which no measurement of the polarization can outperform. When the scattering angle is measured as well, the events carry the classical label $\theta$, distributed as $w(\theta) = \sigma^{-1}\dd\sigma/\dd\!\cos\theta$, and the corresponding informations add:
\begin{equation}
	F = \int \dd\!\cos\theta\; w(\theta)\, F_Q[\rho(\theta)] + \int \dd\!\cos\theta\; \frac{\qty(\partial_p w)^2}{w} ,
	\label{eq:QFItheta}
\end{equation}
in which the second term is the information of the $\cos\theta$ distribution. The same decomposition holds for $I_p$.

We report the resulting bounds in Tables~\ref{tab:sensitivity} and~\ref{tab:couplings}, next to the results of the tomography. At the SM point at one loop, with all the $ZH$ events of each run, the inclusive state gives $\delta\klam \geq 1.6$ at $240\GeV$ and $1.7$ at $365\GeV$, and $1.2$ for the two runs together. Resolving the scattering angle leaves the bound unchanged at $240\GeV$ and lowers it to $1.25$ at $365\GeV$ and $0.98$ for the combination. For $\Re b_Z$ the bound is $0.9\times 10^{-3}$ and $0.35\times 10^{-3}$, whereas for $\tilde b_Z$ it amounts to $0.55\times 10^{-3}$ and $0.46\times 10^{-3}$ for the inclusive state, $0.38\times 10^{-3}$ and $0.29\times 10^{-3}$ with the scattering angle resolved, and $0.23\times 10^{-3}$ for the combination of the two runs. The decay $Z\to f\bar f$ is a fixed, non-optimal measurement, with the analyzing power of Eq.~\eqref{eq:W} and no access to the vector polarization unless the fermion charge is measured. Per event, it realizes $12$--$14\%$ of $F_Q$ for the inclusive state, the same fraction for leptons and hadrons since the information is in the tensor coefficients, and $7$--$8\%$ at $365\GeV$ when the scattering angle is resolved. For the CP-odd coupling the fraction is lower, $9\%$ and $5\%$, as the decay allows access to only one of the two entries carrying the information, $T_{12}$, with full weight, whereas the other, $P_2$, is accessible only through the leptonic channel. The remaining distance between the quantum bounds and the sensitivities of Tables~\ref{tab:sensitivity} and~\ref{tab:couplings} is due to the branching ratio into visible decays. In other words, the $\delta\klam \simeq 3$ obtained with the tomography is not far from what the polarization of the $Z$ boson can yield at all: the limit is in the state, not in the measurement.

\section{Summary and outlook}  
\label{sec:outlook}

In this paper we have assessed the reach of quantum state tomography, applied to the $Z$ boson produced in Higgsstrahlung at the FCC-ee, on the Higgs boson self-coupling and on the anomalous couplings of the $hZZ$ vertex. We summarize here the main results.

The one-loop diagrams of $e^-e^+ \to ZH$ that depend on the Higgs boson self-coupling are three. Their sum is finite and amounts to a correction of the $hZZ$ vertex with two structures only, $a_Z = 1 + \klam c(s)$ and $b_Z = \klam \beta(s)$, Eqs.~\eqref{eq:beta} and~\eqref{eq:c}. The first changes the rate, reproduces the coefficient $C_1$ of the literature and is invisible to the polarization of the $Z$ boson. The second is a genuine $b_Z$, of size $5.6\times 10^{-4}$ at $240\GeV$ and $1.9\times 10^{-4}$ at $365\GeV$. The complete one-loop electroweak correction, computed with all the diagrams and the on-shell counterterms (the photonic corrections to the initial state excluded, as they cancel in the density matrix), has the same Lorentz structures but is not a vertex correction: its coefficients depend on the chirality of the electron and on the scattering angle, and the box diagrams add a structure that no vertex produces. Written in terms of form factors of the same structures, it contains a $b_Z$ of the SM ten times larger than $\beta(s)$ and of opposite sign, as well as a ten percent difference between the two chiralities in the $a_Z$ structure, which is the running of the weak mixing angle.

We have computed the helicity density matrix of the $Z$ boson in closed form, both for the parametrized vertex and for the SM at one loop, by means of the spin-one projector. The CP-even coupling $b_Z$ transfers weight from the longitudinal to the transverse polarizations, and its absorptive part generates the $\hat T$-odd entries. The CP-odd coupling $\tilde b_Z$, whose real part at linear order changes neither the rate nor the angular distribution, sources instead the imaginary parts of the coherences: a normal polarization that survives the average over the scattering angle and an $\hat r\hat n$ alignment that the hadronic decays measure without the fermion charge. Its imaginary part produces a helicity asymmetry and a forward--backward asymmetry of the $Z$ boson. In the SM at one loop the density matrix is known up to the single parameter $\klam$, $\rho_{\rm SM}(\klam) = \rho_{ 0} + \klam\,\delta\rho_{hhh} + \delta\rho_{\rm rest}$, and the two shifts are very different. The one of the self-coupling is of order $10^{-4}$ per unit of $\klam$. The rest moves the transverse polarization $P_1$ in the production plane by half of its tree-level value, an effect of the effective weak mixing angle at $\sqrt s$, $\sin^2\theta_{\rm eff}(240\GeV) = 0.237$, and the tensor coefficients by $13$ to $20$ times the $\klam$ effect, most of it due to the $b_Z$ of the SM. These shifts amount to two to three standard deviations of the FCC-ee samples. The CP-odd entries, instead, are left empty, as any CP-conserving theory requires.

We have then characterized the $Z$ boson as a qutrit through its purity and coherence, its contextuality and its magic. In the SM the state is purest at $\sqrt s = 365\GeV$ in the central region, with $\Tr\rho^2 = 0.674$ at $\theta = \pi/2$, and only there it violates the KCBS contextuality inequality, with $\mathbb{CNTXT}_5 = 2.06$ against the non-contextual bound $2$. The inclusive states, with purity $0.40$ and $0.51$, cannot violate it. The optimized nine-projector inequality is violated by every state but the maximally mixed one, hence it cannot generally be used to discriminate between states. For the magic we have compared three measures in the helicity basis. The inclusive state has a non-negative Wigner function and a robustness of magic equal to one: it is a mixture of stabilizer states and carries no magic, whatever the phase convention of the helicity states. At fixed angle the mana reaches $0.14$ at $240\GeV$ and $0.21$ at $365\GeV$, values that do not depend on the phase convention, unlike its angular distribution. The mana, however, also vanishes in regions where the robustness exceeds one: there the $Z$ boson is in a state with bound magic, which the mana cannot detect. With the phases of Eq.~\eqref{eq:polvectors}, the robustness finds magic at every scattering angle except $\theta = \pi/2$ at $240\GeV$, and reaches $1.53$ at $240\GeV$ and $1.69$ at $365\GeV$. The stabilizer R\'enyi entropy, which is positive also on mixtures of stabilizer states, tracks the mixedness of the state as much as its magic. All these quantities do not depend at linear order on $\Im b_Z$ and $\Re\tilde b_Z$, respond linearly to $\Re b_Z$ and, at fixed angle, to $\Im\tilde b_Z$, and reach their extreme values when $\Re b_Z = -\MZ^2/(q\vdot k)$ cancels the transverse amplitudes. For couplings of the size probed by the tomographic process, however, the change is at most a few units in $10^{-3}$: these quantum properties are robust features of the $Z$ boson in Higgsstrahlung rather than probes of the vertex, which is best constrained by the Fano coefficients themselves.

Finally, in order to constrain $\klam$ and the anomalous couplings via quantum state tomography at the FCC-ee, we have performed a $\chi^2$ test on the Fano coefficients of the reconstructed density matrix. The test is centered on the SM prediction at one loop and accounts for a $10\%$ uncertainty on its $\klam$-independent part. With no other new physics in the vertex, the polarization determines $\klam$ to $\pm 3.9$ for the two FCC-ee runs, and to $\pm 2.9$ with the full angular information, independently of the rate and of the normalization of the vertex. The rate, under the same assumption and with the same uncertainty on its one-loop correction, performs much better. The polarization is sensitive to $\klam$ only through $b_Z = \klam\beta(s)$, hence the natural outcome of the tomography is the determination of the form factors, which requires no assumption on the new physics: $\Re b_Z$ to $1.1\times 10^{-3}$, or $0.8\times 10^{-3}$ with the full angular information, as a deviation from the SM value, which is $-6.7\times 10^{-3}$ at $240\GeV$ and will be probed by the FCC-ee itself. Its absorptive part is constrained to $5\times 10^{-3}$, at the level of the SM absorptive part. The CP-odd form factor $\tilde b_Z$ is bound to $1.6\times 10^{-3}$, or $1.2\times 10^{-3}$ with the scattering angle resolved, with no SM value and no theory uncertainty. Its imaginary part is constrained to $1.4\times 10^{-3}$, through the forward--backward asymmetry. With an energy-independent anomalous $b_Z$ fitted alongside $\klam$, the different energy dependence of $\beta(s)$ and its absorptive part leave $\delta\klam = 7.4$. We have also shown that the projector estimator loses nothing with respect to the full likelihood, that the hadronic channel carries the information and, through the quantum Fisher information, that no measurement of the polarization, however refined, would determine $\klam$ better than $\delta\klam \simeq 1$.

What remains to be assessed is the comparison with the rate in a global fit. The rate alone, with its per-mille precision, determines $\klam$ well, but through $C_1$, which is degenerate with the normalization of the $hZZ$ vertex, with the input scheme and with every other operator that rescales the cross section. The polarization gives instead $b_Z(s)$, free of all these degeneracies. Whether the second determination helps the first in a global fit with the other operators of the same dimension is left to future work.

\section*{Acknowledgments}
This work was supported by the Estonian Research Council grants TARISTU24-TK10, TARISTU24-TK3, RVTT3, PRG1884, and by the CoE TK 202 ``Foundations of the Universe''. The authors wish to thank Torben Lange and Nicola De Filippis for useful discussions.

\begin{appendices}

\section{Dictionary of the normalizations}  
\label{app:dictionary}

The vertex~\eqref{eq:vertex} is the Feynman rule of the effective interaction~\eqref{eq:Leff}, with $v = 2\MZ\cw/g$, the momentum $q$ of the virtual $Z$ entering the vertex and the momentum $k$ of the real one produced in the process. In our conventions $\epsilon^{0123} = +1$.  

Every other normalization of the literature maps onto Eq.~\eqref{eq:Leff} by a comparison of coefficients, and Table~\ref{tab:dictionary} collects the result. The numerical factors use $v = 246.22\GeV$ and $\Lambda = 1\,$TeV. Two remarks follow: first, all the parametrizations below agree on the CP-even and CP-odd structures, but not on the sign conventions, and the sign matters for the interpretation of the asymmetric LHC intervals. We have re-derived each relation to ensure consistency. Second, the overall sign of $\tilde b_Z$ is tied to the sign convention of the Levi-Civita tensor, which not all the sources state. Where they do not, Table~\ref{tab:bounds} quotes $\abs{\tilde b_Z}$. The sign is physical, and quantum state tomography measures it.

\begin{table}[h]
\centering
\small
\setlength{\tabcolsep}{4pt}
\renewcommand{\arraystretch}{1.25}
\begin{tabular}{@{}l p{5.6cm} c c c@{}}
\toprule
Parametrization & Interaction & $a_Z$ & $b_Z$ & $\tilde b_Z$ \\
\midrule
this work & Eq.~\eqref{eq:Leff} & $a_Z$ & $b_Z$ & $\tilde b_Z$ \\
$\kappa$ framework & $\kappa_Z\, (\MZ^2/v)\, h Z_\mu Z^\mu$ & $\kappa_Z$ & -- & -- \\
JHUGen~\cite{Anderson:2013afp} & $v^{-1}\big[a_1 \MZ^2\, \eps_1^*\vdot\eps_2^*$ $+ a_2\, f^{*(1)}_{\mu\nu} f^{*(2)\mu\nu}$ $+ a_3\, f^{*(1)}_{\mu\nu}\tilde f^{*(2)\mu\nu}\big]$ for $H\to Z_1 Z_2$ & $a_1/2$ & $-a_2$ & $a_3$ \\
Higgs basis~\cite{CMS:2021nnc} & $(h/v)\big[\delta c_z \MZ^2 Z_\mu Z^\mu$ $+ \frac{g^2 + g'^2}{4}\big(c_{zz} Z_{\mu\nu}Z^{\mu\nu}$ $+ \tilde c_{zz} Z_{\mu\nu}\tilde Z^{\mu\nu}\big)\big]$ & $1 + \delta c_z$ & $0.274\, c_{zz}$ & $-0.274\, \tilde c_{zz}$ \\
Ref.~\cite{Craig:2015wwr} & $(\hat\alpha_{ZZ}/v)\, h Z_{\mu\nu}Z^{\mu\nu}$ & -- & $2\hat\alpha_{ZZ}$ & -- \\
ILC~\cite{Ogawa:2017bsx} & $(b^{\rm ILC}_Z/2\Lambda)\, h Z_{\mu\nu}Z^{\mu\nu}$ $+ (\tilde b^{\rm ILC}_Z/2\Lambda)\, h Z_{\mu\nu}\tilde Z^{\mu\nu}$ & -- & $0.246\, b^{\rm ILC}_Z$ & $-0.246\, \tilde b^{\rm ILC}_Z$ \\
Warsaw basis & $(c_{H\widetilde W}/\Lambda^2)\, H^\dagger H\, \widetilde W^a_{\mu\nu} W^{a\mu\nu}$ alone & -- & -- & $\pm 0.094\, c_{H\widetilde W}$ \\
HISZ, ATLAS~\cite{ATLAS:2023xmo} & $\tilde d = \tilde d_B$ & -- & -- & $\pm 2\tilde d$ \\
\bottomrule
\end{tabular}
\caption{The normalizations of the $hZZ$ vertex used in the literature, written in the conventions of Eq.~\eqref{eq:Leff} ($\epsilon^{0123} = +1$, $v = 246.22\GeV$, $\Lambda = 1\,$TeV). The derivations are in the text.}
\label{tab:dictionary}
\end{table}

\emph{The $\kappa$ framework} rescales the SM interaction, $(\MZ^2/v)\, h Z_\mu Z^\mu$, so that $a_Z = \kappa_Z$, and has no anomalous structures.

\emph{JHUGen.} The LHC analyses~\cite{CMS:2021nnc,CMS:2022uox,CMS:2026gg} use the amplitude of Ref.~\cite{Anderson:2013afp} for the decay $H\to Z_1 Z_2$, with $f^{(i)\mu\nu} = \eps_i^\mu q_i^\nu - \eps_i^\nu q_i^\mu$, $\tilde f^{(i)}_{\mu\nu} = \frac12\epsilon_{\mu\nu\rho\sigma} f^{(i)\rho\sigma}$ and $q_i$ the outgoing momenta of the two bosons. In these conventions $a_1 = 2$ in the SM. The interaction~\eqref{eq:Leff} gives in the decay $a_1 = 2 a_Z$, $a_2 = -b_Z$ and $a_3 = \tilde b_Z$: the relative sign between $a_2$ and $b_Z$ comes from the momentum of the virtual $Z$, which enters the decay amplitude as an outgoing momentum and our vertex as an incoming one. The measured quantities are the cross-section fractions $f_{ai} = \abs{a_i}^2\sigma_i/\sum_j\abs{a_j}^2\sigma_j$, with the $\sigma_i$ of $H\to ZZ\to 2e2\mu$, $\sigma_2/\sigma_1 = 0.361$ and $\sigma_3/\sigma_1 = 0.153$~\cite{CMS:2021nnc}. For one anomalous coupling at a time $a_i/a_1 = \mathrm{sgn}(f_{ai})\sqrt{\abs{f_{ai}}\,\sigma_1/((1-\abs{f_{ai}})\sigma_i)}$, and with $a_1 = 2$ this is how the $f_{ai}$ intervals of Table~\ref{tab:bounds} become intervals on $b_Z = -a_2$ and $\tilde b_Z = a_3$. The analyses assume $a_i^{WW} = a_i^{ZZ}$ or the SU(2)$\times$U(1) relations, and no anomalous $H\gamma\gamma$ and $HZ\gamma$ couplings.

\emph{Higgs basis.} With $(g^2 + g'^2)/4 = \MZ^2/v^2$ the comparison with Eq.~\eqref{eq:Leff} gives $a_Z = 1 + \delta c_z$, $b_Z = 2(\MZ^2/v^2)\, c_{zz}$ and $\tilde b_Z = -2(\MZ^2/v^2)\, \tilde c_{zz}$, with $2\MZ^2/v^2 = 0.274$. Combined with $a_2 = -2(\MZ^2/v^2)\, c_{zz}$, which follows from the same interaction in the decay, this is the relation between the JHUGen couplings and the Higgs basis used in Ref.~\cite{CMS:2021nnc}. The contact term $c_{z\Box}\, g^2 (h/v) Z_\mu\partial_\nu Z^{\mu\nu}$ of the Higgs basis gives, on the conserved electron current and on the on-shell $Z$, a contribution proportional to $g^{\mu\nu}$ only: it is an energy-dependent $a_Z$, invisible in the density matrix.

\emph{Refs.~\cite{Craig:2015wwr} and~\cite{Ogawa:2017bsx}} write the anomalous terms as $(\hat\alpha_{ZZ}/v)\, h Z_{\mu\nu}Z^{\mu\nu}$ and as $(b_Z^{\rm ILC}/2\Lambda)\, h Z_{\mu\nu}Z^{\mu\nu} + (\tilde b_Z^{\rm ILC}/2\Lambda)\, h Z_{\mu\nu}\tilde Z^{\mu\nu}$, whence $b_Z = 2\hat\alpha_{ZZ}$, $b_Z = (v/\Lambda)\, b_Z^{\rm ILC}$ and $\tilde b_Z = -(v/\Lambda)\, \tilde b_Z^{\rm ILC}$, with $v/\Lambda = 0.246$. Only the sizes of the projected errors enter Table~\ref{tab:bounds}.

\emph{Warsaw basis.} For the CP-odd operator $c_{H\widetilde W}$ alone, the Higgs-basis coupling is $\tilde c_{zz} = 4 g^2 c_{H\widetilde W} v^2/((g^2 + g'^2)^2\Lambda^2) = \MW^2 v^4 c_{H\widetilde W}/(\MZ^4\Lambda^2) = 0.343\, c_{H\widetilde W}$, so that $\abs{\tilde b_Z} = 0.094\abs{c_{H\widetilde W}}$. The same operator also gives an $hWW$ coupling, and the VBF-dominated bounds on it are bounds on a combination of the two. For the HISZ parameter $\tilde d$ of the ATLAS analysis~\cite{ATLAS:2023xmo}, with $\tilde d = \tilde d_B$, we use $\abs{\tilde b_Z} = 2\abs{\tilde d}$, the relation between $\tilde d$ and $a_3$ given there.

\section{Conventions for the density matrix}  
\label{app:conventions}

\emph{Phases of the polarization vectors.} The vectors~\eqref{eq:polvectors} are the helicity vectors in the Condon--Shortley convention, built on the triad $\{\hat r, \hat n, \hat k\}$ attached to the production plane. They satisfy
\begin{equation}
	\eps(k, -\lambda) = (-1)^\lambda\eps(k, \lambda)^*.
\end{equation}
Other conventions differ by phases, $\eps(\lambda)\to e^{i\varphi_\lambda}\eps(\lambda)$, under which $\rho_{\lambda\lambda'}\to e^{-i(\varphi_\lambda - \varphi_{\lambda'})}\rho_{\lambda\lambda'}$: the diagonal entries, the rate, the decay distribution $W(\hat d)$ and the polarization vector and alignment tensor of Eq.~\eqref{eq:rhoPT}, which are covariant objects built on the triad, do not change, and only the coherences in the helicity basis do. Two cases are common. With the opposite sign of $\eps(+1)$, i.e.\ $\eps(\pm 1) = (0, 1, \pm i, 0)/\sqrt2$ for $\vb*k$ along $z$, the entries $\rho_{+0}$ and $\rho_{+-}$ change sign while $\rho_{0-}$ does not, and the standard $S^i$ of Section~\ref{sub:projector} reproduce $P_i$ only after $S^i\to U S^i U$ with $U = \mathrm{diag}(-1, 1, 1)$. In the convention of Haber~\cite{Haber:1994pe}, the vectors for a momentum along $(\theta, \phi)$ are ours multiplied by $e^{i\lambda\phi}$: they are obtained with the rotation $R(\phi, \theta, -\phi)$, which keeps them single valued along the beam axis, instead of $R(\phi, \theta, 0)$. The two coincide at $\phi = 0$, where our kinematics is set, and in general $\rho^{\rm Haber}_{\lambda\lambda'} = e^{-i(\lambda - \lambda')\phi}\rho_{\lambda\lambda'}$. Our $\rho$ does not depend on the azimuth of the event, because the triad turns with the production plane, and this is what allows the average over the events that defines the inclusive $\bar\rho$. With Haber's phases the coherences rotate with the event and the phase has to be removed before averaging.

\emph{The projector.} Equation~\eqref{eq:projectorS} is the complex conjugate of Eq.~(15) of Choi, Lee and Song~\cite{Choi:1989yf}, who write the tensor as $\eps^\mu(\lambda)\eps^{\nu*}(\lambda')$ and index the matrices as $(S^i)_{\lambda'\lambda}$. In a density matrix the conjugate vector carries the row index, so the object we need is the conjugate of theirs, and since the $S^i$ and $S^{ij}$ are hermitian the conjugation turns their $(S^i)_{\lambda'\lambda}$ into our $(S^i)_{\lambda\lambda'}$. The placement of the indices is therefore not a choice: using their placement with our ordering of the vectors would give $\rho^*$ instead of $\rho$, which is invisible in the real entries and flips the sign of the $\hat T$-odd ones, the polarization along $\hat n$ and the $\hat n\hat k$ alignment. One caveat deserves a mention: that paper takes $\epsilon_{0123} = +1$, i.e.\ $\epsilon^{0123} = -1$, so their Eq.~(15) read with $\epsilon^{0123} = +1$, the {\tt FeynCalc} default and our convention, has the sign of its $\epsilon$ term reversed. The conjugation flips that sign back, and Eq.~\eqref{eq:projectorS} ends up with the sign printed there, but for the conjugate tensor. We verified Eq.~\eqref{eq:projectorS} numerically against the explicit vectors~\eqref{eq:polvectors}. 

\section{Contextuality tests for a qutrit} 
\label{app:contextuality}

We follow Refs.~\cite{Fabbrichesi:2025ywl,Fabbrichesi:2025aqp}, where these tests were applied to the polarization of massive spin-one particles at colliders, and to which we refer for further details. On the three-dimensional space of the $Z$ boson helicity, consider a set of $n$ rank-one projectors $P_i = \ket{v_i}\bra{v_i}$, some of which are orthogonal, hence compatible and mutually exclusive. In a non-contextual hidden-variable model every $P_i$ is assigned the value $0$ or $1$, with at most one $1$ among mutually orthogonal projectors, and the sum $\sum_i \expval{P_i}$ is bounded by the independence number $\alpha$ of the orthogonality graph of the set. Quantum mechanics replaces this bound with the largest eigenvalue of $\sum_i P_i$. We consider two sets:
\begin{itemize}
\item the five projectors of the KCBS test~\cite{Klyachko:2008}, with
\begin{equation}
	\begin{gathered}
	\ket{v_j} = \qty(\cos\phi,\ \sin\phi\cos\frac{4\pi j}{5},\ \sin\phi\sin\frac{4\pi j}{5}) , \qquad j = 1,\dots,5 , \\
	\cos^2\phi = \frac{\cos(\pi/5)}{1+\cos(\pi/5)} .
	\end{gathered}
\end{equation}
Consecutive vectors are orthogonal, the orthogonality graph is a pentagon and
\begin{equation}
	\mathbb{CNTXT}_5 = \sum_{j=1}^5 \expval{P_j} , \qquad \begin{cases} \mathbb{CNTXT}_5 \leq 2 & \text{non-contextual} \\ \mathbb{CNTXT}_5 \leq \sqrt5 & \text{quantum} \end{cases}.
\end{equation}
\item the nine projectors, of the kind introduced in Ref.~\cite{Kurzynski:2012}, onto
\begin{equation}
	\begin{gathered}
	\qty(1,0,0),\quad \qty(0,1,0),\quad \qty(0,0,1),\quad \tfrac{1}{\sqrt2}\qty(0,1,-1),\quad \tfrac{1}{\sqrt3}\qty(1,0,-\sqrt2),\\
	\tfrac{1}{\sqrt3}\qty(1,\sqrt2,0),\quad \tfrac12\qty(\sqrt2,1,1),\quad \tfrac12\qty(\sqrt2,-1,-1),\quad \tfrac12\qty(\sqrt2,-1,1),
	\end{gathered}
\end{equation}
with thirteen orthogonal pairs and two complete bases, for which
\begin{equation}
	\mathbb{CNTXT}_9 = \sum_{i=1}^9 \expval{P_i} , \qquad \begin{cases} \mathbb{CNTXT}_9 \leq 3 & \text{non-contextual} \\ \mathbb{CNTXT}_9 \leq 10/3 & \text{quantum} \end{cases}.
\end{equation}
The non-contextual bound does not change when one requires, in addition, exactly one assignment equal to $1$ in each complete basis.
\end{itemize}
The orientation of the projectors with respect to the helicity basis of the $Z$ is a free choice. We optimize it: for a given state, the maximum of $\Tr[\rho\, U \qty(\sum_i P_i) U^\dagger]$ over the unitary transformations $U$ is obtained by aligning the eigenvectors of $\rho$ with those of $\sum_i P_i$ in the same order, and is $\sum_k \rho_k \pi_k$, with $\rho_1\leq\rho_2\leq\rho_3$ the eigenvalues of $\rho$ and $\pi_1\leq\pi_2\leq\pi_3$ those of $\sum_i P_i$. The latter are $\qty{\tfrac{5-\sqrt5}{2}, \tfrac{5-\sqrt5}{2}, \sqrt5}$ for the KCBS set and $\qty{\tfrac83, 3, \tfrac{10}{3}}$ for the nine projectors, so that, with $\Tr\rho = 1$~\cite{Fabbrichesi:2025ywl,Fabbrichesi:2025aqp},
\begin{equation}
	\mathbb{CNTXT}_5 = \frac{5-\sqrt5}{2} + \frac{3\sqrt5 - 5}{2}\, \rho_3 , \qquad \mathbb{CNTXT}_9 = 3 + \frac{\rho_3 - \rho_1}{3} .
	\label{eq:ctxclosed}
\end{equation}
The optimized witnesses depend only on the spectrum of the density matrix. The KCBS inequality is violated if and only if the largest eigenvalue exceeds $(5+\sqrt5)/10 \simeq 0.724$, and it is saturated by pure states. The nine-projector inequality is instead violated by every state but the maximally mixed one: in agreement with Ref.~\cite{Kurzynski:2012}, once the orientation is optimized, $\mathbb{CNTXT}_9 - 3$ measures only the spread of the spectrum and does not discriminate between states. We nonetheless report it, for comparison with the other analyses of contextuality in high-energy processes.

\section{Magic of a qutrit}
\label{app:magic}

\subsection{Magic}

We collect here the definitions used in Section~\ref{sub:magic}. Magic, or non-sta\-bi\-li\-zer\-ness, measures how far a state is from the stabilizer states, which can be prepared and transformed by Clifford operations and are efficiently simulated on a classical computer. It is the resource that allows  quantum computers to outperform their classical counterparts. For a qutrit, let $\omega = e^{2\pi i/3}$, $Z\ket{j} = \omega^j\ket{j}$ and $X\ket{j} = \ket{j+1}$, with $j+1$ taken modulo $3$, and define the displacement operators $D_{p,q} = \omega^{-2pq} Z^p X^q$, with $p,q \in \qty{0,1,2}$. We identify the helicities $(+1, 0, -1)$ with $(\ket{0}, \ket{1}, \ket{2})$, but any other ordering differs by a Clifford operation and gives the same results.

For a qutrit, there are 12 pure stabilizer states: the eigenstates of $Z, X, X Z$ and $X Z^2$, namely $|j\rangle$ and $\left(|0\rangle+\omega^a|1\rangle+\omega^b|2\rangle\right) / \sqrt{3}$.
Their convex hull $\mathcal{S}$ is the set of all classical mixtures,
\begin{equation}
\mathcal{S}=\left\{\sum_s p_s|s\rangle\langle s|: p_s \geq 0, \sum_s p_s=1\right\}:
\end{equation}
pick stabilizer state $s$ with probability $p_s$ and forget which one was picked. Geometrically, a qutrit state has 8 real parameters, so states are points in an 8-dimensional space. The 12 stabilizer states are 12 corners, and taking all mixtures fills in a solid with flat faces, a polytope. It sits inside the larger, curved set of all states, contains $\mathbbm{1} / 3$ near its center, and excludes every pure non-stabilizer state.

States in $\mathcal{S}$ can be made from stabilizer states and classical randomness, and stabilizer circuits acting on them are efficiently classically simulable due to the Gottesman--Knill theorem~\cite{Gottesman:1998hu, Gottesman:1998se}. Non-stabilizer, or magic, states are the resource that makes such circuits universal, via magic-state distillation~\cite{Bravyi:2004isx, Anwar:2012qut}. A state is magic iff $\rho \notin \mathcal{S}$.

\subsection{The discrete Wigner function}

The discrete Wigner function of Gross~\cite{Gross:2006} is
\begin{equation}
	W_\rho(p,q) = \frac{1}{3} \Tr\qty[\rho\, A_{p,q}] , \qquad A_{0,0} = \frac13 \sum_{p,q} D_{p,q} , \quad A_{p,q} = D_{p,q} A_{0,0} D_{p,q}^\dagger .
\end{equation}
The nine phase-point operators $A_{p,q}$ have entries $\delta_{j+k\equiv2u}\,\omega^{v(j-k)}$ for a relabeling $(p,q)\to(u,v)$ of the phase-space points. The Wigner function then takes the form
\begin{equation}
W_\rho(u,v)=\tfrac13\Big[\rho_{uu}+2\,\Re\big(\rho_{u+1,u-1}\,\omega^{v}\big)\Big],\qquad u,v\in\{0,1,2\},
\end{equation}
with indices modulo 3. Each value is one population plus the interference of the other two helicity states. For a real density matrix only two values per row are distinct, $(\rho_{uu}+2\rho_{u+1,u-1})/3$ and $(\rho_{uu}-\rho_{u+1,u-1})/3$.

\subsection{Mana}

The mana~\cite{Veitch:2014} is
\begin{equation}
	\mathbb{M}(\rho) = \ln \sum_{p,q} \abs{W_\rho(p,q)} .
\end{equation}
It vanishes for every mixture of stabilizer states, does not increase under Clifford operations and can be computed exactly for mixed states. Mana, however, depends on the basis used to describe the qutrit and the stabilizer states. In Section~\ref{sub:magic} we use the helicity basis with the Condon--Shortley conventions in Eq.~\eqref{eq:polvectors}. The phases matter: the diagonal Clifford operations of a qutrit multiply the basis states by powers of $\omega$, up to a global phase, so that a relative sign between helicity states, as in the alternative convention of Appendix~\ref{app:conventions}, is not a Clifford operation and changes the mana.

With the other common choice of polarization vectors discussed in Appendix~\ref{app:conventions}, the density matrix is conjugated by ${\rm diag}(-1,1,1)$, which leaves every observable unchanged but is not a Clifford operation for a qutrit: the mana at $\cos\theta = 0.5$ then becomes $0$ instead of $0.135$ at $240\GeV$ and $0.099$ instead of $0.176$ at $365\GeV$. More generally, with this choice the mana is essentially unchanged in the backward hemisphere, $\cos\theta < 0$, but reduced in the forward one, where it vanishes altogether at $240\GeV$. The symmetry in $\cos\theta$ shown in Figure~\ref{fig:magic} is then a property of the convention utilized for the helicity basis, while the maximum values over the angle are the same in the two conventions. The statements that do not depend on the basis choice are that the inclusive state has no magic and that $\Im b_Z$ and $\Re\tilde b_Z$ enter only at second order. The values of the mana at fixed angle should be read as referring to the convention of Eq.~\eqref{eq:polvectors}. The same holds for the robustness of magic: at $240\GeV$ and $\theta = \pi/2$, for instance, it equals $1$ with our convention and $1.43$ with the other.

Another quantity routinely linked to magic is the stabilizer R\'enyi entropy~\cite{Leone:2022}. Its expression for a pure state and its usual extension to mixed states are
\begin{equation}
	M_2 = -\ln \qty[\frac13 \sum_{p,q} \abs{\Tr(\rho D_{p,q})}^4] , \qquad \tilde M_2 = -\ln \frac{\sum_{p,q} \abs{\Tr(\rho D_{p,q})}^4}{\sum_{p,q} \abs{\Tr(\rho D_{p,q})}^2} ,
\end{equation}
which coincides with $M_2$ for pure states. This quantity, however, does not vanish on mixtures of stabilizer states: the state ${\rm diag}(1/2, 1/2, 0)$, for instance, has $\tilde M_2 = 0.29$. For mixed states it therefore tracks the mixedness of the state as well as its magic.

\subsection{Robustness of magic}\label{app:robustness}

In the definition of the robustness of magic~\cite{Howard:2016lyb}, Eq.~\eqref{eq:robustness}, the weights sum to one, so that $\mathcal R=1+2\times$(total negative weight). Hence $\mathcal R=1$ if and only if $\rho$ is a mixture of stabilizer states, and $\mathcal R^2$ sets the cost of simulating $\rho$ classically by quasi-probability sampling. Notice that $\mathcal R$ is the solution of a linear optimization problem with 24 non-negative variables and nine real equality constraints.

Decomposing $W_\rho=\sum_sx_sW_{|s\rangle}$ with $W_{|s\rangle}\ge0$ gives $\mathbb{M}\le\ln\mathcal R$. The mana vanishes on the whole region $W_\rho\ge0$, which strictly contains the stabilizer polytope, while $\mathcal R = 1$ only on the polytope. States in between have bound magic~\cite{Veitch:2012ihl}. 
The stabilizer polytope has 81 facets, all at Hilbert--Schmidt distance $1/(2\sqrt6)$ from $\mathbb 1/3$. This gives the purity bound below which no qutrit state is magic.

It is worth noticing the relationship between Wigner positivity, magic and mana.
For all states, we have that $\rho \in \mathcal{S} \Rightarrow W_\rho \geq 0$. Hence, if some $W_\rho<0$ then $\rho$ is magic, and a positive mana always proves magic. On the other hand, the converse $W_\rho \geq 0 \Rightarrow \rho \in \mathcal{S}$ holds for pure states due to the discrete Hudson theorem~\cite{Gross:2006}, but it fails for mixed states~\cite{Veitch:2012ihl}. Geometrically, $\{W \geq 0\}$ is the state space cut by 9 half-spaces $W(u, v) \geq 0$, a region strictly larger than $\mathcal{S}$. States in the difference are magic with zero mana, or bound magic states. The following table thus proposes a course of action to detect magic.

\begin{center}
\begin{tabular}[t]{|l|l|l|}
\hline what is found & pure state & mixed state \\
\hline some $W<0$ (mana $>0$) & magic & magic \\
\hline all $W \geq 0$ (mana $=0$) & not magic & undecided: check $\mathcal{R}$ \\
\hline
\end{tabular}
\end{center}

\section{Loop functions}  
\label{app:loopfunctions}

The one-loop integrals are normalized as
\begin{equation}
	\frac{(2\pi\mu)^{4-D}}{i\pi^2}\int \dd[D]{\ell}\, \frac{\qty{1,\; \ell^\mu,\; \ell^\mu \ell^\nu}}{\qty[\ell^2 - m_1^2]\qty[(\ell + p_1)^2 - m_2^2]\qty[(\ell + p_2)^2 - m_3^2]} = \qty{C_0,\; C^\mu,\; C^{\mu\nu}},
\end{equation}
with $C^\mu = p_1^\mu C_1 + p_2^\mu C_2$ and $C^{\mu\nu} = g^{\mu\nu} C_{00} + \sum_{ij} p_i^\mu p_j^\nu C_{ij}$, and the arguments written as $C_{\ldots}(p_1^2, (p_1 - p_2)^2, p_2^2; m_1^2, m_2^2, m_3^2)$. The functions $B_0(p^2; m_1^2, m_2^2)$ and $A_0(m^2)$ are defined analogously with two and one propagators. This is the convention of {\tt FeynCalc} and, with the same ordering of the arguments, of {\tt LoopTools}, so that the expressions of the main text can be evaluated directly. The ultraviolet divergent functions are $A_0$, $B_0$ and $C_{00}$, with the pole parts $m^2\Delta$, $\Delta$ and $\Delta/4$, $\Delta = 1/\epsilon - \gamma_E + \log 4\pi$. All the others are finite. In the main text and below, the symbols $A_0$, $B_0$ and $C_{00}$ denote the finite parts, in any fixed convention: since the poles cancel in $c(s)$ and $\beta(s)$, so do the constants and the $\log\mu^2$ that accompany them.

For the triangle diagram with two Higgs bosons and a $Z$ boson in the loop, the arguments are those of the external momenta:
\begin{equation}
	C_{1,11,12} = C_{1,11,12}(\MH^2, s, \MZ^2; \MH^2, \MH^2, \MZ^2), \qquad C_{0,00} = C_{0,00}(\MH^2, \MZ^2, s; \MH^2, \MH^2, \MZ^2),
\end{equation}
the two orderings being related by the symmetries of the functions. The function $B_0 = B_0(\MH^2; \MH^2, \MH^2)$ is the two-point function of the bubble diagram, evaluated at the Higgs boson mass.

\section{The form factors in terms of scalar integrals}  
\label{app:reduced}

Reducing the tensor coefficients of Eqs.~\eqref{eq:beta} and~\eqref{eq:c} to scalar integrals introduces in the denominators the Gram determinant of the two independent momenta of the triangle, $p_1^2 p_2^2 - (p_1\vdot p_2)^2 = -\frac14\lam(p_1^2, p_2^2, (p_1 - p_2)^2)$, which for the external momenta of our triangle is $-\lam/4$ with the K\"all\'en function $\lam$ of Eq.~\eqref{eq:lambda}, and the six scalar functions
\begin{gather}
	A_0^{(h)} = A_0(\MH^2), \qquad A_0^{(Z)} = A_0(\MZ^2), \qquad B_0^{(hhh)} = B_0(\MH^2; \MH^2, \MH^2), \nn\\
	B_0^{(ZhZ)} = B_0(\MZ^2; \MH^2, \MZ^2), \qquad B_0^{(s)} = B_0(s; \MH^2, \MZ^2), \qquad C_0 = C_0(\MH^2, \MZ^2, s; \MH^2, \MH^2, \MZ^2) .
\end{gather}
The result is
\begin{align}
	\beta(s) = \frac{3 g^2 \MH^2}{64 \pi^2 \cw^2 \lam^2}\Big[ & -\lam\, \qty(\MH^2 - \MZ^2 - s) - 2\lam\, \qty(A_0^{(h)} - A_0^{(Z)}) + P_{hhh}\, B_0^{(hhh)} \nn\\
	& - \qty(\qty(\MH^2 + 4 \MZ^2)\lam + 12 \MH^2 \MZ^2 s)\, B_0^{(ZhZ)} - \qty(\MH^2 - \MZ^2 - s)\qty(\lam + 6 \MH^2 s)\, B_0^{(s)} \nn\\
	& - 2 \MH^2 \qty(\MH^2 - \MZ^2 - s)\qty(2\lam + 3 \MH^2 s)\, C_0 \Big],
	\label{eq:beta_reduced}
\end{align}
with $P_{hhh} = 4 \MH^6 - \MH^4\qty(7 \MZ^2 + 3 s) + 2 \MH^2 \MZ^2 \qty(\MZ^2 - s) + \qty(\MZ^2 - s)^3$, and
\begin{align}
	c(s) = \frac{3 g^2 \MH^2}{64 \pi^2 \cw^2 \MZ^2 \lam^2}\Big[ & -2 s \MZ^2 \lam - \qty(\MH^2 - \MZ^2 - s)\lam\, \qty(A_0^{(h)} - A_0^{(Z)}) + \qty(\MH^2 - \MZ^2 + s)\, Q\, B_0^{(hhh)} \nn\\
	& - 2 \MZ^2 \qty(\MH^2 - \MZ^2 - s)\qty(\lam + 3 \MH^2 s)\, B_0^{(ZhZ)} - 2 s\, Q\, B_0^{(s)} \nn\\
	& - \qty(2 \MZ^2 \lam^2 + \MH^2 R)\, C_0 \Big],
	\label{eq:c_reduced}
\end{align}
with
\begin{equation}
	Q = \qty(\MH^2 + \MZ^2)\lam + 6 \MH^2 \MZ^2 s, \qquad R = \lam\qty(\lam + 2 \MH^2 s + 8 \MZ^2 s) + 12 \MH^2 \MZ^2 s^2 .
\end{equation}
The ultraviolet poles cancel separately in each form factor between the $A_0$ and $B_0$ terms, as one can verify by inserting $A_0^{(i)} \to m_i^2\Delta$ and $B_0 \to \Delta$ in the square brackets. The $1/\lam^2$ is the usual artifact of the reduction: at threshold, $\lam\to 0$, the scalar functions conspire to keep $\beta(s)$ and $c(s)$ finite (cf.\ Figure~\ref{fig:formfactors}), but Eqs.~\eqref{eq:beta}--\eqref{eq:c} are the forms to use numerically close to it.


\section{Validation}  
\label{app:validation}

\begin{figure}[h]
\centering
\includegraphics[width=\textwidth]{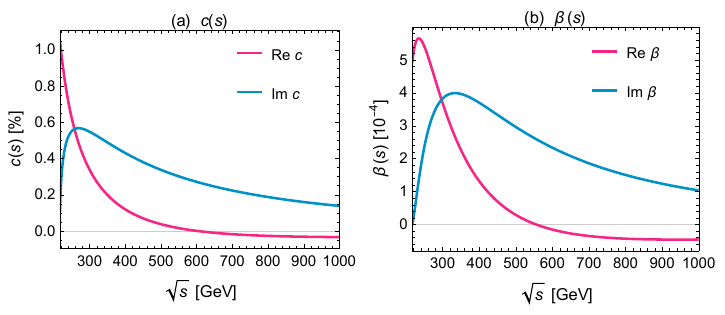}
\caption{(a) The form factor $c(s)$ and (b) the induced $b_Z$ per unit of $\klam$, $\beta(s)$, of the three diagrams containing the Higgs boson self-coupling, as functions of the center-of-mass energy.}
\label{fig:formfactors}
\end{figure}

\emph{The form factors $c(s)$ and $\beta(s)$.} Figure~\ref{fig:formfactors} shows the two form factors of the three diagrams involving the Higgs boson self-coupling, Eqs.~\eqref{eq:beta} and~\eqref{eq:c}, from the $Zh$ threshold to $1\,$TeV. Table~\ref{tab:numbers} gives their values at the energies of interest. Two checks validate the numerical evaluation with {\tt LoopTools}. The numbers do not depend on the renormalization scale $\mu$ of the loop functions, as they must since the poles cancel, which we verified by changing $\mu$ by orders of magnitude. Moreover, the direct evaluation of the tensor coefficients agrees with the reduced expressions of Appendix~\ref{app:reduced}, in terms of scalar integrals only, to machine precision.

\begin{figure}[h]
\centering
\includegraphics[width=0.5\textwidth]{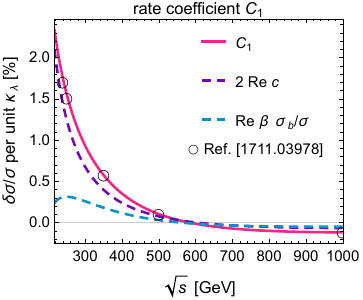}
\caption{The coefficient $C_1$ of the cross section, Eq.~\eqref{eq:C1}, with its two components and, as circles, the values of Ref.~\cite{DiVita:2017vrr}.}
\label{fig:C1}
\end{figure}

\emph{The rate coefficient $C_1$.} The coefficient $C_1$ of Eq.~\eqref{eq:C1} is the one introduced in Ref.~\cite{Degrassi:2016wml} and tabulated for Higgsstrahlung in Ref.~\cite{DiVita:2017vrr}, up to the Higgs wave-function renormalization, which is quadratic in $\klam$: the $h$ loop in the Higgs boson self-energy gives
\begin{equation}
	\delta Z_h = -\frac{9 g^2 \MH^2}{128 \pi^2 \MW^2}\qty(\frac{2\pi}{3\sqrt 3} - 1) = -1.49\times 10^{-3},
	\label{eq:dZh}
\end{equation}
a universal factor $\qty(1 + \klam^2 \delta Z_h/2)$ on the amplitude that does not touch the polarization. In the notation of Refs.~\cite{Degrassi:2016wml,DiVita:2017vrr}, $\sigma = Z_H\, \sigma_{ 0}\qty(1 + \klam C_1)$ with $Z_H = (1 - \klam^2 \delta Z_h)^{-1}$, so that around the SM point
\begin{equation}
	\frac{\sigma}{\sigma_{\rm SM}} - 1 = \qty(C_1 + 2\, \delta Z_h)\, \delta\klam + \order{\delta\klam^2}, \qquad \delta\klam = \klam - 1 .
\end{equation}
Table~\ref{tab:C1} compares our $C_1$ with the published values. The agreement is at the level of the quoted digits over the whole energy range, including the sign change above $500\GeV$. The residual difference is the input scheme, the published numbers being in the $G_\mu$ scheme in which the prefactor $g^2/\cw^2 = 4\sqrt 2 G_\mu \MZ^2$ is $3.6\%$ larger than our $\alpha(0)$ value, as is $\delta Z_h$ ($-1.54\times 10^{-3}$ there). Figure~\ref{fig:C1} shows $C_1$ and its two components as functions of the energy.

\begin{table}[h]
\centering
\begin{tabular}{lccccc}
\toprule
$\sqrt s$ [GeV] & $2\Re c$ [\%] & $\Re\beta\,\sigma_b/\sigma_{ 0}$ [\%] & $C_1$ [\%] & $C_1 + 2\delta Z_h$ [\%] & $C_1$ of Ref.~\cite{DiVita:2017vrr} [\%] \\
\midrule
220  & 1.98 & 0.25 & 2.24 & 1.94 & 2.2 (threshold) \\
240  & 1.43 & 0.31 & 1.74 & 1.44 & 1.7 \\
250  & 1.24 & 0.31 & 1.55 & 1.26 & 1.5 \\
350  & 0.40 & 0.18 & 0.58 & 0.28 & 0.57 \\
365  & 0.34 & 0.16 & 0.50 & 0.20 & -- \\
500  & 0.07 & 0.03 & 0.10 & $-0.20$ & 0.099 \\
1000 & $-0.07$ & $-0.05$ & $-0.12$ & $-0.42$ & $-0.12$ \\
\bottomrule
\end{tabular}
\caption{The shift of the $e^-e^+\to ZH$ cross section per unit of $\klam$, Eq.~\eqref{eq:C1}, split into its two components, the same including the wave-function renormalization of the Higgs boson, and the values of Ref.~\cite{DiVita:2017vrr} ($G_\mu$ scheme).}
\label{tab:C1}
\end{table}

\emph{The complete one-loop correction.} The weak correction to the rate that comes with the computation of Section~\ref{sub:fullloop} is $+3.2\%$ at $240\GeV$ and $+3.0\%$ at $365\GeV$ in the $\alpha(0)$ scheme, the photonic corrections to the initial state excluded. This is the ``pure weak'' correction of the literature. With the inputs of Refs.~\cite{Bondarenko:2018sgg,Arbuzov:2021fmd} ($\MW = 80.4515\GeV$, $m_t = 173.8\GeV$, their light-quark masses) and the one-loop conversion $\delta_{G_\mu} = \delta_{\alpha(0)} - 2\Delta r$ we find $-2.98\%$ in the $G_\mu$ scheme at $240\GeV$ against their $-2.99\%$, and $+2.74\%$ in the $\alpha(0)$ scheme against their $+2.72\%$. The one-loop $\Delta r$ assembled from the renormalization constants of the same run is $0.0286$ for their inputs and $0.0303$ for ours.

\end{appendices}


\bibliographystyle{jhep}   
\bibliography{bib}

\end{document}